\documentclass[english]{article}
\usepackage[T1]{fontenc}
\usepackage[latin9]{inputenc}
\usepackage{geometry}
\usepackage[dvips]{color}
\usepackage{psfrag}
\usepackage{hyperref}
\usepackage{graphicx}
\usepackage{babel}
\usepackage{epsfig}
\usepackage{amsmath,amsfonts}
\usepackage{amssymb}
\usepackage{mathrsfs}
\usepackage{feynmp}

\def\bvev#1{\Big\langle #1\Big\rangle}

\DeclareMathOperator{\Kop}{K}
\DeclareMathOperator{\Rop}{R}

\usepackage{ybymac1}

\usepackage{ybymac2}

\begin{document}

\begin{center}

{\hfill HU-MATH-2013-16, HU-EP-13/46 \\
\hfill ITP-UU-13/23, Spin-13/16 \\
\hfill ITP-Budapest-Report 663 \\
\hfill UMTG-279}

\vskip 12mm

\title{The spectrum of tachyons in $AdS$/CFT}

\vskip 12mm

\renewcommand{\thefootnote}{$\alph{footnote}$}

Zolt\'an Bajnok$^a$,
Nadav Drukker$^b$,  
\'Arp\'ad Heged\H us$^a$,
Rafael I. Nepomechie$^c$,
\\L\'aszl\'o Palla$^d$,
Christoph Sieg$^e$,
and Ryo Suzuki$^f$,

\vskip 3mm
{\it
$^a$ MTA Lend\" ulet Holographic QFT Group, Wigner Research Centre,
H-1525 Budapest 114, P.O.B. 49, Hungary \\
$^b$Department of Mathematics, King's College,
The Strand, WC2R 2LS, London, UK\\
$^c$Physics Department, P.O. Box 248046, University of Miami, 
Coral Gables, FL 33124, USA\\
$^d$Institute for Theoretical Physics, Roland E\"otv\"os University, 
1117 Budapest, P\'azm\'any s. 1/A Hungary\\
$^e$Institut f\"ur Mathematik und Institut f\"ur Physik, 
Humboldt-Universit\"at zu Berlin, 
\\IRIS-Haus, Zum Gro{\ss}en Windkanal 6, 12489 Berlin, Germany\\
$^f$Mathematical Institute, University of Oxford, Andrew Wiles Building, Radcliffe Observatory Quarter,
\\Woodstock Road, Oxford, OX2 6GG, UK
}
\vskip3mm
\small{\href{mailto:bajnok.zoltan@wigner.mta.hu}{\tt bajnok.zoltan@wigner.mta.hu},
\quad\href{mailto:nadav.drukker@gmail.com}{\tt nadav.drukker@gmail.com},
\quad\href{mailto:hegedus.arpad@wigner.mta.hu}{\tt hegedus.arpad@wigner.mta.hu},
\quad\href{mailto:nepomechie@physics.miami.edu}{\tt nepomechie@physics.miami.edu},
\quad\href{mailto:palla@ludens.elte.hu}{\tt palla@ludens.elte.hu},
\quad\href{mailto:sieg@math.hu-berlin.de}{\tt sieg@math.hu-berlin.de},
\quad\href{mailto:Ryo.Suzuki@maths.ox.ac.uk}{\tt Ryo.Suzuki@maths.ox.ac.uk}}

\renewcommand{\thefootnote}{\arabic{footnote}}
\setcounter{footnote}{0}

\end{center}

\vskip 12mm

\abstract{
\normalsize
\noindent

We analyze the spectrum of open strings stretched between a D-brane and an anti-D-brane 
in planar $AdS$/CFT using various tools. We focus on open strings ending on two giant 
gravitons with different orientation in $AdS_5\times S^5$ and study the 
spectrum of string excitations using the following approaches: open spin-chain, 
boundary asymptotic Bethe ansatz and boundary thermodynamic Bethe ansatz (BTBA). 
We find agreement between a perturbative high order diagrammatic calculation  in $\cN=4$ SYM 
and the leading finite-size boundary L\"uscher correction. We study the ground state energy of 
the system at finite coupling by deriving and numerically solving a set of BTBA equations. 
While the numerics give reasonable results at small coupling, they break down at finite coupling 
when the total energy of the string gets close to zero, possibly indicating that 
the state turns tachyonic. 
The location of the breakdown is also predicted analytically.
}

\newpage

\setcounter{tocdepth}{2}
{\addtolength{\parskip}{-1ex}
\tableofcontents}

\section{Introduction}
\label{sec:intro}

Tachyons are ubiquitous in string theory. The ground state of the bosonic string is 
tachyonic, and even for superstrings the tachyons are removed from the spectrum 
only by a carefully chosen GSO projection \cite{gso}. 
The understanding of these tachyonic states has undergone a revolution in the last 15 years. The 
tachyons arise from an expansion around non-minimal saddle points, and in many instances 
the instability that the tachyons represent has been understood and the endpoint of tachyon 
condensation has been identified.

This is particularly true for tachyons in the open-string spectrum, which represent instabilities 
of the D-branes on which they end, rather than of space-time itself. 
For the bosonic string tachyon condensation removes the D-branes and eliminates open strings 
altogether from the spectrum, which was shown by using off-shell cubic string field theory both 
numerically and analytically
\cite{Sen:1999mh, Sen:1999nx, Schnabl:2005gv}. 
Similar considerations led to the understanding of D-brane charges in superstring theory in terms 
of K-theory \cite{Sen:1998sm,Witten:1998cd}: In addition to the usual stable BPS D$p$-branes 
in type IIA and IIB string theories there are unstable ones 
``of the wrong dimensionality'' (odd $p$ in IIA and even $p$ in IIB). Similarly the coincident 
D-brane--anti-D-brane (henceforth D-$\bar{\text{D}}$) system
includes tachyonic states from the strings connecting the two 
\cite{Banks:1995ch}. 
Both are examples of open superstring systems undergoing the wrong GSO projection.

In this paper we study the coincident \DDbar\ system 
within the $AdS$/CFT correspondence. In flat space, when the two D-branes are 
coincident, the ground state of the open string connecting them is tachyonic with a 
mass-squared $-1/(2\alpha')$. If the D-branes are not coincident this mass squared 
is increased and beyond a distance of order the string length all states become massive.

Unstable D-brane systems have been studied within the $AdS$/CFT correspondence initially 
in \cite{dgi}. 
In certain cases it is possible to match instabilities in the field theory to those in string theory. 
Generically, these systems are not amenable to perturbative calculations in either, let 
alone both, weak and strong coupling. In special circumstances it has been possible to take a 
scaling limit to get a match between weak and strong coupling \cite{Gibbons:2009hi, Pomoni:2010et}. 
Here we study an unstable system beyond such a limit.%
\footnote{It is possible in this case too to expand in small angles $\theta_1$ and $\theta_2$ 
defined below.}

Our study relies on the integrability of the \AdSxS\ superstring, 
a property which is conjectured 
to hold beyond the classical string limit. Integrability has led to a great understanding of 
the spectrum of closed string states in $AdS_5\times S^5$,
as well as certain open-string sectors which are conjectured to be integrable.
These correspond 
to strings ending on different types of D-branes
\cite{DeWolfe:2004zt, Berenstein:2005vf, Hofman:2007xp, Correa:2008av, CRY} 
as well as macroscopic open strings extending to the boundary of space and 
representing Wilson loops in the dual 4d gauge theory 
\cite{dk-spinchain, Dru-int-WL, CMS}.
The most studied case is that of a  ``giant graviton'', which is a D3-brane 
carrying $N$ units of angular momentum on $S^5$ \cite{McGreevy:2000cw}.

Integrability of the giant graviton systems can be seen from both sides of the
$AdS$/CFT correspondence.  From the gauge theory side, the dilatation
matrix -- calculated perturbatively -- coincides with an integrable
open spin chain Hamiltonian \cite{Berenstein:2005vf, Hofman:2007xp}.
From the string theory side, integrability is a consequence of the
fact that the classical two-dimensional sigma model with boundary
admits a Lax pair formulation, which leads to an infinite number of
conserved charges \cite{Mann:2006rh, Dekel:2011ja}.

Instead of a single D-brane we consider here a pair of coincident D-branes
with arbitrary orientation. When the two orientations are identical 
this is a BPS system, and when opposite,  
this is a D-$\bar{\text{D}}$ system.
Thanks to integrability, we can compute the asymptotic spectrum of open 
strings on these D-branes by solving the Bethe Ansatz equations with boundaries. To 
be more specific, let us choose the reference ground state of the Bethe Ansatz 
as $Z^L$. There are two important orientations of the D-brane, one carrying the 
angular momentum on $S^5$ in the same direction 
(``$Z=0$ giant graviton''), or the other in a perpendicular direction (``$Y=0$ giant graviton'') 
\cite{Hofman:2007xp}. The 
names reflect the fact that the world-volume of the D3-brane is embedded as 
$S^3\subset S^5\subset \bC^3$, where $\bb{C}^3$ is
parameterized by complex $X,Y,Z$ coordinates satisfying $|X|^2+|Y|^2+|Z|^2=1$. 
In our problem, one brane satisfies $Y=0$ while the other
an arbitrary linear equation 
involving $Y$, $\bar Y$, $X$ and $\bar X$, which we call $\hat Y=0$ with
\beq
\label{y-hat}
\begin{aligned}
\hat Y
=Y\cos\theta_1\cos\theta_2 - X\cos\theta_1\sin\theta_2
+\bar X\sin\theta_1\cos\theta_2 + \bar Y\sin\theta_1\sin\theta_2 \,.
\end{aligned}
\eeq
We will mainly concentrate on the \DDbar\ system, which corresponds 
to $\theta_1=\theta_2=\pi/2$, but many of the calculations can be generalized to arbitrary 
angles.

In the next section we discuss the gauge theory dual of these operators. The $Y=0$ giant graviton is 
a determinant operator made of $N$ of the $Y$ scalar fields 
\cite{Balasubramanian:2001nh}. An open string attached to it is obtained 
by replacing one of the $Y$ fields with an adjoint-valued word made of other fields 
(and covariant derivatives). 
The system studied here should involve two determinants connected by a pair of adjoint valued words 
with mixed indices. A single adjoint valued word can replace one of the letters in one
determinant, but not both, which is why two words are required. 
The dual statement in string 
theory is that a compact D-brane cannot support a single open string, due to the Gauss law 
constraint and must have an even number of strings (with appropriate orientations) attached. 
In the planar approximation the two open strings should not interact, which we verify 
in the gauge theory calculation in the next section. Hence we can consider the spectral 
problem independently for each of the insertions/open strings.

The exact gauge theory description of the \DDbar\ system is not known and it requires solving a mixing problem which is quite complicated, because the operator consists of more than $N$ fields.
At tree level, an orthogonal basis of gauge-invariant scalar operators is constructed by the Brauer algebra \cite{Kimura:2007wy} or the restricted Schur polynomial \cite{Bhattacharyya:2008rb} at any $N$. At loop level, little is known about how to find dilatation eigenstates in the \DDbar\ system using these bases \cite{Kimura:2009jf}.
Nevertheless, the mixing problem of our interest seems to simplify at large $N$. We expect that the \DDbar\ system with or without open strings has the gauge theory dual closely resembling the double determinant.

The mass of the (potentially tachyonic) open-string state should correspond to the dimension of the local operator, or more precisely, the contribution to the dimension from the insertion of the word into the determinant operators as discussed in Section \ref{sec:pert}. 
In the case of the $Y=0$ brane, the insertion of the word $Z^L$ corresponding to 
the ground state of the open string gives a protected operator. 
The system with $Y$, $\bar Y$ 
and $Z^L$ is not protected and we expect that the ground state energy is lifted by 
`wrapping type' graphs which involve the interaction between the $Y$ and $\bar Y$ 
fields at the two boundaries of the word. We identify a set of such graphs 
at order $\lambda^{2L}$ in perturbation theory, which we conjecture to be the first ones to
contribute to the anomalous dimension of these operators. 
In fact, the leading non-vanishing wrapping correction coming from the integrability 
formulation derived in Section~\ref{sec:aba} is exactly of order $\lambda^{2L}$ 
and equal to the UV divergences of the integrals that arise from these graphs.
The UV-divergences of these 
integrals were recently proven to agree with our conjecture \cite{Schnetz:2012nt}.

In the integrable description we identify how the string excitations scatter off from
the D-branes. Combining these reflection factors from both ends of the open string, 
we analyze the finite volume spectrum of excitations via the double-row transfer matrix.
Eigenvalues of this matrix
provide the large $L$ anomalous dimensions together with their leading finite-size 
L\"uscher corrections.

At strong coupling we expect the properties of the open string to be rather similar to those in flat space, and therefore there should be a tachyon in the 
spectrum. The mass-squared of the ground state of the open string in flat space 
is $m^2=-1/(2\alpha')$, which translates to $-\sqrt\lambda/2$ in units of the $AdS$ 
curvature radius.%
\footnote{If we compare to the mass-squared of the Konishi operator $4\sqrt\lambda$, 
\cite{Gromov:2009zb,Frolov:2010wt} 
which matches the first excited closed string state in flat space, a factor of 4 arises 
from replacing closed strings by open strings, and another factor of 2 from taking the 
wrong GSO projection.} 
In the case of arbitrary angles $\theta_1$, $\theta_2$ \cite{Epple:2003xt} 
the expression becomes%
\footnote{Note that for $\theta_1=0$ the rotation in \eqn{y-hat} mixes only 
$Y$ and $X$, so the ground state $Z^L$ is still BPS, hence the mass is zero. 
Likewise for $\theta_2=0$.}
\beq
m^2=-\frac{|\theta_1+\theta_2|-|\theta_1-\theta_2|}{2\pi} \sqrt\lambda \,.
\eeq

The dimension of the operator inserted in the determinants is dominated by the 
charge $L$ at weak coupling, but at strong coupling it should asymptote 
to $m \propto i\sqrt[4]{\lambda}$.
We therefore expect the dimension to turn imaginary at a finite value of the coupling.
To probe this transition we employ boundary 
thermodynamic Bethe ansatz equations (BTBA).

BTBA was derived first in \cite{LeClair:1995uf} for models with diagonal S-matrices.
If the S-matrix is non-diagonal it is difficult to construct BTBA explicitly by applying 
the methods of \cite{LeClair:1995uf}. 
In specific cases, however, it is possible
to overcome the appearing technical problems and derive a BTBA with
non-diagonal S-matrices, as was done for example in \cite{Dru-int-WL, CMS}. 
As our case is more complicated, 
the approach we take here is to use the Y-system equations together with 
their analytic properties to derive the BTBA, as was done in \cite{Pearce:2000dv}.
In Section~\ref{sec:BTBA} we apply this method (following \cite{Balog:2011nm}) to derive a set of BTBA
for the ground state of the $\theta_1=\theta_2=\pi/2$ case.

We develop numerical algorithms to solve these equations and evaluate the anomalous dimensions 
of ground states with different values of $L$ at finite coupling. In all cases we find that 
the anomalous dimension is a monotonously decreasing function of the coupling.
However, when the BTBA energy becomes comparable to $1-L$, namely when the total energy of the open string 
gets close to zero, it becomes extremely difficult to obtain the precise value of the energy from BTBA solutions.
As a result, the evolution of the energy cannot be traced further toward strong coupling.

Such a pathological behavior can arise for states with negative anomalous dimension.
A novel lower bound for the BTBA energy is derived analytically, and the 
violation of this bound makes the BTBA solution inconsistent.
We expect this breakdown 
to signal the transition of the states from 
massive at weak coupling 
to tachyonic beyond the critical value of the coupling.
Beyond this singular point another formalism must be employed to find a continuation 
of the BTBA equations, whose details are beyond the scope of this paper.

\section{The $Y\bar{Y}$ brane system in gauge theory}
\label{sec:pert}

The $Y=0$ giant graviton is described in the gauge theory by a determinant 
operator \cite{Balasubramanian:2001nh}
\begin{equation}
{\cal O}_Y=\det Y=\epsilon^{a_1\cdots a_N}_{b_1\cdots b_N}
Y_{a_1}^{b_1}\cdots Y_{a_N}^{b_N}
\end{equation}
where $a_i$ and $b_i$ are color indices and $\epsilon$ is a product of two 
regular epsilon tensors
$\epsilon^{a_1\cdots a_N}_{b_1\cdots b_N}=
\epsilon^{a_1\cdots a_N}\epsilon_{b_1\cdots b_N}$.

An open string ending on the giant graviton is described by replacing one $Y$ 
with an adjoint valued local operator $\cW$ \cite{Balasubramanian:2002sa}
\begin{equation}
{\cal O}_Y^{\cW}=\epsilon^{a_1\cdots a_N}_{b_1\cdots b_N}
Y_{a_1}^{b_1}\cdots Y_{a_{N-1}}^{b_{N-1}}\cW^{b_N}_{\ a_N}\,.
\end{equation}
The simplest insertion is the vacuum $\cW=Z^L$.

One can consider also two giant gravitons by taking the combination 
${\cal O}_Y{\cal O}_Y$ and likewise add an open string attached to one or to the 
other. But with two giant gravitons we can also consider strings stretched between 
the two D-branes. Having a single such string is impossible, though. The endpoint of 
a string serves as a source of charge on the D-brane world-volume, which is compact, 
and there must be another charge source with the opposite sign. We 
therefore will consider the case of a pair of open strings with opposite orientation 
connecting the two D-branes.

The gauge theory description of this system is the double-determinant operator 
with all fields at a single point
\begin{equation}
\label{YY}
{\cal O}_{Y,Y}^{\cW, {\cV}}=
\epsilon^{a_1\cdots a_N}_{b_1\cdots b_N}
Y_{a_1}^{b_1}\cdots Y_{a_{N-1}}^{b_{N-1}}
\,\epsilon^{c_1\cdots c_N}_{d_1\cdots d_N}
Y_{c_1}^{d_1}\cdots Y_{c_{N-1}}^{d_{N-1}}
\cW^{d_N}_{\ a_N}{\cV}^{b_N}_{\ c_N}
\end{equation}
so one $Y$ was removed from each determinant and then the two 
words $\cW$ and ${\cV}$ are inserted with the indices crossed.

However, we are not interested here in the case with two identical D-branes.
That configuration is BPS and the spectrum of open strings also includes BPS states, 
which belong to the multiplet of the non-abelian gauge fields on the pair of D-branes. 
We want to study 
instead the spectrum of strings stretching between a D-brane and an anti D-brane.

The anti D-brane can be realized by replacing all the $Y$ by $\bar Y$, and we shall 
call it a $\bar Y=0$ giant graviton, as opposed to the original $Y=0$ giant graviton. 
If we parameterize the $S^5$ part of the target space by three complex 
coordinates $Y$, $X$ and $Z$ subject to $|X|^2+|Y|^2+|Z|^2=1$, then the $Y=0$ 
giant graviton wraps an $S^3$ given by $Y=0$. The $\bar Y=0$ brane will 
wrap the same $S^3$ but with the opposite orientation.

The precise gauge theory dual of the coincident maximal $Y\bar{Y}$ branes is not known. 
We will approximate it by a double determinant, one with $Y$s and the other with $\bar Y$s. 
We think that the correct state may have other structures 
involving these $Y$ and $\bar Y$ fields, but should be similar to the double determinant. 
Mixing with other fields seems to be suppressed at large $N$.
This approximation of the state leads to reasonable answers for the 
anomalous dimensions of the open strings we want to study.

Under this assumption, an obvious guess for the generalization of \eqn{YY} is then
\begin{equation}
\label{YYbar}
{\cal O}_{Y\bar Y}^{\cW, {\cV}}=
\epsilon^{a_1\cdots a_N}_{b_1\cdots b_N}
Y_{a_1}^{b_1}\cdots Y_{a_{N-1}}^{b_{N-1}}
\,\epsilon^{c_1\cdots c_N}_{d_1\cdots d_N}
\bar Y_{c_1}^{d_1}\cdots\bar Y_{c_{N-1}}^{d_{N-1}}
\cW^{d_N}_{\ a_N}{\cV}^{b_N}_{\,c_N}\,.
\end{equation}
The simplest insertion is $\cW=Z^L, \cV=Z^{L'}$.

It is clear that we can further generalize the construction where instead of the 
$\bar Y=0$ brane we have $\hat Y=0$ with $\hat Y$ defined in \eqn{y-hat}. 
This allows to smoothly interpolate between $Y$ at $\theta_1=\theta_2=0$ 
and $\bar Y$ at $\theta_1=\theta_2=\pi/2$. 
With this we have constructed a two parameter family of pairs of D-branes interpolating 
between the pair of identical D-branes and the \DDbar\ systems. 

We expect the planar dilatation operator to act on this complicated operator as the 
sum of two independent integrable open spin--chain Hamiltonians, one acting on each of the 
two words:%
\footnote{The dilatation operator actually mixes also the structure of the $Y$ and $\bar Y$ fields, which is an indication that this is not the exact state. 
We assumed that the correct state without insertions to be protected at large $N$. Otherwise the first term of \eqref{dim split} is a certain function of $\lambda$.\label{note:protection}}
\begin{equation}
\Delta [{\cal O}_{Y\bar Y}^{\cW, {\cV}}] = \Delta_{\rm bare} [{\cal O}_{Y\bar Y}^{\cW, {\cV}}] 
+ \delta \Delta [\cW_{Y\bar Y}] + \delta \Delta [\cV_{\bar Y Y}].
\label{dim split}
\end{equation}
We will verify this splitting now at the one loop level in perturbation theory, and 
then assume it holds in general. This allows us to study the spectrum of each of these open strings 
independently by application 
of different integrability tools.

\subsection{Integrable spin-chain}
\label{sec:spin-chain}

To calculate the conformal dimension of an operator, we consider the two point function 
between two similar operators and find the mixing matrix. We therefore study the 
pair of operators $\cO_{Y,\bar Y}^{\cW,{\cV}}(0)$ and 
$\cO_{\bar Y,Y}^{{\bar \cV'},{\bar \cW'}}(x)$.

It is useful to separate the calculation according to how many of the fields $Y$ and $\bar Y$ 
from the determinant operators interact with the $\cW$ and $\cV$ insertions. 
In the case when none do, we perform free field contractions between the $Y$ and 
$\bar Y$ of the two operators. Using
\beq
\label{contractions}
\epsilon^{a_1\cdots a_N}\epsilon_{b_1'\cdots b_N'}
\delta_{a_1}^{b_1'}\cdots\delta_{a_{N-1}}^{b_{N-1}'}
= (N-1)! \, \delta^{a_N}_{b_N'}
\eeq
we find
\beq
\label{tree}
\vev{\cO_{Y,\bar Y}^{\cW,{\cV}}(0)\ \cO_{\bar Y,Y}^{{\bar \cV'},{\bar \cW'}}(x)}
= (N-1)!^6 
\vev{\Tr\big[\cW(0){\bar \cW'}(x)\big]\,\Tr\big[{\cV}(0){\bar \cV'}(x)\big]}
\eeq
where $\vev{Y^b_{\ a}(0)\bar Y(x)^{b'}_{\ a'}}$ is normalized to $\delta_a^{b'}\delta_{a'}^b$\,.
This is a non-local trace, which is not gauge invariant, but this is due to the fact that 
$2(N-1)$ contractions were already done in a specific gauge. The entire 
correlator is of course gauge invariant.

In the planar approximation the expectation value in \eqn{tree} factorizes and 
at tree level we find that $\cW$ and $\bar \cW'$ are conjugate operators, as are 
$\cV$ and $\bar \cV'$. Each trace gives an extra factor of $N$. This statement 
holds as long as the last letter 
in $\cW$ and the first one in $\cV$ are orthogonal to $Y$ and the other 
ends of the words are orthogonal to $\bar Y$. Otherwise there would be extra 
planar tree-level contractions beyond \eqn{contractions} which will mix these 
states with operators made of a sub-determinant and a single trace operator.

There are also interacting graphs contributing to
$\vev{\Tr\big[\cW(0){\bar \cW'}(x)\big]}$, which by construction do not know about the 
rest of the determinant operators. These will give the bulk part of the spin--chain Hamiltonian. 
At one-loop level in the $\so(6)$ sector this is the same as the usual Minahan-Zarembo 
Hamiltonian \cite{Minahan:2002ve}.

We should consider separately the boundary interactions, where the beginning and 
end of $\cW$ and $\cV$ interact with $Y$ and $\bar Y$ from the rest of the determinant. 
The first boundary interactions involve just one pair of $Y$ and $\bar Y$. The 
trace structure arising from contracting all the determinant fields except for one 
$Y(0)$ and one $\bar Y(x)$ is given by \eqn{one-loop-boundary} in 
Appendix~\ref{app:boundary}.

The one-loop interactions arising from these graphs are very similar to the 
boundary interaction for the giant graviton open spin-chain \cite{Berenstein:2005vf}. Again the interaction 
between $\cW$ and $\cV$ completely factorizes at large $N$ and for each 
word one finds the usual one-loop boundary interaction. The only modification 
is that the first letter is projected on states orthogonal to $\bar Y$ and the last 
letter should be orthogonal to $Y$. The one-loop Hamiltonian acting 
on the word $\mathcal{W}$ is
\beq
\label{1-loop-H}
H^{(2)}= \frac{\lambda}{8 \pi^2} \, Q^{\bar Y}_1Q^{Y}_L\left[
\sum_{l=1}^{L-1}\left(I_{l,l+1}-P_{l,l+1}+\frac{1}{2}K_{l,l+1}\right)
+2-Q^{Y}_1-Q^{\bar Y}_L\right]Q^{Y}_L Q^{\bar Y}_1
\eeq
Here $I$, $P$ and $K$ are respectively the usual identity, permutation and trace operators 
on the spin--chain \cite{Minahan:2002ve}. $Q^\phi_l$ is a projector whose kernel are all words 
with the field $\phi$ at location $l$. The Hamiltonian acting on the word $\mathcal{V}$ 
can be obtained by exchanging $Y\leftrightarrow\bar Y$.

Although we do not derive here 
the explicit Hamiltonian at higher loop order, one can still write down 
the Bethe ansatz, as we do in the next section.

\subsection{Wrapping corrections}
\label{sec:wrapping}

One can proceed this way to higher order boundary interactions, but we would like 
to study the first wrapping corrections, where interactions are communicated between 
the two boundaries and the energy of the $Z^L$ ground state is lifted.

Wrapping graphs come from the interaction of the word $\cW$ with 
$Y$ on one side and $\bar Y$ on the other. The leading wrapping corrections will 
arise by choosing one $Y(0)$ and one $\bar Y(0)$ from each operator and requiring 
that they all interact with $\cW$. We analyze this 
in Appendix~\ref{app:boundary} and find that we should include connected graphs 
contributing to
\beq
\label{wrapping1}
\vev{\Tr[Y(x)\cW(0)\bar Y(x)Y(0)\bar \cW(x)\bar Y(0)]}
\eeq

To be more specific, consider the ground state $\cW=Z^L$. Since it shares 
some supercharges with each of the determinants, the interaction with only 
one boundary will not give rise to an anomalous dimension. This is identical 
to the state attached to two $Y$ determinants. Only when the interaction 
involves both a $Y$ and $\bar Y$ determinant will the ground state energy 
be lifted.%
\footnote{This is true even if we change some of the index structure of the 
$Y$ and $\bar Y$ in \eqn{YYbar}, so this statement is quite insensitive to 
the exact details of the state.}
To capture this we can consider the difference between the two cases
\beq
\vev{\Tr[Y(x)Z^L(0)\bar Y(x)Y(0)\bar Z^L(x)\bar Y(0)]}
-\vev{\Tr[\bar Y(x)Z^L(0)\bar Y(x)Y(0)\bar Z^L(x)Y(0)]}
\eeq

\begin{figure}[t]
\begin{center}
 \epsfig{file=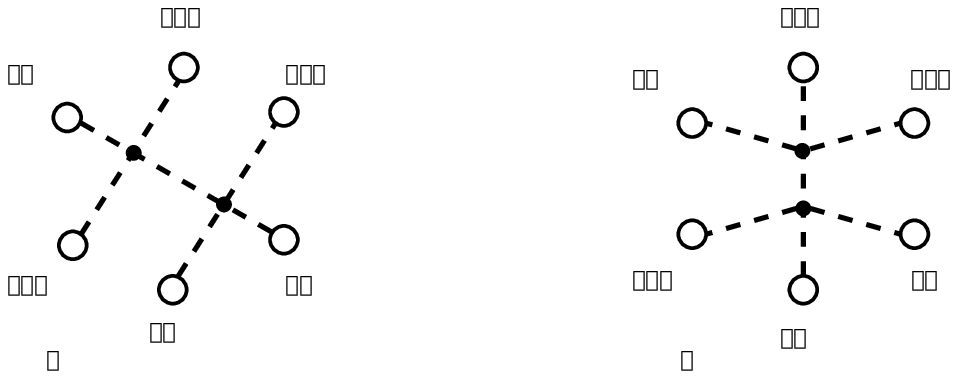,width=110mm
 \psfrag{z0}{\footnotesize$Z(0)$}
 \psfrag{zbx}{\footnotesize$\bar Z(x)$}
 \psfrag{y0}{\footnotesize$Y(0)$}
 \psfrag{yb0}{\footnotesize$\bar Y(0)$}
 \psfrag{yx}{\footnotesize$Y(x)$}
 \psfrag{ybx}{\footnotesize$\bar Y(x)$}
 \psfrag{a}{$(a)$}
 \psfrag{b}{$(b)$}
}
\parbox{15cm}{
\caption{Graphs for insertion of $\cW=Z$ (the case of $L=1$). The fields are ordered 
according to the trace structure in \eqn{wrapping1}.
\label{fig:Z-wrapping}}}
\end{center}
\end{figure}

In the case of $L=1$ there seem to be two types of relevant 2-loop graphs, depicted in 
Figure~\ref{fig:Z-wrapping}. The graph $(a)$ exists also in the case where both 
boundaries are on the $Y=0$ brane (i.e., with $\bar Y(0)\to Y(0)$ and $Y(x)\to\bar Y(x)$).

\begin{figure}[t]
\begin{center}
 \epsfig{file=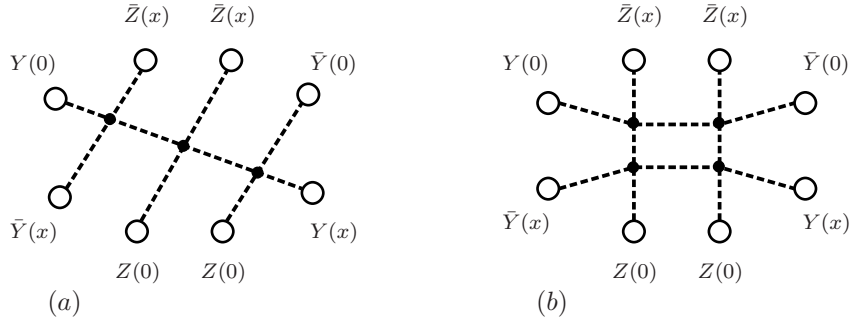,width=110mm
 \psfrag{z0}{\footnotesize$Z(0)$}
 \psfrag{zbx}{\footnotesize$\bar Z(x)$}
 \psfrag{y0}{\footnotesize$Y(0)$}
 \psfrag{yb0}{\footnotesize$\bar Y(0)$}
 \psfrag{yx}{\footnotesize$Y(x)$}
 \psfrag{ybx}{\footnotesize$\bar Y(x)$}
 \psfrag{a}{$(a)$}
 \psfrag{b}{$(b)$}
 }
\parbox{15cm}{
\caption{Graphs for insertion of $\cW=Z^2$. The fields are ordered according 
to the trace structure in \eqn{wrapping1}.
\label{fig:ZZ-wrapping}}}
\end{center}
\end{figure}

We expect wrapping effects to start at order $2L$ in perturbation theory.%
\footnote{Note though that in the quark-antiquark case the first interaction happen 
at ``half'' wrapping order ($L+1$), which can be attributed to the finite density of 
zero momentum single particle states in the mirror BTBA formulation as 
encoded by the poles of the Y-functions \cite{Dru-int-WL, CMS}.}
The generalization of the graph in Figure~\ref{fig:Z-wrapping}$a$ extended to the case of $L>1$ is of order $L+1$ (see Figure~\ref{fig:ZZ-wrapping}$a$), 
therefore its contribution must be equal to the case with the BPS boundary interactions, 
or the difference should cancel against other graphs.

\begin{figure}[t]
\begin{center}
 \epsfig{file=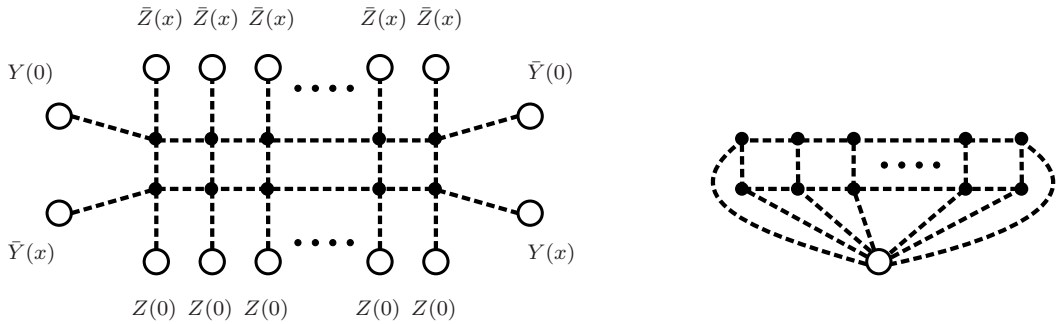,width=140mm
 \psfrag{z0}{\footnotesize$Z(0)$}
 \psfrag{zbx}{\footnotesize$\bar Z(x)$}
 \psfrag{y0}{\footnotesize$Y(0)$}
 \psfrag{yb0}{\footnotesize$\bar Y(0)$}
 \psfrag{yx}{\footnotesize$Y(x)$}
 \psfrag{ybx}{\footnotesize$\bar Y(x)$}
 }
\parbox{15cm}{
\caption{The generalization of Figures~\ref{fig:Z-wrapping}$b$ 
and ~\ref{fig:ZZ-wrapping}$b$ for insertion of $Z^L$. 
The figure on the right is the space-time structure of the diagram 
(the resulting $2L$-loop integral $I_{2L}$) 
with the lines going to $x$ amputated. They can be easily restored 
by adding a line to each trivalent vertex.
\label{fig:ZL-wrapping}}}
\end{center}
\end{figure}

The graph such as in Figure~\ref{fig:Z-wrapping}$b$ exists for all $L$ at order $2L$, see 
Figure~\ref{fig:ZZ-wrapping}$b$ and Figure~\ref{fig:ZL-wrapping} for $L\geq2$. 
For $L=2$ there are many other graphs of the same 
order as this graph. For example when the box is replaced with a fermionic hexagon. 
This graph generalizes to arbitrary $L$ and is of order $L+2$, so again 
it should be the same as for the BPS vacuum and cancel against other graphs.

For general $L$ we conjecture therefore that the first wrapping correction comes 
from the graph in Figure~\ref{fig:ZL-wrapping}. It gives rise to a UV divergent loop integral
depicted on the right of Figure~\ref{fig:ZL-wrapping}, 
where the ellipses correspond to repeating the structure to generate the total number of $2L$ loops in the integrals.

Based on explicit data for $L=1,2,3$ and our conjecture for larger $L$ 
(from the results of the next section),
these integrals were recently shown in \cite{Schnetz:2012nt} to have the 
same divergence structure as the zig-zag integrals \cite{Broadhurst:1995km,Brown:2012ia}.
We present the map of the above integrals to the 
zig-zag integrals in some more detail in appendix \ref{app:integ}.
The results for the overall UV divergences (denoted by calligraphic $\mathcal{I}$) of the integral 
$I_2$ arising in the $L=1$ case from the diagrams in Figure~\ref{fig:Z-wrapping}
and the integrals $I_{2L}$ depicted on the right of 
Figure~\ref{fig:ZL-wrapping},
regularized in $D=4-2\varepsilon$ (and with 
the coupling restored) are
\begin{equation}\label{I2I2L}
\mathcal{I}_2=\frac{\lambda^2}{(4\pi)^{4}}\Big(-\frac{1}{2\varepsilon^2}+\frac{1}{2\varepsilon}\Big)
\,,\qquad
\mathcal{I}_{2L}=\frac{\lambda^{2L}}{(4\pi)^{4L}\varepsilon}\frac{1}{L^2}\binom{4L-2}{2L-1}\zeta(4L-3)
\,,
\end{equation}
where our conjectured result $\mathcal{I}_{2L}$ was proven recently ({\em cf.}, Eqn (3) of \cite{Schnetz:2012nt}).
Note that for $L\ge 2$ the integrals are free of subdivergences, which is
indicated by the absence of all higher order poles in $\varepsilon$. Assuming these 
are indeed the first graphs that contribute, we conclude that the anomalous dimension 
of the ground state is
\beq
\delta\Delta_L=-2\varepsilon\lambda\partial_\lambda I_{2L}+O(\lambda^{2L+1})
=-\frac{4\lambda^{2L}}{(4\pi)^{4L}}\frac{1}{L}\binom{4L-2}{2L-1}\zeta(4L-3)+O(\lambda^{2L+1})\,.
\label{delta DL gauge}
\eeq

At $L=1$, however, the integral contains a one-loop subdivergence which leads to an inconsistency
here: it enters as a simple $\varepsilon$-pole in the anomalous dimension,
which has to be finite and independent of $\varepsilon$.
In this setup, i.e.\ for a gauge invariant composite operator in a theory with 
unrenormalized coupling, 
such a one-loop subdivergence at two loops can only be cancelled either by further
two-loop diagrams or by a one-loop
counter term, both of which are associated with the renormalization of this
operator. We conjecture that the approximation of the state with $L=1$ by \eqn{YYbar} 
is inappropriate and that the correct state will be renormalized at one-loop order.

The fact that short states are subtle and can lead to divergences was seen in the past in the 
integrability-based descriptions \cite{Frolov:2009in,deLeeuw:2011rw,Ahn:2011xq,deLeeuw:2012hp}.
In certain cases (deformed or orbifold setups) a possible resolution of 
this issue in the field theory was given in \cite{Fokken:2013aea,Fokken:2013mza}, though 
the effect observed there has no net effect for 
the $\mathcal{N}=4$ SYM theory. We note that the integrability calculation in 
the next section also gives a divergence for the $L=1$ state - so 
whatever effect lifts this divergence (presumably by a one-loop anomalous 
dimension) should somehow also alter the integrability description of this 
state. We leave it to the future to resolve this issue.

\subsection{General angle}
\label{sec:gen-angle}

As mentioned above, the two D-branes do not have to be coincident, but can be at arbitrary 
angles on $S^5$, which on the gauge theory side amounts to replacing $\bar Y$ with an 
arbitrary linear combination of $Y$, $X$, $\bar X$ and $\bar Y$ which we denoted by 
$\hat Y$ in \eqn{y-hat}. It is easy to see that the wrapping graphs we calculated will 
see only the $\bar Y$ factor in $\hat Y$ and the result of the first wrapping effect will 
be multiplied by $\sin^2\theta_1\sin^2\theta_2$.

\section{Integrable description of the $Y\bar{Y}$ brane system}
\label{sec:aba}

In this section we formulate the integrable description of the $Y\bar{Y}$
brane system.
In the integrable formulation we characterize and solve the system in 
terms of the scattering data of the particle-like excitations of the 
strings. The finite-volume energy spectrum of the particles corresponds
to the sought-for anomalous dimensions on the gauge theory side.

D-branes provide boundary conditions for open strings, which
translate into reflection amplitudes for the elementary particle-like
excitations.  The bulk scattering matrix of these excitations
supplemented by the
reflection factors define the theory and enable one to calculate both
the asymptotic large-volume energy spectrum and all finite-size effects.

In the integrable description we assume the quantum integrability of
the model, and solve the theory in semi-infinite geometry by
determining the scattering (reflection) data.  Integrability forces
the multiparticle reflection process to factorize into pairwise
scatterings and individual reflections.  Integrable boundary
conditions in this point of view can be classified by finding all
one-particle reflection matrices, which are compatible with the given
bulk S-matrix and residual symmetries.

When two boundaries exist their relative orientation is also important, which is used to break the supersymmetry of the vacuum state of the $Y=0$ brane studied in \cite{Bajnok:2012xc}.
The new vacuum state acquires a nontrivial anomalous dimension from finite-size effects.
Our notation in this section is summarized in Appendix \ref{app:notation}.

\subsection{Reflection matrices in $\hat Y=0$ brane systems}

We focus on systems in which the left and right boundary
conditions are not the same, but all of them are related to the $Y=0$
system in a relatively simple way. The $Y=0$ boundary condition preserves
an $\alg{su}(1|2)\oplus \alg{su}(1|2)$ sub-algebra of the full $\alg{su}(2|2)\oplus \alg{su}(2|2)$
symmetry of the bulk S-matrix. If we label the excitations in the
fundamental representation by $(1,2\vert3,4)\otimes(\dot{1},\dot{2}\vert\dot{3},\dot{4})$
then the $\alg{su}(2)$ symmetry, which rotates in the $(1,2)$
or $(\dot{1},\dot{2})$ space 
\begin{equation}
\left(\begin{array}{c}
1\\
2
\end{array}\right)\to\left(\begin{array}{cc}
\cos\theta & \sin\theta\\
-\sin\theta & \cos\theta
\end{array}\right)\left(\begin{array}{c}
1\\
2
\end{array}\right)\label{eq:rotation}
\end{equation}
is broken by the presence of the $Y=0$ brane. The reflection factor
compatible with the unbroken symmetry, which satisfies the boundary
Yang-Baxter equation has the following factorized form \cite{Hofman:2007xp, Ahn:2008df}
\begin{equation}
\mathbb{R}_{Y}^{-}(p)=R_{0}^{-}(p)\, R_{Y}^{-}(p)\otimes\dot{R}_{Y}^{-}(p)
\label{fundrefl1}
\end{equation}
where 
\begin{equation}
R_{Y}^{-}(p)=\dot{R}_{Y}^{-}(p)=\mbox{diag}(e^{-i\frac{p}{2}},-e^{i\frac{p}{2}},1,1)
\,, \qquad R_{0}^{-}(p)=-e^{-ip}\sigma(p,-p) \,,
\label{fundrefl2}
\end{equation}
and $\sigma(p,-p)$ is the BES dressing factor \cite{Beisert:2006ez}.
This reflection factor can be extended for bound-states both in the
string/mirror theories belonging to the totally symmetric/anti-symmetric
atypical representations of $\alg{su}(2|2)\oplus \alg{su}(2|2)$, 
respectively \cite{Ahn:2010xa, Palla:2011eu, Correa:2009mz}. 
The totally anti-symmetric representation describing the bound-states
of $a$ fundamental particles in the mirror theory has a diagonal
reflection factor
\begin{equation}
R_{Y}^{-}(p)=\mbox{diag}(\mathbb{I}_{a}e^{-i\frac{p}{2}},-\mathbb{I}_{a}e^{i\frac{p}{2}},
\mathbb{I}_{a+1},-\mathbb{I}_{a-1}) \,,
\label{eq:Ra}
\end{equation}
and its scalar factor is obtained by fusion.

Although the presence of the D-brane breaks the rotational symmetry,
this symmetry does not completely disappear from the system. Acting with such
a transformation (\ref{eq:rotation}) will rotate the D-brane itself
and acts on the reflection factors in the following way: 
\begin{equation}
R_{\theta}^{-}=O R_{Y}^{-} O^{T}= 
\arraycolsep=0.4em
\begin{pmatrix}
\cos^{2}\theta \, e^{-i\frac{p}{2}}-\sin^{2}\theta \, e^{i\frac{p}{2}} &
\sin\theta\cos\theta \, (e^{-i\frac{p}{2}}+e^{i\frac{p}{2}}) & 0 & 0\\[1mm]
\sin\theta\cos\theta \, (e^{-i\frac{p}{2}}+e^{i\frac{p}{2}}) & 
\sin^{2}\theta\,  e^{-i\frac{p}{2}}-\cos^{2}\theta \, e^{i\frac{p}{2}} & 0 & 0\\[1mm]
0 & 0 & 1 & 0\\
0 & 0 & 0 & 1
\end{pmatrix},
\label{def:Rtheta}
\end{equation}
where $O$ acts as in Eq. (\ref{eq:rotation}) in the $(1,2)$ space
and as identity in the $(3,4)$ space.
We introduce two rotation angles $\theta_1$ and $\theta_2$ for dotted and undotted indices.
The reflection factor has to satisfy unitarity, boundary crossing unitarity and the boundary 
Yang-Baxter equations to maintain integrability. The reflection factor \eqref{def:Rtheta} 
solves these constraints, as the rotation by $O$ is part of the bulk symmetry which 
commutes with the S-matrix.

If we choose $\theta_1=\theta_2=\frac{\pi}{2}$ 
we obtain the reflection factor of the $\bar{Y}=0$ system. This is
the anti-brane counterpart of the $Y=0$ brane, and has a reflection
factor in which the two labels $(1,2)$ are exchanged
\begin{equation}
R_{\bar{Y}}^{-}(p)=\dot{R}_{\bar{Y}}^{-}(p)=\mbox{diag}
(-e^{i\frac{p}{2}},e^{-i\frac{p}{2}},1,1) \,.
\label{ybarrefl}
\end{equation}
This is nothing but the charge conjugated reflection factor:
\begin{equation}
R_{\bar{Y}}^{-}=CR_{Y}^{-}C^{-1}\,, \qquad C=\left(\begin{array}{cccc}
0 & -i & 0 & 0\\
i & 0 & 0 & 0\\
0 & 0 & 0 & 1\\
0 & 0 & -1 & 0
\end{array}\right) \,.
\end{equation}
This picture extends to the reflection factors of the bound states,
too: they can be simply obtained by exchanging the labels $(1,2)$.

\subsection{$Y\bar{Y}$ system in large volume }

In the following we analyze a two-boundary system in finite volume,
namely in the strip geometry. We place $\mathbb{R}_{Y}^{-}$ on the
right boundary but 
\begin{equation}
\mathbb{R}_{\bar{Y}}^{+}(p)=\mathbb{R}_{\bar{Y}}^{-}(-p)
\end{equation}
on the left boundary. We are interested in the asymptotic spectrum
of multiparticle states and an exact description of the ground state.
Both problems can be attacked via double-row transfer matrices and
Y-system. 
The energy of a multiparticle state gives a half of the total anomalous dimension in gauge theory \eqref{dim split}, 
and the other half is obtained in an analogous way.

The boundary Bethe-Yang equations (also called boundary asymptotic Bethe Ansatz equations) are 
valid for large size $L$, the R-charge of the inserted word. They determine
the momenta, $\left\{ p_{i}\right\} $, of a multiparticle state 
by the periodicity of the wave function as follows. 
Pick up any particle (with momentum $p_{k}$ say), scatter through the others to the right, 
reflect back from the right boundary, scatter with momentum $-p_{k}$ on all particles to the left,  
reflect back from the left boundary and scatter back to its original position; 
we have to arrive at the same state
multiplied by $e^{-2ip_{k}L}$. During the scattering processes the
labels of the multiparticle state are mixed up, and the problem is
to find eigenstates of the mixing matrix. This problem is solved by
introducing and diagonalizing the double-row transfer matrix. As the
diagonalization problem factorizes between the two $\alg{su}(2\vert2)$
factors we focus on one copy only, and define this double-row transfer
matrix by
\begin{equation}
T_{Y\bar{Y}}(p)=\mbox{STr}\left(R_{\bar{Y}}^{+}(p)S(p,p_{1})
\dots S(p,p_{M})R_{Y}^{-}(p)S(p_{M},-p)\dots S(p_{1},-p)\right) ,
\label{dr transfer}
\end{equation}
where we used the non-graded
S-matrix $\mathbb{S}=S_{0}\, S\otimes S$, in the $\alg{su}(2)$ normalization
($S_{11}^{11}=1$). 
On the left boundary $R_{\bar{Y}}^{+}(p)$ is introduced in \eqref{dr transfer}, which is different from the standard definition \cite{Sklyanin:1988yz}. Nevertheless the two are equivalent up to an overall normalization \cite{Bajnok:2010ui, Bajnok:2012xc}.
Our choice ensures that a ``test'' particle is brought around the two boundaries in the above sense; 
if we specify the test particle momenta as $p=p_{k}$ we obtain the $k^{th}$ particle's mixing
matrix.

Let us diagonalize the double-row transfer matrix for $Y\bar Y$ states in the $\alg{su}(2)$ sector.
We start by analyzing the ground state of the first level $\vert1,1,\dots,1\rangle$ in algebraic Bethe Ansatz, and denote its eigenvalue by $T(p\vert\{p_{i}\})$ or $T$ for short.
It describes an $M$-particle state in the $\alg{su}(2)$ sector. Interestingly,
similarly to the $YY$ system \cite{Bajnok:2012xc}, the eigenvalue can be expressed
in terms of the \emph{diagonal} elements: 
\begin{equation}
T=\rho_{1}T_{1}+\rho_{2}T_{2}-\rho_{3}T_{3}-\rho_{4}T_{4}\,,
\label{eq:T(p)}
\end{equation}
where 
\bal
T_{1} & = S_{11}^{11}(p,p_{1})\dots S_{11}^{11}(p,p_{M})S_{11}^{11}(p_{M},-p)
\dots S_{11}^{11}(p_{1},-p)=1 \\
T_{2} & = S_{21}^{21}(p,p_{1})\dots S_{21}^{21}(p,p_{M})S_{12}^{12}(p_{M},-p)
\dots S_{12}^{12}(p_{1},-p) \\
T_{3}=T_{4} & = S_{31}^{31}(p,p_{1})\dots S_{31}^{31}(p,p_{M})S_{13}^{13}(p_{M},-p)
\dots S_{13}^{13}(p_{1},-p)
\eal
which explicitly read as:
\bal
T_{3} & = \prod_{i=1}^{M}\frac{(x^{+}-x_{i}^{+})}{(x^{+}-x_{i}^{-})}
\frac{(x^{+}+x_{i}^{-})}{(x^{+}+x_{i}^{+})}=:\frac{R^{(-)+}}{R^{(+)+}}\\
T_{2} & = \prod_{i=1}^{M}\frac{(x^{+}-x_{i}^{+})(x^{+}+x_{i}^{-})(x^{-}
x_{i}^{+}-1)}{(x^{+}-x_{i}^{-})(x^{+}+x_{i}^{+})(x^{-}x_{i}^{-}-1)}
\frac{(x^{-}x_{i}^{-}+1)}{(x^{-}x_{i}^{+}+1)}=:\frac{R^{(-)+}}{R^{(+)+}}
\frac{B^{(-)-}}{B^{(+)-}} \,.
\eal
The $\rho_i \ (i=1,\dots,4)$ can be calculated following the considerations in \cite{Bajnok:2012xc}.
The result turns out to be
\begin{equation}
\rho_{1}=-\rho_{3}=-\frac{(1+(x^{-})^{2})(x^{-}+x^{+})}{2x^{-}(1+x^{+}x^{-})}
\,, \qquad
\rho_{2}=-\rho_{4}=-\frac{(1+(x^{+})^{2})(x^{-}+x^{+})}{2x^{+}(1+x^{-}x^{+})} \,.
\end{equation}
Clearly $x^{-}\leftrightarrow x^{+}$ exchanges $1\leftrightarrow2$
and $3\leftrightarrow4$, and thus acts as conjugation, under which the $Y\bar{Y}$ ground state is invariant. 
The momenta of the state are determined from the following asymptotic
boundary Bethe-Yang equations
\begin{equation}
e^{-2ip_{k}L}T(p_{k})^{2}d_{1,1}(p_{k})=1\,,\label{eq:BBA}
\end{equation}
where we introduced the proper normalization factor 
\begin{equation}
d_{1,1}(p)=R_{0}^{-}(p)\hat{R}_{0}(-p)\prod_{j=1}^{M}S_{0}(p,p_{j})S_{0}(p_{j},-p)
\,,\qquad
\hat{R}_{0}(-p)=\frac{e^{-2ip}R_{0}^{-}(p)}{S_{0}(p,-p)\rho_{1}^{2}(p)} \,,
\end{equation}
which is determined by the boundary Bethe-Yang equations for one-particle states \eqref{bbye}.

The extension from the diagonal sector to the full sector
can be easily done at the level of the generating functional, which
we now introduce. 
The eigenvalues of the double-row transfer matrix, in which mirror
bound-state test particles of charge $a$ are scattered and reflected
through the multiparticle state, are generated as 
\begin{equation}
\tilde{\mathcal{W}}_{\alg{su}(2)}^{-1}=(1-\mathcal{D}\rho_{1}T_{1}\mathcal{D})
(1-\mathcal{D}\rho_{3}T_{3}\mathcal{D})^{-1}
(1-\mathcal{D}\rho_{4}T_{4}\mathcal{D})^{-1}
(1-\mathcal{D}\rho_{2}T_{2}\mathcal{D})=
\sum_{a=0}^{\infty}(-1)^{a}\mathcal{D}^{a}\tilde{T}_{a,1}\mathcal{D}^{a} \,,
\end{equation}
in the $\alg{su}(2)$ grading, where $\mathcal{D}$ is the shift operator \eqref{def:shift operator}.
Technically it is simpler to renormalize the generating functional and the transfer matrices as 
\begin{equation}
\mathcal{W}_{\alg{su}(2)}^{-1}=\Bigl(1+\frac{R^{(+)}}{R^{(-)}}\mathcal{D}^{2}\Bigr)\Bigl(1-
\mathcal{D}^{2}\Bigr)^{-1}\Bigl(1-\mathcal{D}\frac{u^{+}}{u^{-}}\mathcal{D}\Bigr)^{-1}\Bigl(1+
\mathcal{D}\frac{u^{+}}{u^{-}}\mathcal{D}\frac{B^{(-)}}{B^{(+)}}\Bigr)=
\sum_{a=0}^{\infty}(-1)^{a}\mathcal{D}^{a}T_{a,1}\mathcal{D}^{a} \,.
\label{eq:genfun}
\end{equation}
The relation between the normalizations is of the fusion type:
\begin{equation}
\tilde{T}_{a,1}=f^{[a-1]}f^{[a-3]}\dots 
f^{[3-a]}f^{[1-a]}T_{a,1}\,,\qquad f=T_{3}\rho_{3} \,.
\end{equation}
Explicit calculation gives all the antisymmetric transfer matrix 
eigenvalues
\begin{equation}
(-1)^{a}T_{a,1}=(a+1)\frac{u}{u^{[-a]}}+a\frac{u^{-}}{u^{[-a]}}\frac{R^{(+)[a]}}{R^{(-)[a]}}
+a\frac{u^{+}}{u^{[-a]}}\frac{B^{(-)[-a]}}{B^{(+)[-a]}}+(a-1)\frac{u}{u^{[-a]}}
\frac{R^{(+)[a]}}{R^{(-)[a]}}\frac{B^{(-)[-a]}}{B^{(+)[-a]}} \,.
\label{ybyeigenval}
\end{equation}

\subsubsection{Boundary asymptotic Bethe Ansatz equations for generic states}

Comparing eq.(\ref{ybyeigenval}) with the corresponding expression of the 
$YY$ system \cite{Bajnok:2012xc},
we can observe that the result is the same up to signs in front of
the fermionic contributions, as if we had performed the trace instead
of the supertrace. This, however, breaks supersymmetry and allows
a nontrivial ground state energy for the $Y\bar{Y}$ system. This
simple observation allows us to conjecture the generating functional
for the eigenvalue of the double-row transfer matrix for a generic
state
\begin{equation}
\Lambda (p) =\left(\frac{x^{+}(p)}{x^{-}(p)}\right)^{m_{1}}\frac{R^{(-)+}}{R^{(+)+}}
\Biggr[\rho_{1}\frac{R^{(+)+}}{R^{(-)+}}\frac{B_{1}^{-}R_{3}^{-}}{B_{1}^{+}R_{3}^{+}}
-\rho_{3}\frac{B_{1}^{-}R_{3}^{-}}{B_{1}^{+}R_{3}^{+}}\frac{Q_{2}^{++}}{Q_{2}}
-\rho_{4}\frac{R_{1}^{+}B_{3}^{+}}{R_{1}^{-}B_{3}^{-}}\frac{Q_{2}^{--}}{Q_{2}}
+\rho_{2}\frac{B^{(-)-}}{B^{(+)-}}\frac{R_{1}^{+}B_{3}^{+}}{R_{1}^{-}B_{3}^{-}}\Biggl] \,,
\end{equation}
where as in \eqref{def:BRQ}, $B_{1}R_{3}$ and $R_{1}B_{3}$ represent type 1 Bethe roots denoted by $y_{j}$, and $Q_2$ represents type 2 Bethe roots denoted by $\tilde{\mu}_{l}$.
Regularity of the transfer matrix at the roots gives the boundary Bethe-Yang equations.
Type $1$ roots are specified as $x^{+}(p)=y_j$, type $2$ roots when $u(p)=\tilde{\mu}_{l}$, finally type $3$ roots when $x^{-}(p)=y_j^{-1}$. 
The corresponding Bethe equations read as
\begin{equation}
\frac{R^{(+)+}Q_{2}}{R^{(-)+}Q_{2}^{++}}\Bigg\vert_{x^{+}(p)=y_j}=-1,
\quad
\frac{\rho_{3}}{\rho_{4}}\frac{R_{1}^{-}B_{1}^{-}R_{3}^{-}
B_{3}^{-}Q_{2}^{++}}{R_{1}^{+}B_{1}^{+}R_{3}^{+}B_{3}^{+}Q_{2}^{--}}
\Bigg\vert_{u(p)=\tilde{\mu}_{l}}=-1,
\quad
\frac{B^{(-)-}Q_{2}}{B^{(+)-}
Q_{2}^{--}}\Bigg\vert_{x^{-}(p)=y_j^{-1}}=-1.
\end{equation}
The Bethe Ansatz equation which determine the momenta are
\begin{equation}
e^{-2ip_{k}L}\Lambda(p_{k})^{2}d_{1,1}(p_{k})=1.
\label{eq:BBA gen}
\end{equation}

\subsubsection{Asymptotic Y-system for the vacuum state}

From now on we focus only on the unprotected vacuum state. As $M=0$, we have $R=B=1$, and the expression (\ref{ybyeigenval}) for the eigenvalues of 
the transfer matrices simplifies considerably:
\begin{equation}
T_{a,1}=(-1)^{a}\frac{4au}{u^{[-a]}} \,.
\label{asympTa}
\end{equation}
They constitute part of a solution of the T-system
\begin{equation}
T_{a,s}^{+}T_{a,s}^{-}=T_{a-1,s}T_{a+1,s}+T_{a,s-1}T_{a,s+1} \,,
\label{Tsystem}
\end{equation}
on the $\alg{su}(2|2)$-hook. 
For completeness and later applications
we provide here the full solution of the $\alg{su}(2|2)$
T-system. The transfer matrix eigenvalues in the symmetric representations are
generated via the inverse of (\ref{eq:genfun}): 
\begin{equation}
\mathcal{W}_{su(2)}=\left(1+\mathcal{D}\frac{u^{+}}{u^{-}}\mathcal{D}\right)^{-1}
\left(1-\mathcal{D}\frac{u^{+}}{u^{-}}\mathcal{D}\right)\left(1-\mathcal{D}^{2}\right)\left(1+
\mathcal{D}^{2}\right)^{-1}=\sum_{s=0}^{\infty}\mathcal{D}^{s}T_{1,s}\mathcal{D}^{s} \,,
\label{eq:genYbarY}
\end{equation}
which results in
\begin{equation}
T_{1,s}
=(-1)^{s}2\left[1+\frac{u^{[s]}}{u^{[-s]}}+2\sum_{k=1}^{s-1}\frac{u^{[s]}}{u^{[s-2k]}}\right] \,.
\label{eq:T1s}
\end{equation}
The T-functions on the boundary of the $\alg{su}(2|2)$-hook are
\begin{equation}
T_{0,a}=T_{a,0}=1 \quad (a \ge 0), \qquad 
T_{2,Q}=T_{Q,2}=\frac{16u^{[Q]}u^{[-Q]}}{u^{[-Q+1]}u^{[-Q-1]}} \quad (Q \ge 2).
\end{equation}
The asymptotic Y-functions are defined from the T-functions as 
\begin{equation}
Y_{a,s}=\frac{T_{a,s+1}T_{a,s-1}}{T_{a+1,s}T_{a-1,s}}
\label{eq:Yfuncs1}
\end{equation}
for $s>0$. For $s=0$, (in a similar analysis for the $\alg{su}$(2) sector) $Y_{1,0}(p_{k})=-1$ should provide the boundary
Bethe-Yang equations (\ref{eq:BBA}). This allows us to restore the correct normalization:
\begin{equation}
Y_{1,0}=f_{1,1}T_{1,1}^{2}e^{-2iLp}\,, \qquad 
f_{1,1}=d_{1,1}(\rho_{3}T_{3})^{2} \,.
\label{eq:Yfuncs2}
\end{equation}
The normalization of the bound state transfer matrix eigenvalues follow from
the bootstrap: 
\begin{equation}
f_{a,1}=f_{1,1}^{[a-1]}f_{1,1}^{[a-3]}\dots 
f_{1,1}^{[3-a]}f_{1,1}^{[1-a]} \,. 
\label{eq:Yfuncs3}
\end{equation}

\subsubsection{L\"uscher correction for the vacuum state}

In the following we use the asymptotic Y-functions to calculate the
leading finite-size -- so called L\"uscher -- correction for the vacuum
state. For this we analytically continue $Y_{a,0}$ in $u$ to the
mirror plane:
\begin{equation}
Y_{a,0}=f_{a,1}T_{a,1}^{2}\left(\frac{z^{-}}{z^{+}}\right)^{2L}=
\frac{u^{[-a]}}{u^{[a]}}\left(\frac{4au}{u^{[-a]}}\right)^{2}
\left(\frac{z^{-}}{z^{+}}\right)^{2L}=\frac{16a^{2}u^{2}}{u^{[a]}
u^{[-a]}}\left(\frac{z^{-}}{z^{+}}\right)^{2L} \,,
\label{Ya0 transfer}
\end{equation}
where we denote the analytically-continued variables $x^{[\pm a]}$ by $z^{\pm}$,
which can be parametrized by the mirror momenta $q$ as:
\begin{equation}
z^{\pm}=\frac{q+ia}{4g}\left(\sqrt{1+\frac{16g^{2}}{q^{2}+a^{2}}}\pm1\right) \,.
\label{def:zpm q}
\end{equation}
We can compare the $Y_{a,0}$ functions with the integrand of the vacuum
L\"uscher correction \cite{Correa:2009mz,Bajnok:2010ui} calculated directly from the reflection
matrices: 
\begin{equation}
\Delta E(L)=-\sum_{a=1}^{\infty}\int\limits_{-\infty}^\infty\frac{dq}{4\pi}Y_{a,0}
=-\sum_{a=1}^{\infty}\int\limits_{-\infty}^\infty\frac{dq}{4\pi}
\mathbb{R}^{-}{}_{i}^{j}(z^{\pm})\mathbb{C}_{j\bar{j}}
\mathbb{R}^{+}{}_{\bar{i}}^{\bar{j}}(-1/z^{\mp})
\mathbb{C}^{\bar{i}i}\left(\frac{z^{-}}{z^{+}}\right)^{2L} \,.
\label{eq:BLusher}
\end{equation}
As charge conjugation exchanges the $\bar{Y}=0$ boundary with the
$Y=0$ boundary, we simply square the analytically-continued bound state reflection
factor (\ref{eq:Ra}) and perform the trace. This gives for the matrix
part 
\begin{equation}
a\Bigl(2+\frac{z^{+}}{z^{-}}+\frac{z^{-}}{z^{+}}\Bigr)=a\frac{(z^{+}+z^{-})^{2}}{z^{+}z^{-}}\,.
\end{equation}
The prefactor was already calculated in \cite{Bajnok:2010ui}
\begin{equation}
R_{0}(z^{\pm})R_{0}(-1/z^{\mp})=\frac{4(1+z^{+}z^{-})^{2}}{(z^{+}
+\frac{1}{z^{+}})(z^{-}+\frac{1}{z^{-}})(z^{-}+z^{+})^{2}} \,.
\label{altnorm}
\end{equation}
Squaring the matrix part and multiplying with the scalar factor exactly
reproduces the transfer matrix result \eqref{Ya0 transfer}.
A further check on the $Y$ functions obtained with the
aid of the generating functional is described in Appendix \ref{sec:oneparticle}.

It is now easy to evaluate the finite-size correction in the weak
coupling limit. At leading order in $g^{2}$ we find the following
correction for the vacuum: 
\begin{equation}
\Delta E(L)=-\sum_{a=1}^{\infty}\int\limits_{-\infty}^\infty\frac{dq}{4\pi}
\left(4g^{2}\right)^{2L}16a^{2}\frac{q^{2}}{(q^{2}+a^{2})^{2L+1}}=
-\frac{4g^{4L}}{4L-1}{4L \choose 2L} \zeta (4L-3) \,,
\label{Luscher YbarY}
\end{equation}
which agrees precisely with the gauge theory result \eqref{delta DL gauge} for $L \ge 2$, and diverges at $L=1$.

\subsection{Generic angle, the $\hat Y=0$ brane}

Here we analyze the system with generic angles. We keep $\mathbb{R}_{Y}^{-}$
on the right boundary but place $\mathbb{R}_{\theta}^{+}(p)=\mathbb{R}_{\theta}^{-}(-p)$
on the left boundary.  The reflection factor in the totally antisymmetric
representation can be dressed as:%
\footnote{This is not quite the same as fusing the already-dressed reflection matrices.}
\begin{equation}
R_{\theta}^{-}(p)= 
\arraycolsep=0.4em
\begin{pmatrix}
\(\cos^{2}\theta \, e^{-i\frac{p}{2}}-\sin^{2}\theta \, e^{i\frac{p}{2}}\) \mathbb{I}_{a} &
\(\sin\theta\cos\theta \, (e^{-i\frac{p}{2}}+e^{i\frac{p}{2}}) \)\mathbb{I}_{a} & 0 & 0\\[2mm]
\left(\sin\theta\cos\theta \, (e^{-i\frac{p}{2}}+e^{i\frac{p}{2}})\right)\mathbb{I}_{a} & 
\left(\sin^{2}\theta \, e^{-i\frac{p}{2}}-\cos^{2}\theta \, e^{i\frac{p}{2}}\right)\mathbb{I}_{a} & 0 & 0\\
0 & 0 & \mathbb{I}_{a+1} & 0\\
0 & 0 & 0 & -\mathbb{I}_{a-1}
\end{pmatrix}.
\end{equation}

\subsubsection{L\"uscher correction}

In order to calculate the L\"uscher correction for the ground state 
energy, we start from the expression in Eq. $(\ref{eq:BLusher})$. Only the matrix
part is deformed by the angle: 
\begin{eqnarray}
a(2+\sin^{2}\theta(e^{ip}+e^{-ip})-2\cos^{2}\theta) & = & a\sin^{2}\theta
\left[2+\left(\frac{z^{+}}{z^{-}}+\frac{z^{-}}{z^{+}}\right)\right]\nonumber \\
 & = & a\sin^{2}\theta\frac{(z^{+}+z^{-})^{2}}{z^{+}z^{-}} \,,
\end{eqnarray}
which shows that we simply have to include an additional $\sin^{2}\theta$
factor compared to the $Y\bar{Y}$ system for each $\alg{su}(2|2)$
wing. The resulting Y-functions are 
\begin{equation}
Y_{a,0}=\frac{16a^{2}u^{2}}{u^{[a]}u^{[-a]}}\sin^{2}\theta_{1}\sin^{2}
\theta_{2}\left(\frac{z^{-}}{z^{+}}\right)^{2L} \,,
\end{equation}
which at leading order leads to the wrapping correction 
\begin{equation}
\Delta_{\theta}E(L)=-\sin^{2}\theta_{1}\sin^{2}\theta_{2}\frac{4g^{4L}}{4L-1}{4L \choose 2L} \zeta (4L-3) \,.
\end{equation}
This is precisely what we expect from gauge theory calculations for the $Y$-$\hat Y$ brane system. It is \eqref{Luscher YbarY} multiplied by the square of the respective angular dependence in \eqref{y-hat}.

The generating functional for the vacuum in case of a generic angle is analyzed in Appendix \ref{sec:genfun}.

\section{The $Y\olY$ ground state BTBA}
\label{sec:BTBA}

In this section we derive the ground state BTBA equations for the $Y\bar{Y}$ system and analyze them numerically.

TBA equations in the presence of boundaries can be formulated in the same way as in the periodic case, provided that
the S-matrix and the boundary reflection amplitudes are diagonal \cite{LeClair:1995uf}. 
BTBA follows from the mirror trick, which equates the open string worldsheet partition function in the string region 
with the closed string transition amplitude between boundary states in the mirror region. 
In the mirror picture, the boundary state projects the intermediate states to those consisting 
of an even number of particles with the opposite momentum. As a result, a Y-function 
in the ground state BTBA is the ratio of the density of particle pairs to that of hole pairs.

When the S-matrix is non-diagonal, it becomes very difficult to compute the source term in BTBA explicitly,
which comes from the overlap between a boundary state and the bulk state written in terms of the density 
of Bethe roots and holes. Thus, a simple alternative approach is called for.
Recall that the periodic TBA can also be derived by integrating the Y-system assuming appropriate 
discontinuity relations and analyticity of Y-functions \cite{Cavaglia:2010nm,Balog:2011nm}.
In this section we apply this method to derive a set of BTBA equations, and solve them numerically.

\subsection{Boundary TBA from Y-system and discontinuity relations}

The derivation of the equations goes along the lines of ref. \cite{Balog:2011nm} relying on the following assumptions:\footnote{These assumptions are also supported by the asymptotic solution of excited states.}
\begin{itemize}
\item There exist TBA-type integral equations governing the spectrum of the $Y\bar{Y}$ system.
\item The Y-functions of the BTBA equations satisfy the Y-system functional equations \cite{Gromov:2009tv} 
of AdS/CFT.
\item The Y-functions satisfy the discontinuity relations of ref. \cite{Cavaglia:2010nm}, too.
\item The Y-functions are real functions. They are meromorphic in the vicinity of the real axis away from cuts
prescribed by the discontinuity relations.
\item The ground state Y-functions are parity even and left-right symmetric.
\item The Y-functions are smooth deformations of their asymptotic limit, so qualitative 
information on the location of their point-like singularities can be borrowed from the asymptotic solution.
\item The massive Y-functions decay at large rapidity at least as $1/u$, while
the large $u$ behavior of the other Y-functions is the same as that of the asymptotic solution, 
namely in the $u \to \infty$ limit they tend to state- and coupling-independent 
constants.
\end{itemize}  
From the assumptions above it is clear that the BTBA equations presented in this section 
are valid only as long as the analytic structure of Y-functions
agrees with that of the asymptotic solution and the massive Y-functions decay fast enough at infinity.
The value of the coupling constant where one of the previous two assumptions fails is 
called a critical value, and some of our assumptions need to be relaxed.

In this section the notations of refs. \cite{Arutyunov:2009ax,Balog:2011nm} are used so that their results could be 
referred directly. All kernels and source functions of the subsequent BTBA equations can be found
in appendix A of ref. \cite{Balog:2011nm}.

For the ground state of the $Y\bar{Y}$ system the local singularities which affect the 
actual form of the BTBA equations lie in the fundamental strip $-1/g\leq\mbox{Im}\,u \leq 1/g$
and they have fixed positions located at $0$ or $\pm i/g$ in the complex plane. Based on 
the asymptotic solution given in subsection 3.2.2,  the Y-function combinations which have poles or zeroes at the 
$\{ 0,\pm i/g \}$ positions are listed  {below\footnote{For $1+Y$ and $1+1/Y$ type combinations only the 
real singularities relevant for BTBA equations, so only these points are listed.}}.

\begin{itemize}
\item $Y_Q,\, Y_{\pm}, \, Y_{m|vw},$ and $Y_{2m-1|w}$ have double zero at $u=0$ for 
$Q,m=1,2,...$.
\item $Y_{2m|w}, \, 1+Y_{2m|w}, \, 1-\frac{1}{Y_{\pm}},$ and $1+\frac{1}{Y_{m|vw}}$ have 
double pole at $u=0$ for $m=1,2,....$
\item $Y_1$ and $Y_{1|w}$ have simple poles at $u=\pm i/g$.
\item $Y_{2m-1|w}$ have double poles at $u=\pm i/g$ for $m=1,2,....$
\end{itemize}
The derivation of the BTBA equations goes along the lines of ref. \cite{Balog:2011nm}. Most of the 
equations can be derived straightforwardly from the Y-system equations by taking into 
account the residue contributions of the local singularities listed above. The two subtle 
equations are the discontinuity functions of $\log Y_{-}$ and $\log Y_1$ denoted by ${J}$ and 
$\Delta$, respectively \cite{Balog:2011nm}. The derivation of equations for these quantities 
requires the usage of discontinuity relations of \cite{Cavaglia:2010nm}. Since asymptotically $e^{ J}=1$ 
for the ground state it is assumed that it has no local singularities on the whole complex 
plane. So it follows that $\frac{Y_{-}}{Y_{+}}$ is given by (5.30) of \cite{Balog:2011nm} by taking 
the set $\{u_j\}=\emptyset$ or equivalently $R_{p/m}\to 1$ and $B_{p/m}\to 1$.

The computation of $\Delta$ goes along the lines of section 6. of ref. \cite{Balog:2011nm} taking 
into account the different singularity structure and asymptotic behavior of the 
Y-functions. Here we introduce the notations:
\begin{equation}
{\mathscr L}_{\pm}=\log\left[\tau^2 \left( 1-\frac{1}{Y_{\pm}}\right) \right], \qquad 
{\mathscr L}_{m}=\log\left[\tau^2 \left( 1+\frac{1}{Y_{m|vw}}\right) \right], \qquad 
\tau(u)=\tanh(\frac{\pi g u}{4}).
\end{equation}
The discontinuity function $\Delta$ satisfies the following equation:
\begin{equation}
\Delta = 2{\mathscr L}_--2({\mathscr L}_-+{\mathscr L}_+)\ \hat\star\ K-
2\sum_{m=1}^\infty{\mathscr L}_m\star k_m+
2{\cal W}-2 L_{\rm BTBA} \log x^2+\Delta_{{\rm red}},
\label{Deltares}
\end{equation}
where $\Delta_{{\rm red}}$ is given {by\footnote{Here the $\epsilon$ description means, that the integration
contour goes just above the real axis}} 
\begin{equation}
\Delta_{{\rm red}}=2\sum_{N=1}^\infty\,{J}^{[\epsilon]}
\ \check\star\ (K^{[2N]}-K^{[-2N]})
\end{equation}
and 
the source term ${\cal W}$ is given by the integral representation:
\begin{equation}
{\cal W}=\left( \int\limits_{[\gamma]}+\int\limits_{[-\gamma]}\right)
\, dv \log \tau(v)\, K(v,u)+\int\limits_{[\gamma]}  dv \, \log \tau^{-}(v)\, K(v,u)
+\int\limits_{[-\gamma]} dv \, \log \tau^{+}(v)\, K(v,u),
\end{equation}
where $[\pm \gamma]$ means that the integration runs along the lines $v\pm i \gamma$ 
from $-\infty$ to $\infty$, with $\gamma$ being a small positive number.  
Starting from the discontinuity function (\ref{Deltares}) and applying the 
simplification techniques described in sections 7 and 8 of \cite{Balog:2011nm}, the 
hybrid BTBA equations for the massive Y-functions can be derived. Carrying out the
whole process the following BTBA equations were derived:
\begin{eqnarray}
Y_{m\vert vw}&=&\tau^2 \,\exp\left\{
\log\left[\frac{(1+Y_{m+1\vert vw})(1+Y_{m-1\vert vw})}
{(1+Y_{m+1})}\right]\star s\right\},\qquad m\geq2,
\label{TBAmvw}\\
Y_{1\vert vw}&=&\tau^2 \, \exp\left\{
\log\left[\frac{(1+Y_{2\vert vw})}{(1+Y_2)}\right]\star s
+\log\left[\frac{1-Y_-}{1-Y_+}\right]\ \hat\star\ s
\right\},\label{TBA1vw}\\
Y_{m\vert w}&=&\tau^{2 \, (-1)^{m+1}} \, \exp\left\{
\log\left[(1+Y_{m+1\vert w})(1+Y_{m-1\vert w})\right]
\star s\right\},\qquad m\geq2,\label{TBAmw}\\\
Y_{1\vert w}&=&\tau^2 \, \exp\left\{
\log\left[1+Y_{2\vert w}\right]\star s+ 
\log\left[\frac{1-\frac{1}{Y_-}}{1-\frac{1}{Y_+}}\right]
\ \hat\star\ s \right\}\,,\label{TBA1w}\\
Y_Q&=&\tau^2 \, \exp\left\{\log\left[
\frac{Y_{Q+1}\,Y_{Q-1}
(1+Y_{Q-1\vert vw})^2}
{Y^2_{Q-1\vert vw}
(1+Y_{Q+1})(1+Y_{Q-1})}\right]\star s \right\},\quad Q\geq2,\label{TBAQ}\\
\frac{Y_-}{Y_+}&=&
\exp\left\{-\sum_{Q=1}^\infty \log (1+Y_Q)\star K_{Qy}\right\}.
\label{YmperYp} \\
Y_+Y_-&=& \tau^4 
\exp\Bigg\{
2\log\left[\frac{1+Y_{1\vert vw}}{1+Y_{1\vert w}}
\right]\star \! s+\! \! \sum_{Q=1}^\infty \log (1+Y_Q)\star\left[
-K_Q+2K^{Q1}_{xv}\star \! s\right] \Bigg\}\label{YpYm}
\end{eqnarray}
The symbols $\star, \hat\star$ denote the convolutions defined in (A.4) of \cite{Balog:2011nm}.
Equations (\ref{YmperYp}) and (\ref{YpYm}) determine $Y_{\pm}$ up to an overall sign factor.
The sign factor can be fixed from the asymptotic solution and its value is $-1$.
Thus the fermionic Y-functions can be expressed in terms of the LHS of (\ref{YmperYp}) and
(\ref{YpYm}) by the formula:
\begin{equation}
Y_{\mp}=-e^{\frac12 \log Y_{+} Y_{-}\pm \frac12 \log \frac{Y_{-}}{Y_{+}}}.
\end{equation}
For the massive Y-functions we present the hybrid form of the BTBA equations.
\begin{equation}
\begin{split}
\log Y_Q=&-2 L_{\rm BTBA} \tilde{\cal E}_Q+f_Q+2 \log(1+Y_{Q-1|vw}) \star s+
2\log(1+Y_{1|vw})\star s\ \hat\star\ K_{yQ}\\
&-2\log\left[\frac{1-Y_-}{1-Y_+}\right]\ \hat\star\ s\star K^{1Q}_{vwx}
+2{\mathscr L}_-\ \hat\star\ K^{yQ}_-+2{\mathscr L}_+\ \hat\star\ K^{yQ}_+\\
&\qquad+\sum_{Q'=1}^\infty \log(1+Y_{Q'})\star\left[K^{Q'Q}_{{\mathfrak{sl}(2)}}
+2s\star K^{Q'-1,\,Q}_{vwx}
\right], \qquad Q=1,2,...
\end{split}
\label{hybrid}
\end{equation}
where the source term $f_Q$ is given by:
\begin{equation}
f_Q=\log\tau^2-\log \tau^2 \star K_Q-2 \, {\cal{\tilde{W}}}_Q,
\end{equation}
with 
\begin{equation}
\begin{aligned}
{\cal{\tilde{W}}}_Q(u)&=\left( \int\limits_{[\gamma]}+\int\limits_{[-\gamma]}\right) 
\, dv \log \tau(v)\, K^{yQ}_+(v,u)+\int\limits_{[\gamma]}  dv \, \log \tau^{-}(v)\, 
K^{yQ}_+(v,u) 
+\int\limits_{[-\gamma]} dv \, \log \tau^{+}(v)\, K^{yQ}_+(v,u).
\end{aligned}
\end{equation}
The parameter $L_{\rm BTBA}$ in \eqref{Deltares} and the hybrid BTBA equations \eqref{hybrid} is related to
the R-charges of the determinant-like operator denoted by $L$ in the previous sections.
In particular $L_{\rm BTBA}=L$ for the ground state.

The energy of the $Y\bar{Y}$ ground state after subtraction of the bare dimension
is given by the formula
\begin{equation}
E_{\rm BTBA} \equiv \sum_{Q=1}^\infty {\bf E} (Q) = - \sum_{Q=1}^\infty
\int_{0}^\infty \frac{du}{2\pi} \, \frac{d \tilde p^Q}{d u} \log(1+Y_Q).
\label{energy1}
\end{equation}
This BTBA energy corresponds to the energy of a single open string. 
The total dimension of the determinant-like operator \eqref{dim split} with $\cW=Z^L, \cV=Z^{L'}$ 
is written as
\begin{equation}
\Delta [ {\cal O}_{Y\bar Y}^{Z^L, Z^{L'} } ] = \Delta_{\rm bare} + E_{\rm BTBA} (L) + E_{\rm BTBA} (L').
\end{equation}

\bigskip

Before closing the subsection we argue that due to the constraints imposed by the Y-system equations, 
the L-R symmetry and parity, the local singularities
of the Y-functions located at the positions $\{0,\pm \frac{i}{g}\} $ do not receive any wrapping 
corrections. As an example we show that the double zero of $Y_1$ located at the origin remains fixed at any
value of the coupling constant. As a first step let us invoke the Y-system equations
\begin{equation}
Y_1 ^+ Y_1 ^-=\frac{Y_2}{Y_-^2}\frac{(1-Y_-)^2}{1+Y_2} .
\end{equation}
At the level of the asymptotic solution, the double zero of $Y_1$ at the origin is related to a simple 
zero of $1-Y_-(v)$ at $v=\pm \frac{i}{g} $.
Suppose that wrapping corrections change the quantization condition as
\begin{equation}
Y_- (v) = 1 \quad {\rm at} \ \ v=\frac{i}{g} - i \delta \qquad \Leftrightarrow \qquad
Y_1 (v) = 0^2 \quad \text{(double zero) \ at} \ \ v = -i \delta.
\end{equation}
Since $Y_-(v)$ is parity even, this equation should be valid after the parity transformation $v \mapsto -v$, 
which implies $Y_1(v) = 0^2$ at $v=+i \delta$. Now if we take the asymptotic limit, $Y_1$ 
possesses a quartic zero instead of a double zero at the origin, which is a contradiction.
In other words, the origin $v=0$ is a special point where the parity transformation acts trivially, 
thus a double zero is allowed.

\subsection{Lower bounds for TBA energy in \AdSxS}

In this subsection we show that the energy which can be computed from the BTBA equations of the $Y\olY$ ground state 
is bounded from below. This means that the BTBA equations can describe the model 
as long as the energy is real and remains
above the lower bound.

To determine a lower bound, the energy formula (\ref{energy1}) must be studied. 
In order for the energy to be finite,
the individual integrals should stay finite; and having evaluated the integrals, the remaining sum must also converge. 

First consider the case of individual integrals. To see their convergence the large $u$ behavior of the integrand must 
be analyzed. Since $\frac{d\tilde{p}_Q}{du}\sim g$ for large $u$ and at any $Q$, $Y_Q (u)$ must decay faster than $1/u$ at infinity.
This simple remark constrains the range of the BTBA energy, because the large rapidity
behavior of $Y_Q$ is governed by the exact energy through the formula:\footnote{The formula (\ref{BTBA YQ large u}) can be derived from (\ref{hybrid}). 
The $\sim L$ term comes from the $\tilde{\cal E}_Q$ term, while the $\sim E_{\rm BTBA}$ terms
originate from the $\sum\limits_{Q'=1}^\infty \log(1+Y_{Q'})\star K^{Q'Q}_{{\mathfrak{sl}(2)}}$ term by exploiting the following large $u$ expansion of the kernel:
$ K^{Q'Q}_{{\mathfrak{sl}(2)}}(t,u)=-\frac{1}{\pi} \frac{d\tilde{p}_{Q'}}{dt} \log|u|+\cO(1).$}
\begin{equation}
\log Y_Q (u) = - \( 4L + 4 E_{\rm BTBA} \) \log |u| + \cO(1), \qquad (|u| \gg 1).
\label{BTBA YQ large u}
\end{equation}
This gives the large rapidity lower bound for the energy:
\begin{equation}
4L + 4 E_{\rm BTBA} > 1 \qquad \Leftrightarrow \qquad E_{\rm BTBA} > \frac14 - L.
\label{open energy lower bound}
\end{equation}

The convergence of the sum in (\ref{energy1}) imposes a stronger constraint on the energy. Since $\frac{d\tilde{p}_{Q}}{du}=\cO(1)$ at large $Q$,
$Y_Q$ must be sufficiently small for large $Q$.  
Let us investigate the large $Q$ behavior of the summand:
\begin{equation}
{\bf E}(Q)=-\int_{0}^{\infty} \frac{du}{2 \pi} \, \frac{d\tilde{p}_Q}{du} \log(1+Y_Q(u)). 
\label{EQ}
\end{equation}
Since $Y_Q$ is small for large $Q$ in (\ref{EQ}) the $\log$ can be expanded and at leading order one can write:
$${\bf E}(Q)\simeq -\int_{0}^{\infty} \frac{du}{2 \pi} \, \frac{d\tilde{p}_Q}{du} \,Y_Q(u).$$
As it is shown in appendix \ref{app:YQ} the large $Q$ behavior of $Y_Q$ can be deduced from the BTBA equations,  
and it can be expressed in terms of the asymptotic solution as follows: 
\begin{equation}
Y_Q(u)\simeq \xi \, f^{[Q]}(u)\,\bar{f}^{[-Q]}(u) \, Y^{\bullet}_{Q}(u), \label{YQlarge}
\end{equation}
where $\xi$ is a real coupling dependent constant and  $\bar{f}(u)$ is the complex conjugate function of $f(u)$. 
The explicit functional form of $f$ is not important except for its leading large $u$ asymptotics:
\begin{equation}
f(u)=u^{-2 \, E_{\rm BTBA}}(1+\dots),
\end{equation}
where the dots mean contributions negligible for large $u$.
Changing variables $u \rightarrow Q s$: 
\begin{equation}
{\bf E}(Q)\simeq -Q \, \int\limits_{0}^{\infty} \, \frac{ds}{2 \pi} \frac{d\tilde{p}_Q}{du}(Qs) \,\, 
f \(Q \Big(s+\frac{i}{g} \Big) \) \, \bar{f} \(Q \Big(s-\frac{i}{g} \Big) \) \, Y_Q^{\bullet}(Q s) \nonumber
\end{equation}
and expanding all terms for large $Q$ one gets:
\begin{equation}
{\bf E}(Q)\simeq -\frac{16 g}{2 \pi} \xi \, Q^{3-4L-4 E_{\rm BTBA}} \, \int\limits_{0}^{\infty} ds 
\frac{s^2}{(s^2+\frac{1}{g^2})^{2L+1+2 E_{\rm BTBA}}}(1+\dots)
\ \propto \ Q^{3-4L-4 E_{\rm BTBA}} \,, 
\label{energy EQ}
\end{equation}
where the dots stand for subleading corrections in $Q$.
The sum of the energy formula is convergent only if
the energy satisfies the inequality as follows:
\begin{equation}
4L -3 + 4 E_{\rm BTBA} > 1 \qquad \Leftrightarrow \qquad E_{\rm BTBA} > E_{\rm cr} \equiv 1-L \,.
\label{lower bound}
\end{equation}
This formula gives the lower bound for $E_{\rm BTBA}$, which is stronger than the bound given in (\ref{open energy lower bound}).

This result is valid for $L>1$. At $L=1$ the asymptotic solution is not trustable even at weak coupling, 
as can be seen by the divergent L\"uscher correction \eqref{Luscher YbarY}.
Concerning the $L=1$ state, it is not clear if either the BTBA equations \eqref{TBAmvw}-\eqref{hybrid} must be modified, 
or if there exists another BTBA solution, whose small $g$ expansion is different from that of the asymptotic Y-functions
given by (\ref{asympTa},\ref{eq:T1s}).

Finally we argue that the lower bound (\ref{lower bound}) can never be saturated. This follows from (\ref{energy1}).
If the energy reached the lower bound, the LHS of (\ref{energy1})  
would take finite values, i.e. $1-L$. On the other hand the sum on the RHS of
(\ref{energy1}) would tend to infinity which leads to a contradiction. 
Thus we conclude that the BTBA description of the system breaks down and needs to be modified -- if it is possible at all -- 
even before the energy would reach the lower bound given in (\ref{lower bound}).

The numerical results presented in the next section show that the critical energy
is reached at finite values of the coupling constant. 
This prevents us from extending our BTBA solutions to the strong coupling region and extracting 
the large-coupling behavior of the ground state energy.

Our argument concerning the lower bound of the energy is applicable to any other TBA systems 
containing the dressing kernel. 
It becomes particularly important if non-BPS ground states are investigated. There the standard derivation of the TBA equations
is valid, which guarantees the positivity of the Y-functions and so the negativity of the ground state energy.

\subsection{Numerical results}\label{sec:results}

We solved the BTBA equations for the $Y\olY$ ground state numerically for various $(g,L)$ and computed the 
BTBA energy by using the methods explained in Appendix \ref{sec:solving BTBA}. Our numerical results 
are presented in Appendix \ref{sec:raw data} and Figure \ref{fig:YbarY gnd data}, which we now explain in detail.

\begin{figure}
\begin{center}
\includegraphics[scale=0.8]{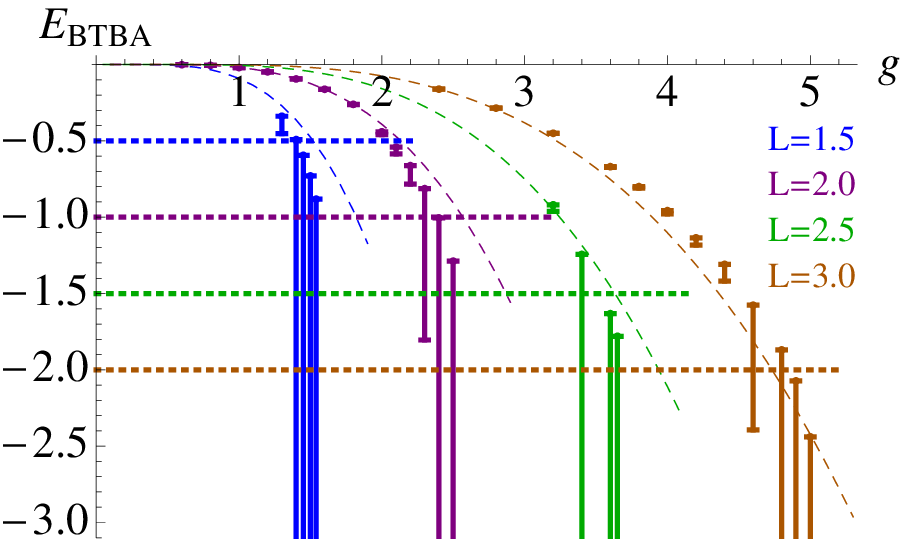}
\hspace{5mm}
\includegraphics[scale=0.8]{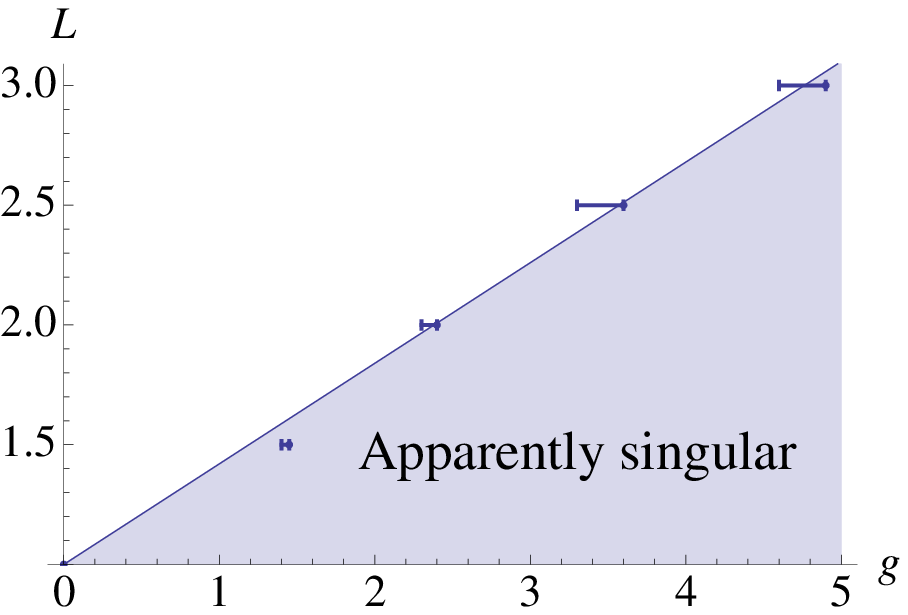}

\vskip 1mm
\parbox{15cm}{
\caption{The exact energy $E_{\rm BTBA} (g)$ of the $Y\olY$ ground state with various $L$ (left) and the phase diagram (right). The thick lines in the left figure represent the numerical data including error bars, the dashed lines the L\"uscher energy, and the dotted lines $E_{\rm cr} = 1-L$. The blue, purple, green, brown curves correspond to $L=\frac32,2, \frac52,3$, respectively. The straight line in the right figure is $L^{\rm (fit)} (g_{\rm cr}) = 1+0.42 \, g_{\rm cr}$.}
\label{fig:YbarY gnd data}
}
\end{center}
\end{figure}

The left figure shows that the BTBA energy for the states with $L=\frac32, 2, \frac52, 3$. 
We examined half-integer values of $L$ to study how the energy depends on $L$, although they 
do not correspond to the determinant-like operator \eqref{YYbar} with $\cW=Z^{L}$ or $\cV=Z^L$.
One sees that the energy loses precision at some critical value $g_{\rm cr}$ of the coupling, 
as indicated by huge error bars spreading out toward $E_{\rm BTBA}=-\infty$.
The right figure shows a plot of $g_{\rm cr}$ as a function of $L$. 
A linear fit is also drawn under the assumption that $g_{\rm cr}=0$ for the $L=1$ state, 
because the L\"uscher energy at $L=1$ diverges logarithmically.

The data points and the error bars of the left figure are computed in the following way.
Once we obtained a numerical solution of the BTBA for each $(g,L)$, we calculate the BTBA energy by
\begin{equation}
E_{\rm BTBA}^{\rm (data)} = - \sum_{Q=1}^{Q_{\rm max}} \int_0^\infty \frac{dv}{2 \pi} \, \frac{d \tp_Q}{dv}  \log (1+Y_Q (v))
- \sum_{Q=Q_{\rm max}+1}^{100} \int_0^\infty \frac{dv}{2 \pi} \, \frac{d \tp_Q}{dv} \log (1+Y_Q^\bullet (v)), 
\label{data energy}
\end{equation}
with $Q_{\rm max}=6$, instead of \eqref{energy1}. Here we truncate the $Y_Q$ functions at $Q=Q_{\rm max}$ to obtain a 
numerical solution of BTBA. Unfortunately, the truncation can induce large errors particularly 
around the critical point $E_{\rm BTBA} \gtrsim E_{\rm cr}$\,.

To estimate the order of truncation errors,
we extrapolate the energy integrals ${\bf E}_Q$ using the large $Q$ behavior \eqref{energy EQ}. 
The extrapolation function is given by
\begin{equation}
{\bf E}^{\rm (fit)} (Q) = \( \frac{Q_{\rm max}}{Q} \)^{4L+4 E_{\rm BTBA}^{\rm (fit)} (Q)-3} \, {\bf E} (Q_{\rm max}),
\label{extrapolate EQ1}
\end{equation}
where the extrapolated BTBA energy is given by
\begin{equation}
E_{\rm BTBA}^{\rm (fit)} (Q_{\rm max}+1) = \sum_{Q=1}^{Q_{\rm max}} {\bf E} (Q) + {\bf E}^{\rm (fit)} (Q_{\rm max}+1)
- \sum_{Q=Q_{\rm max}+2}^{100} \int_0^\infty \frac{dv}{2 \pi} \, \frac{d \tp_Q}{dv} \log (1+Y_Q^\bullet (v)).
\label{extrapolate EQ2}
\end{equation}
We solve these two equations simultaneously to determine ${\bf E}^{\rm (fit)} (Q_{\rm max}+1)$ 
and $E_{\rm BTBA}^{\rm (fit)} (Q_{\rm max}+1)$. By repeating this procedure we obtain 
$E_{\rm BTBA}^{\rm (fit)} (Q_{\rm max}=100)$. It turns out
\begin{equation}
\delta E_{\rm BTBA} \equiv E_{\rm BTBA}^{\rm (data)} - E_{\rm BTBA}^{\rm (fit)} (Q_{\rm max}=100) > 0,
\qquad \text{for any} \ \ (g, L).
\end{equation}
The upper edge of the error bars in the left of Figure \ref{fig:YbarY gnd data} represents 
$E_{\rm BTBA}^{\rm (data)}$, and the lower edge represents $E_{\rm BTBA}^{\rm (fit)} (Q_{\rm max}=100)$.
The right edge of the error bars in the right of Figure \ref{fig:YbarY gnd data} represents the value of 
$g$ at which $E_{\rm BTBA}^{\rm (data)} \simeq E_{\rm cr}$\,, and the left edge represents the value of
$g$ where $E_{\rm BTBA}^{\rm (fit)} (Q_{\rm max}=100) \simeq E_{\rm cr}$\,.

\section{Summary and Discussion}

We have studied the scaling dimension of determinant-like operators which corresponds to the energy of an 
open string stretching between giant graviton branes from different points of view: 
gauge theory perturbation, boundary asymptotic Bethe Ansatz,  boundary L\"uscher formula and boundary TBA equations.
At weak coupling we computed the L\"uscher corrections to the dimension of general $Y\olY$ states. 
For the $Y\olY$ ground state we identified Feynman diagrams which reproduce the L\"uscher corrections.
At general coupling we studied the dimension of the $Y\olY$ ground state by proposing boundary 
TBA equations and solving them numerically. 

We have shown analytically that the ground state energy of TBA have an $L$ dependent lower bound called critical energy $E_{\rm cr} = 1-L$. 
This is actually the point where the physical energy of the open string reaches zero. 
To see this, let us compare the total dimension of the state in gauge theory, $\Delta_{\rm total} $ to the 
total energy in string theory, $E_{\rm total} $.
In gauge theory, the $Y\olY$ operator \eqref{YYbar} is constructed by removing one $Y$ and one $\bar Y$ 
from the determinant and inserting $\cW$ and $\cV$, which means that the total dimension is given 
by $\Delta_{\rm total} = 2N-2 + {\rm dim} \, \cW + {\rm dim} \, \cV$.
In string theory, we measure the energy of a pair of open strings ending on the $Y\olY$ branes. 
Since the open strings do not lower the energy of D-branes, the total energy of this system is
$E_{\rm total} =2N+ E_{\rm open} + E_{\rm open}^{\prime}$\,.
Therefore, when $\cW =Z^L$, the physical energy of the open string is
\begin{equation}
E_{\rm open} = E_{\rm open}^{(0)} + E_{\rm BTBA} [\cW] = L-1+E_{\rm BTBA} (L),
\end{equation}
which is equal to zero at the critical point.

Our numerical studies show that the critical energy is reached at a finite value of the coupling
constant $g_{\rm cr}$ and beyond the critical point our TBA description breaks down. 
At the critical point the open string energy becomes zero, so one can think that this is the 
transition point, where the energy square of the ground state changes sign and the energy becomes complex.
The existence of this behaviour for determinant-like $Y\olY$ operators is 
very natural from the AdS/CFT point of view. We expect that these determinant-like
operators are dual to the open tachyons between $Y=0$ and $\olY=0$ branes in string theory, and the energy of 
an open tachyon is not real. Thus,  there must be a value of the coupling constant at which the BTBA solution exhibits
an exotic behavior, making $E_{\rm BTBA} (g)$ from real to complex valued.
This interpretation of the critical point explains the break down of the TBA description: Our TBA equations 
by construction can account only for real values of the energy, so it should not describe complex energies. 

The emergence of tachyonic instability depends on the value of $L$.
Indeed all wrapping corrections are exponentially small at any coupling if $L$ is sufficiently large, 
showing that the angular momentum $L$ of an open string controls the stability of the \DDbar\ 
system in \AdSxS. 
We can compare this situation with tachyons in flat spacetime. If we add an 
excitation to the tachyonic ground state of an open string in flat spacetime or 
separate the \DDbar\ branes far enough, the resulting state 
is usually no longer tachyonic.
In curved spacetime we expect an excitation to still remove 
the tachyon. From our results we infer that also adding enough R-charge to the ground 
state makes it no longer tachyonic.

There are many interesting open problems: Analytic continuation of the BTBA solution beyond the critical coupling is 
certainly one of them, which should be tackled in the future. 
Further study on the spectrum of an open string ending on $Y\olY$ branes from string theory is also called for.

On top of that, the divergence of the $L=1$ state is another challenging problem.
On one hand, once the precise determinant-like operator is constructed, its dimension 
must be finite in perturbation theory of $\cN=4$ SYM.
On the other hand the BTBA description based on the L\"uscher Y-functions gives an insensible value of the energy, 
as commonly found in the non-BPS state with small $L$ \cite{Frolov:2009in,deLeeuw:2011rw,Ahn:2011xq,deLeeuw:2012hp}.
We would like to conjecture that this divergence in the $Y\olY$ system 
and the breakdown of BTBA at $L>1$ are the same phenomena like a manifestation of the tachyon, 
and that only for $L=1$ this breakdown happens already at $\lambda=0$.

\section*{Acknowledgments}

Z.B., N.D.\ and R.N.\ thank the Israel Institute for Advanced Study, where this collaboration 
was initiated, for its lovely hospitality and financial support.

N.D.\ also thanks the hospitality and financial support of the 
University of Hamburg and the DESY Theory Group through SFB 676, 
the Tokyo University Kavli IPMU and Komaba particle theory group, 
Nagoya University and the Hungarian Academy of Sciences. 
The research of N.D. is underwritten by an advanced 
fellowship of the Science \& Technology Facilities Council and by STFC grant number ST/J002798/1.

C.S.\ thanks the Hungarian Academy of Sciences for hospitality and financial 
support.
The work of C.S.\ is supported by Deutsche Forschungsgemeinschaft (DFG), 
Sonderforschungsbereich SFB 647 \emph{Raum--Zeit--Materie. Analytische und Geometrische Strukturen}.

The work of \'A.H. was supported by the J\'anos Bolyai Research Scholarship of the Hungarian 
Academy of Sciences and OTKA K109312. Z.B and  \'A.H. were supported by an MTA-Lend\"ulet Grant and 
Z.B. and L.P. were supported by OTKA K81461.

R.N. also acknowledges the Wigner Research Centre for hospitality and financial support, and the National
Science Foundation grant PHY-1212337.

Numerical computation in this work was carried out by using Mars beowulf cluster in Utrecht University and  Sushiki server at the Yukawa Institute Computer Facility.
The work of RS is supported by the Netherlands Organization for Scientific Research (NWO) under the VICI grant 680-47-602 and by a Marie Curie Intra-European Fellowship of the European Community's Seventh Framework Programme under the grant agreement number 327996.

Z.B, N.D., R.N., L.P. and R.S. thank the IEU at Ewha University for hospitality and financial support. 

We thank Gleb Arutyunov, Burkhard Eden, Jan Fokken, Sergey Frolov, Sanjaye Ramgoolam, Yusuke Kimura, Stijn van Tongeren and Matthias Wilhelm for useful discussions.

\appendix

\section{Notation}\label{app:notation}

\subsection{Notation for Section \ref{sec:aba}}

The notation of the paper \cite{Bajnok:2012xc} is used in Section \ref{sec:aba} and 
Appendices \ref{sec:oneparticle}, \ref{sec:genfun}, namely
\begin{equation}
g = \frac{\sqrt\lambda}{4\pi} \,, \quad
f^\pm = \bar f \Big( v \pm \frac{i}{2} \Big), \quad
x(v) = \frac{v}{2 g} + i \sqrt{1 - \frac{v^2}{4 g^2} }\,, \quad 
x_s(u) = \frac{u}{2 g} + \sqrt{\frac{u}{2 g} - 1}\sqrt{\frac{u}{2 g} + 1} \,.
\label{BJ notation}
\end{equation}
The rapidity $v$ (or $u$) and the momentum $p$ are defined by
\begin{equation}
\frac{v}{g} = x + \frac{1}{x} \,, \qquad
e^{ip} = \frac{x^{+}}{x^{-}} \,,
\end{equation}
and the mirror energy is given by $\frac{x^{[+Q]}}{x^{[-Q]}} = e^{\tilde \epsilon_Q}$.
In addition, the variable $q$ is defined in \eqref{def:zpm q}.
We use the shift operator 
\begin{equation}
\mathcal{D}f(u)=f \(u-\frac{i}{2} \) \mathcal{D}=:f^{-}\mathcal{D}.
\label{def:shift operator}
\end{equation}

In the integrable description, generic states are specified by $M$ momenta $\{ p_1 \,, \dots \,, p_M \}$, $m_1$ type 1 fermionic roots $\{ y_1 \,, \dots \,, y_{m_1} \}$ and $m_2$ type 2 bosonic roots $\{ \tilde \mu_1 \,, \dots \,, \tilde \mu_{m_2} \}$. The eigenvalue of the double-row transfer matrices can be expressed by the following functions:
\begin{gather}
R^{(\pm)} =\prod_{i=1}^{M}\left(x(p)-x^{\mp}(p_{i})\right)\left(x(p)+
x^{\pm}(p_{i})\right)\,, \quad
B^{(\pm)} =\prod_{i=1}^{M}
\left(\frac{1}{x(p)}-x^{\mp}(p_{i})\right)\left(\frac{1}{x(p)}+
x^{\pm}(p_{i})\right)\,,
\notag \\
B_{1}R_{3}=\prod_{j=1}^{m_{1}}\left(x(p)-y_{j}\right)\left(x(p)+y_{j}\right)
\,, \quad R_{1}B_{3}=\prod_{j=1}^{m_{1}}\left(\frac{1}{x(p)}-y_{j}
\right)\left(\frac{1}{x(p)}+y_{j}\right)\,,
\label{def:BRQ} \\
Q(u)=\prod_{i=1}^{M}(u-u_{i})(u+u_{i})\,, \quad 
Q_{2}(u)= \prod_{l=1}^{m_{2}}(u-\tilde{\mu}_{l})(u+\tilde{\mu}_{l})\,,
\notag
\end{gather}
with $x_i^\pm = x_s (p_i)$.

\subsection{Notation for Section \ref{sec:BTBA}}\label{app:change notation}

In Section \ref{sec:BTBA} and Appendices \ref{app:YQ}, \ref{sec:solving BTBA}, we start using another notation commonly used in the TBA equations \AdSxS\ (e.g. \cite{Arutyunov:2009ur}),
\begin{equation}
g = \frac{\sqrt\lambda}{2\pi} \,, \quad
f^\pm = f \Big( v \pm \frac{i}{g} \Big), \quad
x(v) = \frac{v - i \sqrt{4-v^2}}{2} \,,
\label{AF notation}
\end{equation}
which enables the direct comparison with the literature.
It should be kept in mind that the finite-size corrections are computed using the mirror region in \eqref{AF notation} or the anti-mirror region in \eqref{BJ notation} for the respective notations. The two conventions are related by the $\pm$-flip, $x^\pm \to x^\mp$, and 
\begin{equation}
\bar g = \frac{g }{2} \,,\qquad
\bar v = \frac{g }{2} \, v \,,\qquad
\bar x = x \,,\qquad
\bar x (\bar v=2 \bar g) = x (v=2) = 1,
\end{equation}
where we write a bar on the variables of the notation \eqref{BJ notation}.

The Y-functions are labeled in different ways between two sections as  
\begin{equation}
Y_Q = Y_{Q,0}\,,\quad Y_-  =- 1/Y_{1,1} \,, \quad Y_+ = - Y_{2,2} \,, \quad
Y_{Q|vw} = 1/Y_{Q+1,1} \,, \quad Y_{Q|w} = Y_{1,Q+1} \,.
\end{equation}
We may assume the left-right symmetry $Y_{a,s} = Y_{a,-s}$ for the states of our concern.

\section{Boundary and wrapping interactions}
\label{app:boundary}

We consider here the index structures arising from the partial contraction of the $Y$ and 
$\bar Y$ fields in the determinants leaving behind one or two pairs which give the 
boundary and wrapping interactions.

\subsection{Boundary interactions}

The first boundary interactions involve just one pair of $Y$ and $\bar Y$. To 
evaluate it we need to know the contraction of only $N-2$ indices from each 
determinant
\beq
\label{N-2-cont}
\epsilon^{a_1\cdots a_N}\epsilon_{b_1'\cdots b_N'}
\delta_{a_2}^{b_2'}\cdots\delta_{a_{N-1}}^{b_{N-1}'}
=(N-2)! \, \epsilon^{a_1a_N}_{b_1'b_N'}\,,
\qquad
\epsilon^{a_1a_N}_{b_1'b_N'}
=\delta^{a_1}_{b_1'}\delta^{a_N}_{b_N'}
-\delta^{a_1}_{b_N'}\delta^{a_N}_{b_1'}\,.
\eeq
If we take $Y(0)$ and $\bar Y(x)$ we get a combinatorial factor of $(N-1)^2$ from 
choosing the two fields and the resulting expression is
\bal
\label{one-loop-boundary}
&\frac{(N-1)!^6}{(N-1)}
\epsilon^{a_1a_N}_{b_1'b_N'}\,\delta^{c_N}_{d_N'}
\,\epsilon^{a_1'a_N'}_{b_1b_N}\,\delta^{c_N'}_{d_N}
\bvev{Y(0)_{\ a_1}^{b_1}\cW(0)_{\ a_N}^{d_N}\cV(0)_{\ c_N}^{b_N}
\bar Y(x)_{\ a_1'}^{b_1'}\bar \cV'(x)_{\ a_N'}^{d_N'}\bar \cW'(x)_{\ c_N'}^{b_N'}}
\\&=
\frac{(N-1)!^6}{(N-1)}\epsilon^{a_1a_N}_{b_1'b_N'}
\,\epsilon^{a_1'a_N'}_{b_1b_N}
\bvev{Y(0)_{\ a_1}^{b_1}(\cV(0)\bar \cV'(x))_{\ a_N'}^{b_N}
\bar Y(x)_{\ a_1'}^{b_1'}(\bar \cW'(x)\cW(0))_{\ a_N}^{b_N'}}
\\&=
\frac{(N-1)!^6}{(N-1)}\Big[
\bvev{\!\Tr[\cV(0)\bar \cV'(x)]\,\Tr[\cW(0)\bar \cW'(x)]\,\Tr[Y(0) \bar Y(x)]}
-\bvev{\!\Tr[\bar Y(x)\cV(0)\bar \cV'(x)Y(0)]\,\Tr[\cW(0)\bar \cW'(x)]}
\\&\hphantom{{}={}\frac{(N-1)!^6}{(N-1)}\Big[}
-\bvev{\!\Tr[\cV(0) \bar \cV'(x)]\,\Tr[Y(0)\bar \cW'(x)\cW(0)\bar Y(x)]}
+\bvev{\!\Tr[\cV(0) \bar \cV'(x)Y(0)\bar \cW'(x)\cW(0)\bar Y(x)]}
\Big]
\eal
These terms come with different powers of $N$. In the large $N$ limit the 
interactions factorize to the individual traces in the product. 
When contracting $Y(0)$ and $\bar Y(x)$ with free propagators, 
the first term in brackets scales like $N^{L+L'+4}$, 
the second and third like $N^{L+L'+3}$, 
and the last one like $N^{L+L'+2}$. The first seems to dominate, 
but the combinatorics assumed that $\Tr[Y(0)\bar Y(x)]$ interacts with one of the other 
traces, otherwise it was accounted for already in \eqn{tree} (and gets multiplied 
by the factor $(N-1)^{-1}$). We should therefore consider only graphs with 
interactions that involve the distinguished $Y(0)$ and 
$\bar Y(x)$ and some fields in the adjoint words. In this case, the 
first term will involve connected graphs, which do not give additional powers of $N$, 
similar to the last term. In the second and third term, however, 
these interactions generate planar contributions at leading order in $N$.

The second and third terms on the r.h.s of \eqn{one-loop-boundary} give 
an interaction on one side of $\cV(0)$ and on one 
side of $\cW(0)$. The interaction with $\bar Y(0)$ and $Y(x)$ 
will lead to the interaction on the other sides of these open spin--chains. 
Each of these terms is identical to some of those which arise when considering a single word inside the usual $Y=0$ brane. 
Similar terms with $Y\leftrightarrow\bar Y$ and the interaction at the other
end of the words are identical to the case of the $\bar Y=0$ brane. This ensures that the
one loop boundary interaction is the same as in those cases, which with the appropriate
projections on the two sides of the open spin-chain leading to \eqn{1-loop-H}.

\subsection{Wrapping corrections}
\label{app:wrapping}

Wrapping graphs come from the interaction of one of the words, like $\cW$ with 
$Y$ on one side and $\bar Y$ on the other. The leading wrapping corrections will 
arise by choosing one $Y(0)$ and one $\bar Y(0)$ from each operator and requiring 
that they all interact with either $\cW$ or $\cV$. With two copies of \eqn{N-2-cont} we 
get the index soup%
\footnote{Other graphs of the same order (or lower) will involve interactions of two $Y$s 
taken from one operator and two $\bar Y$s from the other. These will 
not lead to wrapping effects, as they will all sit on one side of $\cW$ or $\cV$ 
and will not lift the energy of the vacuum.}
\bal
\label{wrapping}
\frac{(N-1)!^6}{(N-1)^2}\,
\epsilon^{a_1a_N}_{b_1'b_N'}\,\epsilon^{c_1c_N}_{d_1'd_N'}
\,\epsilon^{a_1'a_N'}_{b_1b_N}\,\epsilon^{c_1'c_N'}_{d_1d_N}
\bvev{Y(0)_{\ a_1}^{b_1}\bar Y(0)_{\ c_1}^{d_1}\cW(0)_{\ a_N}^{d_N}\cV(0)_{\ c_N}^{b_N}
\bar Y(x)_{\ a_1'}^{b_1'}Y(x)_{\ c_1'}^{d_1'}\bar \cV(x)_{\ a_N'}^{d_N'}\bar \cW(x)_{\ c_N'}^{b_N'}}
\eal
There are 16 possible contractions arising from this expression. 
Focusing on the wrapping corrections to $\cW$ we take the free contractions of $\cV$ 
and $\bar\cV$ giving
\beq
\label{wrapping2}
\frac{(N-1)!^6N^{L'+1}}{(N-1)^2}\,
\epsilon^{a_1a_N}_{b_1'b_N'}\,\epsilon^{c_1'c_N'}_{d_1d_N}
\bvev{
\cW(0)_{\ a_N}^{d_N}\left(\bar Y(x)Y(0)\right)_{\ a_1}^{b_1'}
\bar \cW(x)_{\ c_N'}^{b_N'} \left(\bar Y(0)Y(x)\right)^{d_1}_{\ c_1'}
}
\eeq
The piece with the maximum number of traces is of the form
\beq
\bvev{\!\Tr[Y(0)\bar Y(x)]\,\Tr[\bar Y(0)Y(x)]\,\Tr[\cW(0)\bar \cW(x)]}.
\eeq
The planar contractions of this expression will give another factor of $N^{L+5}$, but the 
combinatorics assume that there are interactions between all $Y$ and $\bar Y$ 
(otherwise we have to multiply the whole expression by $(N-1)^{-2}$ and recover \eqn{tree} again). 
The connected correlator $\vev{\Tr[Y(0)\bar Y(x)]\,\Tr[\bar Y(0)Y(x)]}$ 
is independent of both $\cW$ and $\cV$. Such contractions arise also in the 
absence of the insertions and lead to a mixing of the 
determinant operators themselves through the action of the full non-planar 
dilatation operator \cite{Beisert:2003jj} starting at 1-loop. As mentioned in the main text, the mixing problem 
for such determinant-like operators with a total of $2N$ fields, half $Y$ and half $\bar Y$ 
has not been solved and we will ignore this interaction term, which is not directly 
related to the insertions $\cW$ and $\cV$.

The leading wrapping correction comes from the single trace term in 
\eqn{wrapping2}, which leads to \eqn{wrapping1}.

\section{Solution of the integrals}
\label{app:integ}

\begin{fmffile}{YbarYgraph}
\unitlength=0.95mm

The loop integrals for the wrapping corrections are 
obtained by unifying in Figures \ref{fig:Z-wrapping}
and \ref{fig:ZL-wrapping} all fields $Y(0)$, $\bar Y(0)$ and
$Z(0)$ at the space-time point $0$ and regarding the fields
$Y(x)$, $Y(x)$ and $\bar Z(x)$ as external. 
This means, one removes the composite operator at $x$ and the 
propagators that are
connected to it and thus obtains the integral $I_{2L}$, which for
generic $L\ge 2$ is shown on the right in Figure
\ref{fig:ZL-wrapping}. Since $I_{2L}$ has an overall UV divergence, it
contributes to the renormalization of the composite operator. 
The divergence has to be absorbed into the renormalization constant 
$\mathcal{Z}$, which contains the negative of the sum of the UV divergencies 
of the diagrams. The anomalous dimension is determined by this constant,
which is a matrix if mixing with other operators has to be taken into 
account. Here, i.e.\ in a CFT and for gauge invariant operators, the
anomalous dimension is extracted as 
$\delta\Delta=2\varepsilon\lambda\partial_\lambda\ln\mathcal{Z}$.
Consistency of renormalization requires 
that $\ln\mathcal{Z}$ is free of higher-order $\varepsilon$-poles.

For $L=1$ the two-loop integral $I_2$ that is found from 
Figure \ref{fig:Z-wrapping} and its overall UV divergence $\mathcal{I}_2$ read
\begin{equation}
\label{I2}
I_2=
\settoheight{\eqoff}{$\times$}%
\setlength{\eqoff}{0.5\eqoff}%
\addtolength{\eqoff}{-6.75\unitlength}%
\raisebox{\eqoff}{%
\fmfframe(-0.5,-5.5)(1,4){%
\begin{fmfchar*}(15,15)
  \fmfleft{in}
  \fmfright{out1}
\fmf{phantom}{in,v1}
\fmf{phantom}{out,v2}
\fmfforce{(0,0h)}{in}
\fmfforce{(w,0h)}{out}
\fmffixed{(0,0.75h)}{v2,v3}
\fmfpoly{phantom}{v1,v2,v3}
\fmf{plain}{v1,v2}
\fmf{plain}{v1,v3}
\fmf{plain,right=0.25}{v2,v3}
\fmf{plain,left=0.25}{v2,v3}
\end{fmfchar*}}}
\col\qquad
\mathcal{I}_2=\Kop\Rop[I_2]=\Kop[I_2-\Kop[I_1]I_1]=
\frac{\lambda^2}{(4\pi)^4}\Big(-\frac{1}{2\varepsilon^2}+\frac{1}{2\varepsilon}\Big)\col
\end{equation}
where in $D=4-2\varepsilon$ dimensions
the operator $\Kop$ extracts the poles in $\varepsilon$,
while $\Rop$ subtracts the one-loop subdivergence. 
The subdivergence is given by the simple one-loop integral
$I_1$ built from two propagators, which has an overall 
UV divergence $\Kop[I_1]=\frac{\lambda}{(4\pi)^2\varepsilon}$.
In the expression
for the overall UV divergence $\mathcal{I}_2$ in \eqref{I2} 
the presence of the subdivergence is indicated by 
the occurrence of a quadratic $\varepsilon$-pole.
If $\mathcal{Z}$ contains no one-loop contribution, then
this subdivergence is in contradiction with the consistency requirement 
that $\ln\mathcal{Z}$  
is free of higher order $\varepsilon$-poles.
Hence, the $L=1$ state must either have a one-loop divergence, e.g.\ by 
mixing with other states, or at two-loops there must be further 
diagrams which cancel the quadratic $\varepsilon$-pole or even the entire
contribution from the diagram in Figure \ref{fig:Z-wrapping}. 
Such a one-loop mixing or additional two-loop contributions 
could e.g.\ be related to the 
fact that the considered gauge theory state admits interactions of the 
$Y$ and $\bar Y$ fields at large $N$. As mentioned in the main text, 
this state is not known and it is possible that for this state
$\ln\mathcal{Z}$ is free of higher order $\varepsilon$-poles.

For generic $L\ge2$, the integrals are given by the second 
of the figures \ref{fig:ZL-wrapping}. 
They are free of subdivergences and a unifying expression 
can be found for its pole parts, giving them as a function of 
$L$.\footnote{For other examples of such integrals see e.g.\
\cite{Broadhurst:1985vq} and \cite{Fiamberti:2008sn}.}
Our conjecture together with analytical and numerical data and 
the discussion with one of the authors led to 
\cite{Schnetz:2012nt}, in which
a map also of these integrals to the 
zig-zag integrals $I_{\text{Z}_n}$ with 
$n$ loops was presented. 
A conjecture for the pole parts of $I_{\text{Z}_n}$
was made almost 20 years ago \cite{Broadhurst:1995km} and it was proven recently
in \cite{Brown:2012ia}. 
In our conventions, the result of \cite{Broadhurst:1995km,Brown:2012ia} reads
\begin{equation}
\label{Izn}
I_{\text{Z}_n}=
\settoheight{\eqoff}{$\times$}%
\setlength{\eqoff}{0.5\eqoff}%
\addtolength{\eqoff}{-9.5\unitlength}%
\raisebox{\eqoff}{%
\fmfframe(-1,6)(-2,-2){%
\begin{fmfchar*}(30,15)
  \fmfleft{in}
  \fmfright{out1}
\fmf{phantom}{in,v1}
\fmf{phantom}{out,v5}
\fmfforce{(0,h)}{in}
\fmfforce{(w,0h)}{out}
\fmffixed{(0.333w,0)}{v2,v3}
\fmfpoly{phantom,label=$\scriptstyle 1$}{v1,v2,v3}
\fmfpoly{phantom,label=$\scriptstyle 2$}{v1,v3,v4}
\fmfpoly{phantom,label=$\scriptstyle $}{v4,v3,v5}
\fmfpoly{phantom,label=$\scriptstyle n-1$}{v4,v5,v6}
\fmf{plain}{v1,v2}
\fmf{plain}{v1,v3}
\fmf{plain}{v2,v3}
\fmf{plain}{v1,v4}
\fmf{plain}{v3,v4}
\fmf{dots}{v3,v5}
\fmf{dots}{v4,v5}
\fmf{plain}{v4,v6}
\fmf{plain}{v5,v6}
\fmf{plain,left=1}{v2,v6}
\end{fmfchar*}}}
\col\qquad
\mathcal{I}_{\text{Z}_n}=\Kop[I_{\text{Z}_n}]
=\frac{\lambda^n}{(4\pi)^{2n}\varepsilon}\frac{4}{n^2}\binom{2n-2}{n-1}\Big(1-\frac{1}{2n-3}(1-(-1)^n)\Big)\zeta(2n-3)
\pnt
\end{equation}

In the following, we will summarize in brief the
argument presented in \cite{Schnetz:2012nt}, i.e.\ the mapping 
of the integrals in Figure \ref{fig:ZL-wrapping} to the zig-zag integrals.
We set $n=L-1\ge2$ such that the respective loop integral contains $2L=2n+2$
loops and consider the dual graph\footnote{Taking the dual means going from the coordinate-space to the momentum-space representation of the diagram.}
\begin{equation}
\begin{aligned}
I_{2n+2}
&=
\settoheight{\eqoff}{$\times$}%
\setlength{\eqoff}{0.5\eqoff}%
\addtolength{\eqoff}{-15\unitlength}%
\raisebox{\eqoff}{%
\fmfframe(0,-15)(0,5){%
\begin{fmfchar*}(50,40)
\fmftop{vt}
\fmfbottom{vb}
\fmfforce{(0.5w,h)}{vt}
\fmfforce{(0.5w,0)}{vb}
\fmffixed{(whatever,0.25h)}{vsb2,vst2}
\fmf{phantom,left=0.25}{vst2,vt}
\fmf{phantom}{vst3,vt}
\fmf{phantom,left=0.25}{vst4,vt}
\fmf{plain,right=0.25}{vsb2,vb}
\fmf{dots}{vsb3,vb}
\fmf{plain,left=0.25}{vsb4,vb}
\fmf{phantom,left=0.75}{vsb1,vt}
\fmf{phantom,left=0.4}{vst1,vt}
\fmf{plain,right=0.75}{vst1,vb}
\fmf{plain,right=0.4}{vsb1,vb}
\fmf{phantom,right=0.75}{vsb5,vt}
\fmf{phantom,right=0.4}{vst5,vt}
\fmf{plain,left=0.75}{vst5,vb}
\fmf{plain,left=0.4}{vsb5,vb}
\fmfpoly{phantom,label=$\scriptstyle 1$}{vsb1,vsb2,vst2,vst1}
\fmfpoly{phantom}{vs1,vsb2,vst2}
\fmfpoly{phantom,label=$\scriptstyle 2$}{vsb2,vsb3,vst3,vst2}
\fmfpoly{phantom,label=$\scriptstyle\dots$}{vsb3,vsb4,vst4,vst3}
\fmfpoly{phantom,label=$\scriptstyle n$}{vsb4,vsb5,vst5,vst4}
\fmfpoly{phantom}{vs5,vst4,vsb4}
\fmf{plain}{vst1,vst2}
\fmf{plain}{vst2,vsb2}
\fmf{plain}{vsb2,vsb1}
\fmf{plain}{vsb1,vst1}
\fmf{plain}{vst2,vst3}
\fmf{dots}{vst3,vsb3}
\fmf{plain}{vsb3,vsb2}
\fmf{dots}{vst3,vst4}
\fmf{plain}{vst4,vsb4}
\fmf{dots}{vsb4,vsb3}
\fmf{plain}{vst4,vst5}
\fmf{plain}{vst5,vsb5}
\fmf{plain}{vsb5,vsb4}
\end{fmfchar*}}}
=
\settoheight{\eqoff}{$\times$}%
\setlength{\eqoff}{0.5\eqoff}%
\addtolength{\eqoff}{-15\unitlength}%
\raisebox{\eqoff}{%
\fmfframe(0,-15)(0,5){%
\begin{fmfchar*}(50,40)
\fmftop{vt}
\fmfbottom{vb}
\fmfforce{(0.5w,h)}{vt}
\fmfforce{(0.5w,0)}{vb}
\fmffixed{(whatever,0.25h)}{vsb2,vst2}
\fmf{phantom,left=0.25}{vst2,vt}
\fmf{phantom}{vst3,vt}
\fmf{phantom}{vst4,vt}
\fmf{phantom,right=0.25}{vst5,vt}
\fmf{plain,right=0.25}{vsb2,vb}
\fmf{plain}{vsb3,vb}
\fmf{dots}{vsb4,vb}
\fmf{plain,left=0.25}{vsb5,vb}
\fmf{phantom,left=0.4}{vs1,vt}
\fmf{plain,right=0.4}{vs1,vb}
\fmf{phantom,right=0.4}{vs6,vt}
\fmf{plain,left=0.4}{vs6,vb}
\fmfpoly{phantom}{vsb1,vsb2,vst2,vst1}
\fmfpoly{phantom}{vs1,vsb2,vst2}
\fmfpoly{phantom,label=$\scriptstyle 1$}{vsb2,vsb3,vst3,vst2}
\fmfpoly{phantom,label=$\scriptstyle\dots$}{vsb3,vsb4,vst4,vst3}
\fmfpoly{phantom,label=$\scriptstyle n-1$}{vsb4,vsb5,vst5,vst4}
\fmfpoly{phantom}{vsb5,vsb6,vst6,vst5}
\fmfpoly{phantom}{vs6,vst5,vsb5}
\fmf{plain}{vs1,vst2}
\fmf{plain}{vst2,vsb2}
\fmf{plain}{vsb2,vs1}
\fmf{plain}{vst2,vst3}
\fmf{plain}{vst3,vsb3}
\fmf{plain}{vsb3,vsb2}
\fmf{dots}{vst3,vst4}
\fmf{dots}{vst4,vsb4}
\fmf{dots}{vsb4,vsb3}
\fmf{plain}{vst4,vst5}
\fmf{plain}{vst5,vsb5}
\fmf{plain}{vsb5,vsb4}
\fmf{plain}{vs6,vsb5}
\fmf{plain}{vst5,vs6}
\end{fmfchar*}}}
\pnt
\end{aligned}
\end{equation}
Note that in the following we understand equal signs as equalities only of 
the overall divergencies on both sides.
For $n=1$ the pole part is directly given by the one of the $4$-loop 
zig-zag integral $\mathcal{I}_{4}=\mathcal{I}_{\text{Z}_4}$. 
For $n=2$ we see that the dual diagram is the zig-zag integral, and 
hence we find $\mathcal{I}_{6}=\mathcal{I}_{\text{Z}_6}$.

For $n\ge3$, following \cite{Schnetz:2012nt},  
we add a vertex at $\infty$, and connect it with propagators
to the three-valent vertices such that the 
integral becomes conformally invariant. This involves
adding a line of negative weight between the two $(n+3)$-valent vertices
at zero and infinity. 
Then, we apply the twist identity of \cite{Schnetz:2008mp} as follows:
first, we identify four vertices subject to the
condition that the integral decomposes into two disconnected 
pieces when these points are erased. Moreover, these vertices should not 
be connected each by one propagator only to a common further vertex.
The selected vertices will be depicted in blue.
Then, we add auxiliary 
lines connecting the four chosen vertices in a particular way 
\cite{Schnetz:2012nt}. They form two sets of paired lines as will be
indicated by using different colors for them. 
In a next step, all lines which belong to the left part of the diagram
and enter the chosen points are rearranged: their endpoints are 
permuted among the upper and lower two pairs of selected points.
This may lead to vertices that are no longer four-valent and hence
conformal invariance appears to be broken. It is, however, restored in the 
next step where the paired auxiliary lines come into play:
propagators running parallel to the auxiliary lines can be shifted
to run along the respective paired auxiliary lines, thereby
ensuring that the vertices become again four-valent.
One step of applying this procedure, i.e.\ the twist identity of \cite{Schnetz:2008mp}, can be visualized as follows
\begin{equation}
\begin{aligned}
\settoheight{\eqoff}{$\times$}%
\setlength{\eqoff}{0.5\eqoff}%
\addtolength{\eqoff}{-25\unitlength}%
\raisebox{\eqoff}{%
\fmfframe(0,5)(20,5){%
\begin{fmfchar*}(50,40)
\fmftop{vt}
\fmfbottom{vb}
\fmfforce{(0.5w,h)}{vt}
\fmfforce{(0.5w,0)}{vb}
\fmffixed{(whatever,0.25h)}{vsb2,vst2}
\fmf{plain,left=0.25}{vst2,vt}
\fmf{plain}{vst3,vt}
\fmf{dots}{vst4,vt}
\fmf{plain,right=0.25}{vst5,vt}
\fmf{plain,right=0.25}{vsb2,vb}
\fmf{plain}{vsb3,vb}
\fmf{dots}{vsb4,vb}
\fmf{plain,left=0.25}{vsb5,vb}
\fmf{plain,left=0.4}{vs1,vt}
\fmf{plain,right=0.4}{vs1,vb}
\fmf{plain,right=0.4}{vs6,vt}
\fmf{plain,left=0.4}{vs6,vb}
\fmfpoly{phantom}{vsb1,vsb2,vst2,vst1}
\fmfpoly{plain,full,filled,fore=(0.8,,0.8,,0.8),back=(0.8,,0.8,,0.8),smooth,pull=1.5,label=$\scriptstyle z_{2k-2}$}{vs1,vsb2,vst2}
\fmfpoly{phantom,label=$\scriptstyle k$}{vsb2,vsb3,vst3,vst2}
\fmfpoly{phantom,label=$\scriptstyle\dots$}{vsb3,vsb4,vst4,vst3}
\fmfpoly{phantom,label=$\scriptstyle n-1$}{vsb4,vsb5,vst5,vst4}
\fmfpoly{phantom}{vsb5,vsb6,vst6,vst5}
\fmfpoly{phantom}{vs6,vst5,vsb5}
\fmf{plain}{vs1,vst2}
\fmf{plain}{vst2,vsb2}
\fmf{plain}{vsb2,vs1}
\fmf{plain}{vst2,vst3}
\fmf{plain}{vst3,vsb3}
\fmf{plain}{vsb3,vsb2}
\fmf{dots}{vst3,vst4}
\fmf{dots}{vst4,vsb4}
\fmf{dots}{vsb4,vsb3}
\fmf{plain}{vst4,vst5}
\fmf{plain}{vst5,vsb5}
\fmf{plain}{vsb5,vsb4}
\fmf{plain}{vs6,vsb5}
\fmf{plain}{vst5,vs6}
\fmfv{decor.shape=circle,decor.size=4,fillstyle=empty,fore=blue}{vt}
\fmfv{decor.shape=circle,decor.size=4,fillstyle=empty,fore=blue}{vst3}
\fmfv{decor.shape=circle,decor.size=4,fillstyle=empty,fore=blue}{vsb3}
\fmfv{decor.shape=circle,decor.size=4,fillstyle=empty,fore=blue}{vb}
\fmf{plain,l.dist=2,label=$\scriptstyle k-n+1$,left=1.75}{vt,vb}
\fmf{plain,fore=red,left=0.125}{vt,vsb3}
\fmf{plain,fore=green,left=0.5}{vsb3,vst3}
\fmf{plain,fore=red,left=0.125}{vst3,vb}
\fmf{plain,fore=green,right=1.6}{vb,vt}
\end{fmfchar*}}}
&=
\settoheight{\eqoff}{$\times$}%
\setlength{\eqoff}{0.5\eqoff}%
\addtolength{\eqoff}{-25\unitlength}%
\raisebox{\eqoff}{%
\fmfframe(0,5)(20,5){%
\begin{fmfchar*}(50,40)
\fmftop{vt}
\fmfbottom{vb}
\fmfforce{(0.5w,h)}{vt}
\fmfforce{(0.5w,0)}{vb}
\fmffixed{(whatever,0.25h)}{vsb2,vst2}
\fmf{phantom,left=0.25}{vst2,vt}
\fmf{plain}{vst3,vt}
\fmf{dots}{vst4,vt}
\fmf{plain,right=0.25}{vst5,vt}
\fmf{phantom,right=0.25}{vsb2,vb}
\fmf{plain}{vsb3,vb}
\fmf{dots}{vsb4,vb}
\fmf{plain,left=0.25}{vsb5,vb}
\fmf{phantom,left=0.4}{vs1,vt}
\fmf{plain,left=0.5}{vs1,vst3}
\fmf{phantom,right=0.5}{vs1,vb}
\fmf{plain,right=0.5}{vs1,vsb3}
\fmf{plain,right=0.4}{vs6,vt}
\fmf{plain,left=0.4}{vs6,vb}
\fmfpoly{phantom}{vsb1,vsb2,vst2,vst1}
\fmfpoly{plain,full,filled,fore=(0.8,,0.8,,0.8),back=(0.8,,0.8,,0.8),smooth,pull=1.5,label=$\scriptstyle z_{2k-2}$}{vs1,vsb2,vst2}
\fmfpoly{phantom,label=$\scriptstyle k$}{vsb2,vsb3,vst3,vst2}
\fmfpoly{phantom,label=$\scriptstyle\dots$}{vsb3,vsb4,vst4,vst3}
\fmfpoly{phantom,label=$\scriptstyle n-1$}{vsb4,vsb5,vst5,vst4}
\fmfpoly{phantom}{vsb5,vsb6,vst6,vst5}
\fmfpoly{phantom}{vs6,vst5,vsb5}
\fmf{plain}{vs1,vst2}
\fmf{plain}{vst2,vsb2}
\fmf{plain}{vsb2,vs1}
\fmf{plain}{vst2,vst3}
\fmf{plain}{vst3,vsb3}
\fmf{plain}{vsb3,vsb2}
\fmf{dots}{vst3,vst4}
\fmf{dots}{vst4,vsb4}
\fmf{dots}{vsb4,vsb3}
\fmf{plain}{vst4,vst5}
\fmf{plain}{vst5,vsb5}
\fmf{plain}{vsb5,vsb4}
\fmf{plain}{vs6,vsb5}
\fmf{plain}{vst5,vs6}
\fmfv{decor.shape=circle,decor.size=4,fillstyle=empty,fore=blue}{vt}
\fmfv{decor.shape=circle,decor.size=4,fillstyle=empty,fore=blue}{vst3}
\fmfv{decor.shape=circle,decor.size=4,fillstyle=empty,fore=blue}{vsb3}
\fmfv{decor.shape=circle,decor.size=4,fillstyle=empty,fore=blue}{vb}
\fmf{plain,l.dist=2,label=$\scriptstyle k-n+1$,left=1.75}{vt,vb}
\fmf{plain,fore=red,left=0.125}{vt,vsb3}
\fmf{plain,fore=green,left=0.5}{vsb3,vst3}
\fmf{plain,fore=red,left=0.125}{vst3,vb}
\fmf{plain,fore=green,right=1.6}{vb,vt}
\fmffreeze
\fmfposition
\fmfipath{p[]}
\fmfiset{p1}{vpath(__vst2,__vt)}
\fmfiset{p2}{vpath(__vsb2,__vb)}
\fmfcmd{draw (subpath (0,0.15) of p1) withpen pencircle scaled 1 withcolor black;}
\fmfcmd{draw (subpath (0.35,2) of p1) withpen pencircle scaled 1 withcolor black;}
\fmfcmd{draw (subpath (0,0.15) of p2) withpen pencircle scaled 1 withcolor black;}
\fmfcmd{draw (subpath (0.35,2) of p2) withpen pencircle scaled 1 withcolor black;}
\end{fmfchar*}}}
\\
=
\settoheight{\eqoff}{$\times$}%
\setlength{\eqoff}{0.5\eqoff}%
\addtolength{\eqoff}{-25\unitlength}%
\raisebox{\eqoff}{%
\fmfframe(5,5)(20,5){%
\begin{fmfchar*}(50,40)
\fmftop{vt}
\fmfbottom{vb}
\fmfforce{(0.5w,h)}{vt}
\fmfforce{(0.5w,0)}{vb}
\fmffixed{(whatever,0.25h)}{vsb2,vst2}
\fmf{plain,left=0.25}{vst2,vt}
\fmf{plain}{vst3,vt}
\fmf{dots}{vst4,vt}
\fmf{plain,right=0.25}{vst5,vt}
\fmf{plain,right=0.25}{vsb2,vb}
\fmf{plain}{vsb3,vb}
\fmf{dots}{vsb4,vb}
\fmf{plain,left=0.25}{vsb5,vb}
\fmf{plain,left=0.4}{vst1,vt}
\fmf{plain,right=0.5}{vsb1,vb}
\fmf{plain,right=0.4}{vs6,vt}
\fmf{plain,left=0.4}{vs6,vb}
\fmfpoly{phantom}{vsb1,vsb2,vst2,vst1}
\fmfpoly{phantom,label=$\scriptstyle k+1$}{vsb2,vsb3,vst3,vst2}
\fmfpoly{phantom,label=$\scriptstyle\dots$}{vsb3,vsb4,vst4,vst3}
\fmfpoly{phantom,label=$\scriptstyle n-1$}{vsb4,vsb5,vst5,vst4}
\fmfpoly{phantom}{vsb5,vsb6,vst6,vst5}
\fmfpoly{phantom}{vs6,vst5,vsb5}
\fmf{plain}{vst1,vst2}
\fmf{phantom}{vst2,vsb2}
\fmf{plain}{vsb2,vsb1}
\fmf{plain}{vsb1,vst1}
\fmfpoly{phantom}{vs1,vst1,vsb1}
\fmfpoly{plain,full,filled,fore=(0.8,,0.8,,0.8),back=(0.8,,0.8,,0.8),smooth,pull=1.5,label=$\scriptstyle z_{2k-2}$}{vs1,vst1,vsb1}
\fmf{plain}{vsb1,vs1}
\fmf{plain}{vst1,vs1}
\fmf{plain}{vs1,vsb2}
\fmf{plain}{vs1,vst2}
\fmf{plain}{vst2,vst3}
\fmf{plain}{vst3,vsb3}
\fmf{plain}{vsb3,vsb2}
\fmf{dots}{vst3,vst4}
\fmf{dots}{vst4,vsb4}
\fmf{dots}{vsb4,vsb3}
\fmf{plain}{vst4,vst5}
\fmf{plain}{vst5,vsb5}
\fmf{plain}{vsb5,vsb4}
\fmf{plain}{vs6,vsb5}
\fmf{plain}{vst5,vs6}
\fmfv{decor.shape=circle,decor.size=4,fillstyle=empty,fore=blue}{vt}
\fmfv{decor.shape=circle,decor.size=4,fillstyle=empty,fore=blue}{vst2}
\fmfv{decor.shape=circle,decor.size=4,fillstyle=empty,fore=blue}{vsb2}
\fmfv{decor.shape=circle,decor.size=4,fillstyle=empty,fore=blue}{vb}
\fmf{plain,l.dist=2,label=$\scriptstyle k-n+2$,left=1.75}{vt,vb}
\fmffreeze
\end{fmfchar*}}}
&=
\settoheight{\eqoff}{$\times$}%
\setlength{\eqoff}{0.5\eqoff}%
\addtolength{\eqoff}{-25\unitlength}%
\raisebox{\eqoff}{%
\fmfframe(0,5)(20,5){%
\begin{fmfchar*}(50,40)
\fmftop{vt}
\fmfbottom{vb}
\fmfforce{(0.5w,h)}{vt}
\fmfforce{(0.5w,0)}{vb}
\fmffixed{(whatever,0.25h)}{vsb2,vst2}
\fmf{plain,left=0.25}{vst2,vt}
\fmf{plain}{vst3,vt}
\fmf{dots}{vst4,vt}
\fmf{plain,right=0.25}{vst5,vt}
\fmf{plain,right=0.25}{vsb2,vb}
\fmf{plain}{vsb3,vb}
\fmf{dots}{vsb4,vb}
\fmf{plain,left=0.25}{vsb5,vb}
\fmf{plain,left=0.4}{vs1,vt}
\fmf{plain,right=0.4}{vs1,vb}
\fmf{plain,right=0.4}{vs6,vt}
\fmf{plain,left=0.4}{vs6,vb}
\fmfpoly{phantom}{vsb1,vsb2,vst2,vst1}
\fmfpoly{plain,full,filled,fore=(0.8,,0.8,,0.8),back=(0.8,,0.8,,0.8),smooth,pull=1.5,label=$\scriptstyle z_{2k}$}{vs1,vsb2,vst2}
\fmfpoly{phantom,label=$\scriptstyle k+1$}{vsb2,vsb3,vst3,vst2}
\fmfpoly{phantom,label=$\scriptstyle\dots$}{vsb3,vsb4,vst4,vst3}
\fmfpoly{phantom,label=$\scriptstyle n-1$}{vsb4,vsb5,vst5,vst4}
\fmfpoly{phantom}{vsb5,vsb6,vst6,vst5}
\fmfpoly{phantom}{vs6,vst5,vsb5}
\fmf{plain}{vs1,vst2}
\fmf{plain}{vst2,vsb2}
\fmf{plain}{vsb2,vs1}
\fmf{plain}{vst2,vst3}
\fmf{plain}{vst3,vsb3}
\fmf{plain}{vsb3,vsb2}
\fmf{dots}{vst3,vst4}
\fmf{dots}{vst4,vsb4}
\fmf{dots}{vsb4,vsb3}
\fmf{plain}{vst4,vst5}
\fmf{plain}{vst5,vsb5}
\fmf{plain}{vsb5,vsb4}
\fmf{plain}{vs6,vsb5}
\fmf{plain}{vst5,vs6}
\fmfv{decor.shape=circle,decor.size=4,fillstyle=empty,fore=blue}{vt}
\fmfv{decor.shape=circle,decor.size=4,fillstyle=empty,fore=blue}{vst3}
\fmfv{decor.shape=circle,decor.size=4,fillstyle=empty,fore=blue}{vsb3}
\fmfv{decor.shape=circle,decor.size=4,fillstyle=empty,fore=blue}{vb}
\fmf{plain,l.dist=2,label=$\scriptstyle k-n+2$,left=1.75}{vt,vb}
\fmf{plain,fore=red,left=0.125}{vt,vsb3}
\fmf{plain,fore=green,left=0.5}{vsb3,vst3}
\fmf{plain,fore=red,left=0.125}{vst3,vb}
\fmf{plain,fore=green,right=1.6}{vb,vt}
\end{fmfchar*}}}
\col
\end{aligned}
\end{equation}
where the subgraph $z_{2k}$ is obtained from $z_{2k-2}$ as follows
\begin{equation}
\begin{aligned}
\settoheight{\eqoff}{$\times$}%
\setlength{\eqoff}{0.5\eqoff}%
\addtolength{\eqoff}{-7.5\unitlength}%
\raisebox{\eqoff}{%
\fmfframe(0,0)(0,0){%
\begin{fmfchar*}(15,15)
\fmftop{vt}
\fmfbottom{vb}
\fmfforce{(0.725w,h)}{vt}
\fmfforce{(0.725w,0)}{vb}
\fmffixed{(0,-0.166h)}{vtl,vs1}
\fmffixed{(0,0.166h)}{vbl,vs1}
\fmffixed{(0,-0.166h)}{vt2,vst2}
\fmffixed{(0,0.166h)}{vb2,vsb2}
\fmffixed{(0.166w,0)}{vst2,vtr}
\fmffixed{(0.166w,0)}{vsb2,vbr}
\fmffixed{(whatever,0.666h)}{vsb2,vst2}
\fmf{phantom}{vst2,vt}
\fmf{phantom}{vsb2,vb}
\fmfpoly{plain,full,filled,fore=(0.8,,0.8,,0.8),back=(0.8,,0.8,,0.8),smooth,pull=1.5,label=$\scriptstyle z_0$}{vs1,vsb2,vst2}
\fmf{plain}{vs1,vtl}
\fmf{plain}{vs1,vbl}
\fmf{plain}{vst2,vt2}
\fmf{plain}{vst2,vtr}
\fmf{plain}{vsb2,vb2}
\fmf{plain}{vsb2,vbr}
\end{fmfchar*}}}
=
\settoheight{\eqoff}{$\times$}%
\setlength{\eqoff}{0.5\eqoff}%
\addtolength{\eqoff}{-7.5\unitlength}%
\raisebox{\eqoff}{%
\fmfframe(0,0)(0,0){%
\begin{fmfchar*}(15,15)
\fmftop{vt}
\fmfbottom{vb}
\fmfforce{(0.725w,h)}{vt}
\fmfforce{(0.725w,0)}{vb}
\fmffixed{(0,-0.166h)}{vtl,vs1}
\fmffixed{(0,0.166h)}{vbl,vs1}
\fmffixed{(0,-0.166h)}{vt2,vst2}
\fmffixed{(0,0.166h)}{vb2,vsb2}
\fmffixed{(0.166w,0)}{vst2,vtr}
\fmffixed{(0.166w,0)}{vsb2,vbr}
\fmffixed{(whatever,0.666h)}{vsb2,vst2}
\fmf{phantom}{vst2,vt}
\fmf{phantom}{vsb2,vb}
\fmfpoly{phantom}{vs1,vsb2,vst2}
\fmf{plain}{vs1,vsb2}
\fmf{plain}{vsb2,vst2}
\fmf{plain}{vst2,vs1}
\fmf{plain}{vs1,vtl}
\fmf{plain}{vs1,vbl}
\fmf{plain}{vst2,vt2}
\fmf{plain}{vst2,vtr}
\fmf{plain}{vsb2,vb2}
\fmf{plain}{vsb2,vbr}
\end{fmfchar*}}}
\col
\qquad
\settoheight{\eqoff}{$\times$}%
\setlength{\eqoff}{0.5\eqoff}%
\addtolength{\eqoff}{-7.5\unitlength}%
\raisebox{\eqoff}{%
\fmfframe(0,0)(0,0){%
\begin{fmfchar*}(15,15)
\fmftop{vt}
\fmfbottom{vb}
\fmfforce{(0.725w,h)}{vt}
\fmfforce{(0.725w,0)}{vb}
\fmffixed{(0,-0.166h)}{vtl,vs1}
\fmffixed{(0,0.166h)}{vbl,vs1}
\fmffixed{(0,-0.166h)}{vt2,vst2}
\fmffixed{(0,0.166h)}{vb2,vsb2}
\fmffixed{(0.166w,0)}{vst2,vtr}
\fmffixed{(0.166w,0)}{vsb2,vbr}
\fmffixed{(whatever,0.666h)}{vsb2,vst2}
\fmf{phantom}{vst2,vt}
\fmf{phantom}{vsb2,vb}
\fmfpoly{plain,full,filled,fore=(0.8,,0.8,,0.8),back=(0.8,,0.8,,0.8),smooth,pull=1.5,label=$\scriptstyle z_{2k}$}{vs1,vsb2,vst2}
\fmf{plain}{vs1,vtl}
\fmf{plain}{vs1,vbl}
\fmf{plain}{vst2,vt2}
\fmf{plain}{vst2,vtr}
\fmf{plain}{vsb2,vb2}
\fmf{plain}{vsb2,vbr}
\end{fmfchar*}}}
=
\settoheight{\eqoff}{$\times$}%
\setlength{\eqoff}{0.5\eqoff}%
\addtolength{\eqoff}{-12.5\unitlength}%
\raisebox{\eqoff}{%
\fmfframe(0,5)(0,5){%
\begin{fmfchar*}(15,15)
\fmftop{vt}
\fmfbottom{vb}
\fmfforce{(0.725w,h)}{vt}
\fmfforce{(0.725w,0)}{vb}
\fmffixed{(0,-0.166h)}{vtl,vst2}
\fmffixed{(0,0.166h)}{vbl,vsb2}
\fmffixed{(0,-0.166h)}{vt2,vst1}
\fmffixed{(0,0.166h)}{vb2,vsb1}
\fmffixed{(0.166w,0)}{vst1,vtr}
\fmffixed{(0.166w,0)}{vsb1,vbr}
\fmffixed{(whatever,0.666h)}{vsb2,vst2}
\fmf{phantom}{vst1,vt}
\fmf{phantom}{vsb1,vb}
\fmfpoly{phantom}{vs1,vst1,vst2}
\fmfpoly{plain,full,filled,fore=(0.8,,0.8,,0.8),back=(0.8,,0.8,,0.8),smooth,pull=1.5,label=$\scriptstyle z_{2k-2}$}{vs1,vst2,vsb2}
\fmfpoly{phantom}{vs1,vsb2,vsb1}
\fmf{plain}{vs1,vst1}
\fmf{plain}{vst1,vst2}
\fmf{plain}{vst2,vs1}
\fmf{plain}{vs1,vsb2}
\fmf{plain}{vsb2,vsb1}
\fmf{plain}{vsb1,vs1}
\fmf{plain}{vst2,vtl}
\fmf{plain}{vsb2,vbl}
\fmf{plain}{vst1,vt2}
\fmf{plain}{vst1,vtr}
\fmf{plain}{vsb1,vb2}
\fmf{plain}{vsb1,vbr}
\end{fmfchar*}}}
\col
\end{aligned}
\end{equation}
i.e.\ $z_{2k}$ is given by a zig-zag line with $2k+1$ triangles.
The above procedure can be applied $n-2$ times. After the last step, 
we obtain
\begin{equation}
\begin{aligned}
I_{2n+2}
&=
\settoheight{\eqoff}{$\times$}%
\setlength{\eqoff}{0.5\eqoff}%
\addtolength{\eqoff}{-20\unitlength}%
\raisebox{\eqoff}{%
\fmfframe(0,0)(0,0){%
\begin{fmfchar*}(30,40)
\fmftop{vt}
\fmfbottom{vb}
\fmfforce{(0.5w,h)}{vt}
\fmfforce{(0.5w,0)}{vb}
\fmffixed{(whatever,0.25h)}{vsb2,vst2}
\fmf{plain}{vst2,vt}
\fmf{plain}{vst3,vt}
\fmf{plain,right=0.25}{vs4,vt}
\fmf{plain}{vsb2,vb}
\fmf{plain}{vsb3,vb}
\fmf{plain,left=0.25}{vs4,vb}
\fmf{plain,left=0.25}{vs1,vt}
\fmf{plain,right=0.25}{vs1,vb}
\fmfpoly{phantom}{vsb1,vsb2,vst2,vst1}
\fmfpoly{plain,full,filled,fore=(0.8,,0.8,,0.8),back=(0.8,,0.8,,0.8),smooth,pull=1.5,label=$\scriptstyle z_{2n-4}$}{vs1,vsb2,vst2}
\fmfpoly{phantom}{vsb2,vsb3,vst3,vst2}
\fmfpoly{phantom}{vs4,vst3,vsb3}
\fmf{plain}{vs1,vst2}
\fmf{plain}{vst2,vsb2}
\fmf{plain}{vsb2,vs1}
\fmf{plain}{vst2,vst3}
\fmf{plain}{vst3,vsb3}
\fmf{plain}{vsb3,vsb2}
\fmf{plain}{vst3,vs4}
\fmf{plain}{vs4,vsb3}
\end{fmfchar*}}}
=
\settoheight{\eqoff}{$\times$}%
\setlength{\eqoff}{0.5\eqoff}%
\addtolength{\eqoff}{-20\unitlength}%
\raisebox{\eqoff}{%
\fmfframe(0,0)(0,0){%
\begin{fmfchar*}(30,40)
\fmftop{vt}
\fmfbottom{vb}
\fmfforce{(0.5w,h)}{vt}
\fmfforce{(0.5w,0)}{vb}
\fmffixed{(whatever,0.25h)}{vsb2,vst2}
\fmf{phantom}{vst2,vt}
\fmf{phantom}{vst3,vt}
\fmf{phantom,right=0.25}{vs4,vt}
\fmf{plain}{vsb2,vb}
\fmf{plain}{vsb3,vb}
\fmf{plain,left=0.25}{vs4,vb}
\fmf{phantom,left=0.25}{vs1,vt}
\fmf{plain,right=0.25}{vs1,vb}
\fmfpoly{phantom}{vsb1,vsb2,vst2,vst1}
\fmfpoly{plain,full,filled,fore=(0.8,,0.8,,0.8),back=(0.8,,0.8,,0.8),smooth,pull=1.5,label=$\scriptstyle z_{2n-4}$}{vs1,vsb2,vst2}
\fmfpoly{phantom}{vsb2,vsb3,vst3,vst2}
\fmfpoly{phantom}{vs4,vst3,vsb3}
\fmf{plain}{vs1,vst2}
\fmf{plain}{vst2,vsb2}
\fmf{plain}{vsb2,vs1}
\fmf{plain}{vst2,vst3}
\fmf{plain}{vst3,vsb3}
\fmf{plain}{vsb3,vsb2}
\fmf{plain}{vst3,vs4}
\fmf{plain}{vs4,vsb3}
\end{fmfchar*}}}
=
\settoheight{\eqoff}{$\times$}%
\setlength{\eqoff}{0.5\eqoff}%
\addtolength{\eqoff}{-7.5\unitlength}%
\raisebox{\eqoff}{%
\fmfframe(0,0)(3,0){%
\begin{fmfchar*}(15,15)
\fmftop{vt}
\fmfbottom{vb}
\fmfforce{(0.725w,h)}{vt}
\fmfforce{(0.725w,0)}{vb}
\fmffixed{(0,-0.166h)}{vtl,vs1}
\fmffixed{(0,0.166h)}{vbl,vs1}
\fmffixed{(0,-0.166h)}{vt2,vst2}
\fmffixed{(0,0.166h)}{vb2,vsb2}
\fmffixed{(0.166w,0)}{vst2,vtr}
\fmffixed{(0.166w,0)}{vsb2,vbr}
\fmffixed{(whatever,0.666h)}{vsb2,vst2}
\fmf{phantom}{vst2,vt}
\fmf{phantom}{vsb2,vb}
\fmfpoly{plain,full,filled,fore=(0.8,,0.8,,0.8),back=(0.8,,0.8,,0.8),smooth,pull=1.5,label=$\scriptstyle z_{2n}$}{vs1,vsb2,vst2}
\fmf{plain}{vs1,vtl}
\fmf{plain}{vs1,vbl}
\fmf{plain}{vst2,vt2}
\fmf{plain}{vsb2,vb2}
\fmf{plain,left=1}{vst2,vsb2}
\end{fmfchar*}}}
=I_{\text{Z}_{2n+2}}
\col
\end{aligned}
\end{equation}
where in the second step we have removed the vertex at $\infty$
in the conformally invariant integral. 
Then, the remaining 
four triangles extend the zig-zag line $z_{2n-4}$ with $2n-3$ triangles
to a zig-zag line $z_{2n}$ with $2n+1$ triangles.
The upper horizontal propagator
connects the two two-valent vertices of this zig-zag 
line such that one obtains the zig-zag integral $I_{\text{Z}_{2n+2}}$ with 
the known overall UV-divergence given in \eqref{Izn}.
Note that the result also extends to $n=1$.

\section{One-particle L\"uscher correction}\label{sec:oneparticle}

In order to provide further support for the correctness of the
asymptotic Y-functions obtained from the generating functional,  
we compute in this Appendix the L\"uscher correction for the $1\dot{1}$ particle reflecting
between the $Y\bar{Y}$ boundaries 
in two different ways: first from the Y-functions obtained by using the generating functional,
and second by appropriately modifying the ``direct'' L\"uscher 
computation done for 
the $YY$ case in \cite{Bajnok:2010ui}. These two computations should 
give identical results,
and it is shown below that this is indeed the case.
  
We start by solving the boundary Bethe-Yang equations determining the
momentum of the reflecting particles
\begin{equation}
1=e^{-2ipL}\mathbb{R}_{\bar{Y}}^{+}(-p)\mathbb{R}_{Y}^{-}(p) \,,
\label{bbye}
\end{equation}
where $\mathbb{R}_{\bar{Y}}^{+}(p)$ and $\mathbb{R}_{Y}^{-}(p)$ are 
given by Eqs. (\ref{fundrefl1}, \ref{fundrefl2}, \ref{ybarrefl}). Using them in
Eq. (\ref{bbye}) gives  
\begin{equation}
1=e^{-2ip(L+1)}\sigma(p,-p)^2
\end{equation}
for the $1\dot{1}$ particle (in fact for all the bosonic ones i.e. for
$1\dot{2}\ 2\dot{1}\ 2\dot{2}\ 3\dot{3}\ 3\dot{4}$ etc.), while for the
fermionic ones ($1\dot{3}\ 3\dot{1}\ 2\dot{3}\ 3\dot{2}\ 2\dot{4}$ etc.)    
\begin{equation}
1=-e^{-2ip(L+1)}\sigma(p,-p)^2
\end{equation}
is obtained. Therefore, in the weak coupling limit 
\begin{equation}
p_n=\frac{\pi}{L+1}n\quad {\rm for\ bosons}\,, \qquad\quad
p_n=\frac{\pi}{L+1} \( n+\frac{1}{2} \)
\quad {\rm for \ fermions}\,, \qquad\quad n=1,\dots,L \,.
\end{equation}
Note that this is consistent with supersymmetry being broken.

For the computation using the $Y_{a,0}$ functions, one can repeat the 
same steps taken for the $YY$ case.   
Since $S_0(p,p_1)$ is the same as for the $YY$ case, eventually the complete
normalization becomes (see Eqs. (4.9-4.11) of \cite{Bajnok:2012xc})
\begin{equation}
f_{a,1} =\left(  \frac{z^{[-a]}}{z^{[a]}} \right)^{2}\frac{u^{[-a]}}{u^{[a]}}
\frac{Q^{[a-1]}Q^{[1-a]}}{Q^{[a+1]}Q^{[-1-a]}}\simeq  \left(\frac{4g^{2}}{q^{2}+
{a^{2}}}\right)^{2}\frac{q-{i}a}{q+{i}a}
\frac{Q^{[a-1]}Q^{[1-a]}}{Q^{[a+1]}Q^{[-1-a]}} \,.
\end{equation}
From the weak coupling limit of (\ref{ybyeigenval}) we find for the eigenvalue
of the double-row transfer matrix \eqref{ybyeigenval}
\begin{equation}
T_{a,1} = (-1)^a \frac{aq P(q,a,u(p))}{4(q-ia) Q^{[a-1]}Q^{[1-a]}} \,,
\end{equation}
where
\begin{equation}
P(q,a,u)=(-3+q^{4}+2a^{2}-8 u^{2}+2q^{2}(-1+a^{2}-4 u^{2})+(a^{2}+4 u^{2})^{2})\,.
\end{equation}
Thus
\begin{equation}
Y_{a,0}\simeq
f_{a,1} \, T_{a,1}^2 \, e^{-2\tilde{\epsilon}_aL}=g^4\frac{a^2q^2P^2(q,a,u(p))}
{(q^2+a^2)^3Q^{[a+1]}Q^{[-1-a]}Q^{[a-1]}Q^{[1-a]}}\left(\frac{4g^2}{q^2+a^2}\right)^{2L} \,,
\label{Ya0 weak}
\end{equation}
and we can compute the leading weak coupling correction to the energy of the $1\dot{1}$ particle as 
\begin{equation}
\Delta E(L)=-\sum\limits_{a=1}^\infty\int\limits_0^\infty\frac{dq}{2\pi}Y_{a,0}=
\frac{-i}{2}\sum\limits_{a=1}^\infty {\rm Res} \, Y_{a,0}. 
\label{1dot1eq}
\end{equation}

As for the ``direct'' L\"uscher computation, recall that the approach 
in \cite{Bajnok:2010ui} works 
directly in the mirror theory where -- using the boundary state 
formalism -- 
this contribution can be depicted as

\vskip 4mm
\begin{minipage}{0.3\linewidth}
\centering
\includegraphics[height=4cm]{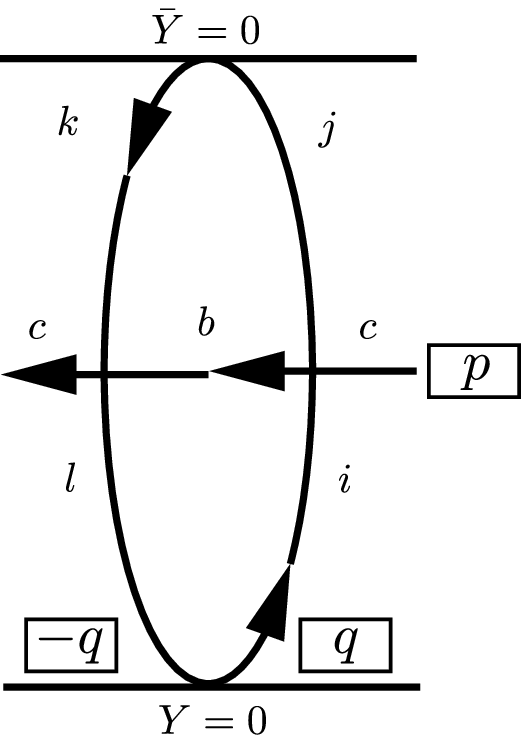}
\end{minipage}
\begin{minipage}{0.66\linewidth}
\begin{equation}
\Delta E (L)= \sum_a \int_{0}^{\infty}\frac{dq}{2\pi} \, \mathbb{K}^{\bar{l}i}(q)
\mathbb{S}_{i c}^{j b}(q,p)\bar{\mathbb{K}}_{j\bar{k}}(q)
\mathbb{S}_{\bar{l} b}^{\bar{k} c}(-q,p) e^{-2\tilde{\epsilon}_a L} .
\end{equation}
\end{minipage}
\vskip 4mm
\noindent
Here $c$ refers to the particle type whose energy correction we are
calculating (which, in the present case, is $1\dot{1}$), 
and the other indices run over $(4a)^2$ components of the $a$-th atypical representation of $\alg{su}(2|2) \oplus \alg{su}(2|2)$.
The boundary state amplitudes, $\mathbb{K}^{\bar{l}i}$ for the $Y=0$ and $\bar{\mathbb{K}}_{j\bar{k}}$ for the $\bar Y=0$, 
are related to the reflection factors (Eq. (\ref{eq:Ra})) by analytical continuation 
$\mathbb{K}^{ij}({\sf z}(q))=\mathbb{C}^{i\bar{i}}\mathbb{R}_{\bar{i}}^{j}(\frac{\omega}{2}-{\sf z}(q))$,
where ${\sf z} (q)$ is the uniformization parameter on the rapidity torus \cite{Arutyunov:2007tc}.
These boundary state amplitudes are the only ones in the whole computation we have to change 
in the $Y\bar{Y}$ case. Since both the bulk S-matrix and the
boundary state amplitudes factorize as $\mathbb{S}=S_{0}\, S\otimes S$ and $\mathbb{K}=K_{0}\, K \otimes K$, 
the energy correction can be written as 
\begin{equation}
\Delta
E(L)=-\sum_a \int_{0}^{\infty}\frac{dq}{2\pi}K_{0}(q)S_{0}(q,p)\bar{K}_{0}(q)S_{0}(-q,p)
\left[\mbox{Tr}\left(\bar{K}(q)S(q,p)K(q)S(-q,p)\right)
\right]^{2}e^{-2\tilde{\epsilon}_{a}L} \,.
\label{directlusch}
\end{equation}
In this expression we have to change only the matrix part compared to
\cite{Bajnok:2010ui} when making the summation over the bound state polarizations. 
Decomposing the $4a$ dimensional bound state representation 
$4a=(a+1)+(a-1)+a+a$ as in \cite{Bajnok:2010ui}, the non-vanishing components of the boundary
state amplitudes that correspond to the $Y\bar{Y}$ boundaries are
\begin{alignat}{9}
K^{11}_{j,a-j} &=\bar{K}^{11}_{j,a-j}=(-1)^j\,, &\qquad 
K^{22}_{j,a-2-j}&=\bar{K}^{22}_{j,a-2-j}=-(-1)^j\,,
\notag \\
K^{34}_{j,a-1-j} &=\bar{K}^{34}_{j,a-1-j}=-i(-1)^je^{-\tilde{\epsilon}_a/2}\,, &\qquad
K^{43}_{j,a-1-j} &=\bar{K}^{43}_{j,a-1-j}=-i(-1)^je^{\tilde{\epsilon}_a/2}\,.
\end{alignat}
Substituting these into the sums over the bound state polarizations given in
\cite{Bajnok:2010ui} leads to 
\begin{equation}
\mbox{Tr}\left(\bar{K}(q)S(q,p)K(q)S(-q,p)\right)=
-\frac{a\, P(q,a,u(p))}{(q+2 u(p) +i(a-1))(i(a-1)-q+2 u(p))(2 u(p) +i)^2}\,.
\end{equation}
Using this together with the fact that 
$\bar{K}_0(q)K_0(q) \equiv R_{0}({\sf z} (q) -\frac{\omega_{2}}{2})R_{0}(-{\sf z} (q)-\frac{\omega_{2}}{2})$ 
in place of Eq. (\ref{altnorm})  
and the explicit form of $S_0(q,p)$ given in \cite{Bajnok:2010ui}
\begin{equation*}
S_{0}(q,p) = \frac{(2u(p)+i)^{2}(2u(p)-q+i(a-1))}{(2u(p)-q-i(a+1))(2u(p)-q-i(a-1))(2u(p)-q+i(a+1))}
+O(g^{2}),
\end{equation*}
we find that the expression (\ref{directlusch}) exactly reproduces 
the integrand, $Y_{a,0}$, of the previous computation \eqref{Ya0 weak}.

For completeness we list here the first few cases of leading L\"uscher corrections  
computed from Eq. (\ref{1dot1eq})
\begin{align}
&L=1,\quad p=\frac{\pi}{2}: &&\Delta E=2^3g^8\left(4\zeta(3)-5\zeta(5)\right)\,,
\\
&L=2,\quad p=\frac{\pi}{3}: &&\Delta E=-2^7g^{12} \, \cdot\,  \frac{3}{8}\zeta(5)\,,
\\
&L=2,\quad p=\frac{2\pi}{3}: &&\Delta E=2^7g^{12}\left(\frac{81}{8}\zeta(5)
-\frac{21}{2}\zeta(9)\right)\,,
\\
&L=3,\quad p=\frac{\pi}{2}: &&\Delta E=2^{11}g^{16}\left(-\frac{15}{16}\zeta(7)
+\frac{165}{64}\zeta(11)-\frac{429}{256}\zeta(13)\right)\,.
\end{align}

\section{Generating function in the rotated case}\label{sec:genfun}

In this Appendix we analyze the generating functional for the 
vacuum state with generic angle (\ref{eq:gentheta}).

\subsection{Asymptotic solution of the T-system}

We now calculate the solution of the T-system \eqref{Tsystem} in the rotated case 
from the already explicitly calculated anti-symmetric transfer matrix eigenvalues:
\begin{equation}
T_{a,1}=(-1)^{a}\frac{4au}{u^{[-a]}}\sin^{2}\theta \,.
\label{eq:Ta1theta}
\end{equation}
Choosing the same boundary condition we had before $T_{0,s}=T_{a,0}=1$,
we can easily find 
\begin{equation}
T_{2,s}=T_{s,2}=\frac{16u^{[s]}u^{[-s]}}{u^{[-s+1]}u^{[-s-1]}}\sin^{4}\theta ,
\qquad (s \ge 2).
\end{equation}
The expression for the symmetric transfer matrix eigenvalues has a complicated
form (cf. Eq. (\ref{eq:T1s})): 
\begin{equation}
T_{1,s} = 2(-1)^{s}\left[a_{0,s}\Bigl(1+\frac{u^{[s]}}{u^{[-s]}}\Bigr)+
 2\sum_{k=1}^{s-1}\frac{a_{k,s}u^{[s]}}{u^{[s-2k]}}\right] \,,
\end{equation}
where 
\begin{equation}
a_{0,s}=\sum_{k=0}^{s-1}(-1)^{k}{s-1 \choose k}{s+k \choose s}
\cos^{2k}\theta\sin^{2}\theta \quad (s>0)\,, \qquad 
a_{l,s}=a_{s-l,s}=a_{0,l}a_{0,s-l} \,,
\end{equation}
and $a_{0,0}=1$

Remarkably, these transfer matrices can be generated from the following
generating functional 
\begin{equation}
\mathcal{W}_{\alg{su}(2)} =F\left(\mathcal{D}\frac{u^{+}}{u^{-}}\mathcal{D}\right)^{-1}\,
\left(1-\mathcal{D}\frac{u^{+}}{u^{-}}\mathcal{D}\right)\, \left(1-\mathcal{D}^{2}\right)\,
F\left(\mathcal{D}^{2}\right)^{-1} 
= \sum_{s=0}^{\infty} \mathcal{D}^{s}T_{1,s}  \mathcal{D}^{s} \,,
\label{eq:gentheta}
\end{equation}
where $F(z)$ is given by
\begin{equation}
F(z)=\sqrt{1-2\cos(2\theta)z+z^{2}} \,.    
\end{equation}
Evidently $F(z)$ can also be written as 
\begin{equation}
F(z)=\sqrt{(1-e^{i2\theta}z)(1-e^{-i2\theta}z)}\,,
\end{equation}
which is a simple deformation of $1-z$.  Indeed, for $\theta=0$, the
vacuum generating functional (\ref{eq:gentheta}) reduces to
$\mathcal{W}=1$ (i.e., $T_{1,s}=0$ for all $s>0$, which is consistent 
with the supersymmetry of the $YY$
system); while for $\theta=\pi/2$, $F(z)=1+z$ and therefore
(\ref{eq:gentheta}) reduces to the $Y{\bar Y}$ generating functional
(\ref{eq:genYbarY}).

We now verify that the inverse of this generating functional reproduces 
our previous result (\ref{eq:Ta1theta}) for 
the transfer matrix eigenvalues for anti-symmetric representations: 
\begin{equation}
\mathcal{W}_{\alg{su}(2)} =F\left(\mathcal{D}^{2}\right)\, \left(1-\mathcal{D}^{2}\right)^{-1}\,
\left(1-\mathcal{D}\frac{u^{+}}{u^{-}}\mathcal{D}\right)^{-1}\,
F\left(\mathcal{D}\frac{u^{+}}{u^{-}}\mathcal{D}\right)=
\sum_{a=0}^{\infty}(-1)^{a}\mathcal{D}^{a}T_{a,1}\mathcal{D}^{a} \,. 
\label{eq:invgentheta}
\end{equation}
To this end, we expand $F(z)$ in powers of $z$:
\begin{equation}
F(z)=\sum_{n=0}^{\infty}\alpha_{n}z^{n} \,.
\end{equation}
By expanding the inverse operators, we can write
\bal
\mathcal{W}_{\alg{su}(2)} &=  \sum_{n=0}^{\infty}\alpha_{n}\mathcal{D}^{2n}
\sum_{k=0}^{\infty}\mathcal{D}^{k}\frac{(k+1)u}{u^{[-k]}}\mathcal{D}^{k}
\sum_{m=0}^{\infty}\alpha_{m}\mathcal{D}^{m}\frac{u^{[m]}}{u^{[-m]}}\mathcal{D}^{m} \\
 & =  \sum_{n,m,k=0}^{\infty}\alpha_{n}\alpha_{m}(k+1)
 \mathcal{D}^{n+m+k}\frac{u^{[m-n]}}{u^{[-n-m-k]}}\mathcal{D}^{n+m+k} \,.
\eal
Comparing with (\ref{eq:invgentheta}), we can read off the following transfer matrix eigenvalues 
\beq
T_{a,1} = (-1)^{a}\frac{2u}{u^{[-a]}}\sum_{n=0}^{a}
\sum_{m=0}^{a-n}\alpha_{n}\alpha_{m}(a+1-n-m)
= (-1)^{a}\sin^{2}\theta\frac{4au}{u^{[-a]}} \,,
\eeq
which coincides with our previous result  (\ref{eq:Ta1theta}).
In passing to the last line we used that
\bal
s(\theta) & \equiv 2\sum_{n=0}^{a}\sum_{m=0}^{a-n}\alpha_{n}\alpha_{m}(a+1-n-m) 
=-\lim_{x\to1}PP\frac{d}{dx}\Big(\sum_{n=0}^{\infty}\sum_{m=0}^{\infty}\alpha_{n}\alpha_{m}x^{n+m-a-1}\Big) 
\\
& =-\lim_{x\to1}PP\frac{d}{dx}((1-2 \cos(2\theta) x+x^{2})x^{-a-1}) 
=-\lim_{x\to1}\frac{d}{dx}(x^{-a-1}-2 \cos(2\theta) 
x^{-a}+x^{-a+1})=4a\sin^{2}\theta \,,
\eal
where $PP$ denotes the principal part of the Laurent series, i.e.
terms with negative powers of $x$.

The generating functional can be rewritten in the following form.
\begin{equation}
\mathcal{W}^{-1} \equiv \mathcal{W}_{\alg{su}(2)} =\left(1-\mathcal{D}^{2}\right)^{-1}g\left(\mathcal{D}^{2}\right)\tilde{g}
\left(\mathcal{D}\frac{u^{+}}{u^{-}}\mathcal{D}\right)\left(1-\mathcal{D}\frac{u^{+}}{u^{-}}
\mathcal{D}\right)^{-1}
=\sum_{a=0}^{\infty}(-1)^{a}\mathcal{D}^{a}T_{a,1}(\theta)\mathcal{D}^{a} \,.
\end{equation}
We conjecture that this quantity is equal to the `dual' generating functional in the $\alg{sl}(2)$ grading, $\mathcal{W}^{-1}=\mathcal{W}_{\alg{sl}(2)}^{-1}$\,, just as in the $YY$ case (see Appendix C of \cite{Bajnok:2012xc}).

We calculate the middle $g\tilde{g}$ term as
\begin{equation}
g(\mathcal{D}^{2})\tilde{g}(\mathcal{D}\frac{u^{+}}{u^{-}}\mathcal{D})=
(1-\mathcal{D}^{2})\mathcal{W}^{-1}(1-\mathcal{D}\frac{u^{+}}{u^{-}}\mathcal{D})
\label{eq:gtildeg}
\end{equation}
We now use the relation between the generic $\theta$ case and 
the $\theta={\pi/2}$ case:
\begin{equation}
T_{a,1}(\theta)=\sin^{2}\theta\, T_{a,1}({\pi/2}) \,, \qquad a > 0 \,,
\end{equation}
which implies that
\beq
\mathcal{W}^{-1}(\theta)-1 
=\sum_{a=1}^{\infty}(-1)^{a}\mathcal{D}^{a}T_{a,1}(\theta)\mathcal{D}^{a} 
=
\sin^{2}\theta\sum_{a=1}^{\infty}(-1)^{a}\mathcal{D}^{a}T_{a,1}({\pi/2})\mathcal{D}^{a} 
= \sin^{2}\theta \left[\mathcal{W}^{-1}({\pi/2}) -1 \right] \,.
\eeq
Substituting this result for $\mathcal{W}^{-1}(\theta)$ into 
(\ref{eq:gtildeg}), and recalling that $\mathcal{W}^{-1}(\frac{\pi}{2})$
is given by the inverse of (\ref{eq:genYbarY}), we obtain
\bal
g\left(\mathcal{D}^{2}\right)\tilde{g}\left(\mathcal{D}\frac{u^{+}}{u^{-}}\mathcal{D}\right) 
&= 
\left(1-\mathcal{D}^{2}\right)\bigg[\cos^{2}\theta+\sin^{2}\theta\left(1+\mathcal{D}^{2}\right)
\left(1-\mathcal{D}^{2}\right)^{-1}
\\&\hskip1.6in\times
\left(1-\mathcal{D}\frac{u^{+}}{u^{-}}\mathcal{D}\right)^{-1}
\left(1+\mathcal{D}\frac{u^{+}}{u^{-}}\mathcal{D}\right)\bigg]\left(1-\mathcal{D}\frac{u^{+}}{u^{-}}
\mathcal{D}\right) \\
& =\cos^{2}\theta\left[\left(1-\mathcal{D}^{2}\right)\left(1-\mathcal{D}\frac{u^{+}}{u^{-}}
 \mathcal{D}\right)\right]+\sin^{2}\theta\left[\left(1+\mathcal{D}^{2}\right)\left(1+\mathcal{D}
 \frac{u^{+}}{u^{-}}\mathcal{D}\right)\right] \\
 & =1-\cos (2\theta) \mathcal{D}\left(1+\frac{u^{+}}{u^{-}}\right)
 \mathcal{D}+\mathcal{D}^{2}\mathcal{D}\frac{u^{+}}{u^{-}}\mathcal{D} \,.
\eal
Clearly the square roots have disappeared. 
This function can be written as $(1-\mathcal{D}^{2}A)(1-A^{-1}\mathcal{D} \frac{u^{+}}{u^{-}} \mathcal{D})$ where $A$ is the solution of the following equation
\begin{equation}
A^- + \frac{1}{A^+} \frac{u^{+}}{u^{-}} = \cos(2\theta) \left(1+\frac{u^{+}}{u^{-}}\right).
\end{equation}
Although we have not managed to find its solution for general angle, explicit solutions can be obtained at special values of $\theta$, such as
\begin{equation}
A \left(u,\theta=\frac{\pi}{4} \right) = \alpha \frac{\Gamma \left(-\frac{iu}{2} \right) 
\Gamma \left(\frac12 + \frac{iu}{2} \right)}{\Gamma \left(\frac{iu}{2}\right) 
\Gamma \left(\frac12 -\frac{iu}{2}\right)} \,, \qquad \alpha^+\alpha^-=1,
\end{equation}
or
\begin{equation}
A(u,\theta) = 1 + \frac{i \theta}{u} - 2 \left( 1 - \frac{r(u)}{4u^2} \right) \theta^2 + \cO (\theta^3),
\quad r(u) =  \frac{u}{2 i} \left( 3 + 8iu + \frac{u^2}{2}+4 \Phi \left(-1,1,2-2iu\right)-4 \psi\left(2-2iu\right) \right),
\end{equation}
where $\psi$ is the digamma function and $\Phi (-1,1,x) \equiv \sum_{k=0}^\infty (-1)^k/(k+x)$ is the Lerch transcendent.

We expect the generating functional for states in the $\alg{sl}(2)$ sector to be of
the form: 
\begin{equation}
\mathcal{W}_{\alg{sl}(2)}^{-1}=\left(1-\mathcal{D}^{2} \frac{R^{(-)}}{R^{(+)}} \right)^{-1}
\left[1-\cos (2\theta) \mathcal{D}\left(1+\frac{u^{+}}{u^{-}}\right)\mathcal{D}
+\mathcal{D}^{2}\mathcal{D}\frac{u^{+}}{u^{-}}\mathcal{D}\right]
\left(1-\frac{B^{(+)}}{B^{(-)}}\mathcal{D}\frac{u^{+}}{u^{-}}\mathcal{D}\right)^{-1} \,.
\end{equation}

\section{The large $Q$ behavior of $Y_Q$}\label{app:YQ}

In this appendix the large $Q$ behavior of the $Y_Q$ functions is determined.
As a first step the large $m$ behavior of the $Y_{m|vw}$ functions is investigated
with the help of (\ref{TBAmvw}).
For large $m$ the factor $(1+Y_m)$ becomes unity and $Y_{m \pm 1|vw}\simeq Y_{m|vw} $
substitution can be done. Thus for lager $m$ at leading order $Y_{m|vw}$ satisfies the equation:
\begin{equation}
Y_{m|vw}=\tau^2 \, \exp\left[2  \log(1+Y_{m|vw})\star s\right],
\end{equation}
such that the large $u$ asymptotics is given by:
$$\lim_{u\to \infty} Y_{m|vw} (u) =m^2 \, \(1+\cO(\frac{1}{m})\).$$
The solution of this equation modulo $\frac{1}{m}$ corrections is given by the asymptotic $Y_{m|vw}^{\circ}$ functions.
Thus for large index $Y_{m|vw}$ tend to its asymptotic counterpart.

Now we can turn to investigate the large $Q$ behavior of the ``massive'' $Y_Q$ functions.
The relevant equation to be studied is (\ref{TBAQ}). For our considerations it is worth
to convert it into its $Y$-system form:
\begin{equation}
\frac{Y_Q^+ \, Y_{Q}^-}{Y_{Q-1}\, Y_{Q+1}}=\frac{\Big(1+\frac{1}{Y_{Q-1|vw}} \Big)^2}{(1+Y_{Q-1}) \, (1+Y_{Q+1})}.
\end{equation}
In the large $Q$ limit it becomes:
\begin{equation}
\frac{Y_Q^+ \, Y_{Q}^-}{Y_{Q-1}\, Y_{Q+1}}=\left(1+\frac{1}{Y_{Q-1|vw}^{\circ}}\right)^2. \label{largeQYQ}
\end{equation}
Now, that solution of (\ref{largeQYQ}) should be found which has the properties as follows:
\begin{itemize}
\item The structure and positions of the local singularities within the fundamental strip
\footnote{Here the fundamental strip means $1/g\leq \mbox{Im} \, u \leq 1/g $.} are the same
as those of the asymptotic counterpart $Y_Q^{\bullet}$.
\item The large $u$ behavior of the solution is given by (\ref{BTBA YQ large u}).
\end{itemize}
To satisfy the first requirement and solve the nontrivial part of the $Y$-system equation, we
search the solution of (\ref{largeQYQ}) in the form:
$Y_Q=\sigma_Q \, Y_Q^{\bullet}$, where $\sigma_Q$ is introduced to connect the different 
large $u$ behavior of the exact and asymptotic $Y_Q$s. Then it follows that $\sigma_Q$
must be a zero mode of the LHS of (\ref{largeQYQ}):
\begin{equation}
\frac{\sigma_Q^+ \, \sigma_Q^-}{\sigma_{Q-1} \, \sigma_{Q+1}}=1,
\end{equation}
with the properties as follows:
\begin{itemize}
\item The large $u$ asymptotics is governed by the energy: $\sigma_Q(u)= \xi \, u^{-4 \, E_{\rm BTBA}} \, (1+\dots)$.
\item $\sigma_Q$ is real and even function.
\item $\sigma_Q$ has no zeroes or poles in the fundamental strip.
\end{itemize}
Thus $\sigma_Q$ can be represented as a product of left and right mover modes:
\begin{equation}
\sigma_Q(u)=\xi \, f\Big(u+\frac{i \, Q}{g}\Big) \, \bar{f}\Big(u-\frac{i \, Q}{g}\Big),
\end{equation}

where $\xi$ is a real and coupling dependent constant, the function $f$ has the large $u$ expansion: \newline
$f(u)=u^{-2 \, E_{\rm BTBA}}(1+\dots)$ with dots denoting negligible terms for large $u$ and $\bar{f}$
is the complex conjugate of $f$.
Putting everything together we get that the large $Q$ behavior of $Y_Q$ is given by formula (\ref{YQlarge}).

\section{Solving the $Y\olY$ BTBA}\label{sec:solving BTBA}

The details of numerical computation which yielded the results in Figure \ref{fig:YbarY gnd data} will be given below.

We want to solve the BTBA equations, which consist of a set of nonlinear integral equations. It is convenient to divide the whole problem into two subproblems, 
nonlinear root-finding and numerical integration. The nonlinear part of the problem is solved by relaxed iteration, and 
the integration part by interpolation and extrapolation of the integrand.
Our algorithms are implemented as {\tt Mathematica} scripts which are executed by CPU clusters.

For notational simplicity, we write the BTBA equations (\ref{TBAmvw}-\ref{hybrid}) as
\begin{equation}
\scrL (Y_a) = \scrL (1 \pm Y_b) \star K_{ba} \,,
\label{def:int eq}
\end{equation}
where we introduce the symbol
\begin{alignat}{9}
\scrL (1 \pm Y_a) &= \log \frac{1 \pm Y_a}{1 \pm Y_a^\circ} \,, &\qquad
\scrL (Y_a) &= \log \frac{Y_a}{Y_a^\circ} \,, \qquad (a \neq Q),
\notag \\
\scrL (1 + Y_Q) &= \log (1 + Y_Q) \,, &\qquad
\scrL (Y_Q) &= \log \frac{Y_Q}{Y_Q^\bullet} \,, \qquad (Y_Q^\circ=0).
\label{def:regularized Y}
\end{alignat}
In the first line we take the positive sign for bosonic Y's and the negative sign for fermionic Y's.\footnote{The $Y_\mp$ functions appear also in the form of $\scrL (1 - 1/Y_\mp)$, which should be defined in accordance with \eqref{def:regularized Y}.}
The normalized variables $\scrL (1+Y)$ have better analytic and numerical behavior than $\log (1+Y)$.

\subsection{Algorithm for nonlinear problems}\label{app:alg nl}

Iteration is one of the simplest methods for nonlinear root-finding problems. We solve the equation \eqref{def:int eq} by iteration of one-dimensional integrals as
\begin{equation}
\scrL ( Y_a^{(n+1)} ) = F_a [Y_b^{(n)}] \equiv  \scrL (1\pm Y_{b}^{(n)}) \star K_{ba} \,.
\label{iter subs}
\end{equation}
$Y_a^{(n)}$ is close to the exact solution provided that the initial conditions $Y_{b}^{(0)}$ are appropriate, $n$ is large enough, and that all eigenvalues of the linear infinite-dimensional integration operator $\delta F_a/\delta Y_b$ stay nonzero and inside the unit circle during the iteration.

Slower iteration algorithms are generally more stable. If one wants to get a reasonable solution from inexact initial data, a large number of iteration steps 
are needed. 
Relaxation is an example of slower algorithms, and used to solve nonlinear integral equations in the literature \cite{Dorey:1996re,Beccaria:2010gq}. In the relaxed iteration, instead of \eqref{iter subs} we update the solution by
\begin{align}
\scrL (Y_a^{(n+1)}) &= \mu^{(n)} \scrL (\widetilde Y_a^{(n+1)}) + (1-\mu^{(n)}) \scrL (Y_a^{(n)}) \,,
\label{def:relaxation n} \\[1mm]
\scrL (\widetilde Y_a^{(n+1)}) &\equiv \scrL (1 \pm Y_b^{(n)}) \star K_{ba} \,,
\label{def:tilde Yan}
\end{align}
where $\mu^{(n)}>0$ is a relaxation parameter. For simplicity we choose the same relaxation parameter for all Y-functions.

The updating rule \eqref{def:relaxation n} says $Y^{(n+1)}$ is related to $Y^{(n)}$, $Y^{(n)}$ to $Y^{(n-1)}$ and so on, which is repeated until one reaches $Y^{(0)}$. However, the computation using recursively-defined variables demands large memory. Thus we use the updating rule \eqref{def:relaxation n} only for the first $\rho$ steps, and use a truncated rule later on:
\begin{alignat}{9}
\scrL (Y_a^{(n+1)}) &= (1-\mu^{(n)})^{n+1} \, \scrL (Y_a^{(0)}) +
\mu^{(n)} \sum_{k=0}^{n} (1-\mu^{(n)})^k \, \scrL (\widetilde Y_a^{(n+1-k)})
&\qquad &(0 \le n \le \rho),
\label{recursive relax} \\[1mm]
\scrL (Y_a^{(n+1)}) &= (1-\mu^{(n)} )^\rho \, \scrL (\widetilde Y_a^{(n+1-\rho)})
+ \mu^{(n)} \sum_{k=0}^{\rho-1} (1-\mu^{(n)})^k \, \scrL (\widetilde Y_a^{(n+1-k)})
&\qquad &(n \ge \rho+1).
\label{non-recursive relax}
\end{alignat}
We used $\rho=4$ or 5.

The Y-functions should be updated carefully in the beginning because they change a lot after one step of iteration. This means that the relaxation parameter should be small. In fact, when we plot the energy at each step of iteration, we find that the energy typically increases for the first few steps, and then decreases monotonically. We used $0.1 \le \mu \le 0.3$ at the beginning of iteration, $0.3 \le \mu \le 0.75$ in later steps, depending on $(g,L)$.

\subsection{Algorithm for integration}\label{app:alg int}

Our next problem is to evaluate the set of one-dimensional integrals \eqref{def:tilde Yan}. 
In numerical analysis, one needs to approximate integrals by finite sums by choosing an appropriate distribution of sampling points $\{ t_1 \,, t_2 \,, \dots t_{N_p} \}$.

If one wants to achieve the best precision at a fixed number of sampling points $N_p$\,, one should look for the best distribution of sampling points $\{ t_i \}$ for each integral. 
However, since \eqref{def:tilde Yan} consists of numerous one-dimensional integrals, it is impractical to construct different sampling points $\{ t_i \}$ for different integrals.

One solution is to choose different sampling points for different Y-functions. With this method, we
construct a fitting function $\scrL (Y_b^{\rm (fit)} (t))$ based on $\{t_i\}$ and use it to compute the integrals \eqref{def:tilde Yan}. This method is similar to the one used in \cite{Frolov:2010wt}, and has the advantage that we can easily keep track of the explicit shape of the Y-functions.

Let us explain our numerical integration scheme in detail. For each $Y_b$ at fixed $(g,L)$ we introduce a rapidity cutoff $M_b$\,. Then we construct a piecewise continuous interpolation function for $t \le M_b$ and an extrapolation function for $t \ge M_b$, and combine them together as
\begin{equation}
\scrL (Y_b^{\rm (fit)} (t)) = 
\Bigl\{ \scrL (Y_b^{\rm (in)} (t)) \ \Big| \ t \in [0, M_b ] \Bigr\} \ \cup \ 
\Bigl\{ \scrL (Y_b^{\rm (ex)}(t)) \ \Big| \ t \in (M_b \,, +\infty) \Bigr\},
\label{Y inter/extra}
\end{equation}
and compute the integral $\scrL (1+Y_b^{\rm (fit)}) \star K_{ba} (v)$ through
\begin{equation}
\scrL (1 \pm Y_b^{\rm (fit)} (t) ) = \log \( \frac{1 \pm Y_b^\circ (t) \exp \, \bigl[ \scrL (Y_b^{\rm (fit)} (t)) \bigr] }{1 \pm Y_b^\circ (t) } \).
\end{equation}
Recall that all Y-functions are even under the parity transformation $t \mapsto -t$. As for $Y_\mp$ functions we only need the interpolation function for $t \in [0,2]$. We also construct fitting functions for some kernels in the hybrid BTBA equation \eqref{hybrid}, namely the dressing phase kernel $K^\Sigma_{Q'Q} (t,v)$ and $s\star K^{Q'-1,\,Q}_{vwx} (t,v)$ to accelerate computation.

We used the third-order spline to obtain $\scrL (Y_b^{\rm (in)} (t))$ from the data points $\{ Y_b (t_i) \}$ for $t_i \le M_b$\,. The extrapolation was constructed via the ansatz
\begin{equation}
\scrL (Y_b^{\rm (ex)}(t)) = \sum_{i=1}^{n_p} \frac{c_b^{(i)}}{t^i} \quad (b=M|vw, M|w), \qquad
\scrL (Y_Q^{\rm (ex)}(t)) = c_Q^{(-1)} \, \log (t) + \sum_{i=0}^{n_p-2} \frac{c_Q^{(i)}}{t^i} \,.
\label{YbQ extern}
\end{equation}
The order of extrapolation $n_p$ should not be too large, as the extrapolation tends to oscillate wildly around the cutoff $t \gtrsim M_b$\,. We mostly used $n_p=3$ or 4.

The rapidity cutoff $M_b$ is determined as follows.
The fitting Y-functions should be a good approximation of the actual Y-functions as far as the number of sampling points $N_p$ is sufficiently large. Since we know that the normalized Y-functions $\scrL (Y_b(t))$ go to zero for at large $t$, we want to choose the value of rapidity $t=M_b$ such that $\abs{ \scrL (Y_b(t)) }$ decays monotonically for $t > M_b$.
To estimate a good choice of $M_b$ we define the ``width'' of Y-function $W_b$ by
\begin{equation}
Y_Q (W_Q) = \frac{1}{20} \, \mathop {\rm Max} \limits_{u \in \bb{R}} \[ Y_Q (u) \], \quad
Y_{M|vw} (W_{M|vw}) = \frac{19}{20} \, M (M+2), \quad
Y_{M|w} (W_{M|w}) = \frac{21}{20} \, M (M+2).
\label{def:width Y}
\end{equation}
If there are multiple solution to these equations, we use the largest one as the width. 
Then we choose the rapidity cutoff within the range from $W_b \lesssim M_b \lesssim 8 W_b$\,.\footnote{$M_Q > W_Q$ means that the numerical coefficient in \eqref{def:width Y} is smaller than $\frac{1}{20}$\,, and similarly for other $W_b$\,.}

The distribution of sampling points for a bosonic Y-function is determined from its width.\footnote{As for $Y_\mp (t)$ the uniform distribution over $t \in [0,2]$ is used.}
Concretely, we used the following semi-uniform distribution for $t \in [0, W_b]$:
\begin{equation}
\Delta t_k = \Biggl\{ \frac{W_b}{4 N_p} \ \(0 \le t < \frac{W_b}{16} \) \Biggr\} \ \cup \ 
\Biggl\{ \frac{3 W_b}{4 N_p} \ \(\frac{W_b}{16} \le t < \frac{W_b}{4} \) \Biggr\} \ \cup \ 
\Biggl\{ \frac{3 W_b}{2 (N_p - 2n_p)} \ \(\frac{W_b}{4} \le t < W_b \) \Biggr\},
\label{distribute in}
\end{equation}
where $\Delta t_k = t_{k+1} - t_k$ is the distance between the two adjacent sampling points.
There are $(N_p-n_p)$ points in total for $0 \le t < W_b$\,.
The outermost $n_p$ points are used to construct the extrapolation,
\begin{equation}
t_k = W_b \times 2^{3/4 (k-N_p+n_p)} \qquad (k=N_p-n_p, \dots N_p-1).
\label{distribute ex}
\end{equation}
For small $g$ we used $t_k = W_b \times 2^{(k - N_p + n_p)}$ for $Y_Q$\,.

\subsection{Numerical parameters}

For clarity the numerical value of various cutoff parameters is given below.

We must truncate the number of Y-functions appearing in the BTBA equations. Let us denote the index cutoff for $Y_Q \,, Y_{M|vw} \,, Y_{M|w}$ by $Q_{\rm max} \,, M_{vw|{\rm max}} \,, M_{w|{\rm max}}$\,, respectively. We used
\begin{equation}
Q_{\rm max} = 6, \quad M_{vw|{\rm max}} = 14, \quad M_{w|{\rm max}} = 10.
\end{equation}
The Y-functions beyond the index cutoffs are fixed at the asymptotic values.

Recall that $N_p$ is the number of total sampling points used to construct the fit of a Y-function, and $n_p$ is the number of sampling points greater than the width as in \eqref{distribute ex}. For the first argument of the kernels $K^\Sigma_{Q'Q} (t,v)$ and $s\star K^{Q'-1,\,Q}_{vwx} (t,v)$,
we constructed suitable distributions of sampling points in a manner similar to Appendix \ref{app:alg int}. 
As for $(N_p, n_p)$, the following values were used for each distribution:\footnote{Usually the outermost 3 or 4 points are not used, because the optimal order of extrapolation is $n_p=3$ or 4.}
\begin{gather}
(N_p, n_p) = (36, 8) \ \ {\rm for} \ \ Y_{Q \ge 1} \,,\qquad
(N_p, n_p) = (128,8) \ \ {\rm for} \ \ Y_{M|vw} \,, Y_{M|w} \,, Y_\mp \,,
\label{cutoff YbarY} \\
(N_p, n_p) = (128, 8) \ \ {\rm for} \ t \ {\rm in} \ s * K_{vwx}^{Q'-1,Q} (t,v),
\notag \\
(N_p, n_p) = (128, 16) \ \ {\rm for} \ t \ {\rm in} \ K^\Sigma_{11} (t,v),
\qquad
(N_p, n_p) = (64, 8) \ \ {\rm for} \ t \ {\rm in} \ K^\Sigma_{Q'Q} (t,v) \quad (Q'Q>1).
\notag
\end{gather}

For small $g$ we start iteration from the asymptotic Y-functions. This part of the computation is completely parallelizable. However, as $g$ increases the finite-size corrections get larger, and one needs to start iteration from the solution at the previous step, i.e. smaller $g$.\footnote{For example, our iteration started from the asymptotic solution for $g \le 2.4$ and $L=2$.}  The total number of iteration steps is typically 20 to 40 depending on $(g,L)$.

With this parameter choice and using a node of CPU clusters with 48 cores, it took around 90 minutes to generate the dressing kernel data, and 6 hours to finish 20 steps of iteration.

\paragraph{Error estimates.}

There are three sources of errors:
(i) truncation of BTBA by a finite number of Y-functions,
(ii) finite number of iterations,
(iii) discretization of the integrals.

The first source of errors is significant around $g=g_{\rm cr}$\,, as represented by the huge error bars in Figure \ref{fig:YbarY gnd data}.

The second source of errors makes our results unreliable at the third or fourth digits. The larger number of iterations does not always indicate the more precise results, because errors may accumulate during iterations.

The third source of errors is negligible compared to the first two.
We carefully choose the distribution of sampling points for each Y-function at each $(g,L)$, and set {\tt PrecisionGoal} no less than 6 in computing various integrals in {\tt Mathematica}.

\subsection{Table of numerical results}\label{sec:raw data}

In Table \ref{tab:raw} we present the numerical results drawn in Figure \ref{fig:YbarY gnd data}, in which the top end of error bars corresponds to raw data and the bottom end of error bars to the fitted data.

The raw data ($Q_{\rm max}=6$) are not sensitive to the large $Q$ singularity \eqref{energy EQ}, and thus contain the points $g > g_{\rm cr}$\,. When the raw data hits the large $u$ singularity \eqref{open energy lower bound}, we cannot compute the finite BTBA energy further in a reliable way. Moreover, when the energy is close to $\frac14 - L$, we always find that ${\bf E}(Q)$ defined in \eqref{EQ} for different $Q$'s have comparable order of magnitude. 

Note that our data also include some points with $E_{\rm BTBA} < \frac14-L$, because we determined the large $u$ behavior of $Y_Q(u)$ by fitting the numerical data instead of using \eqref{BTBA YQ large u}.

\begin{table}[t]
\begin{center}
\small
\begin{tabular}{ccc}
\multicolumn{3}{c}{$L=\frac32$} \\\hline
$g$ & $E_{\rm BTBA}^{\rm (data)}$ & $E_{\rm BTBA}^{\rm (fit)}$ \\\hline
1.3&$-3.39\times 10^{-1}$&$-4.53\times 10^{-1}$\\
1.4&$-4.92\times 10^{-1}$&$-$\\
1.45&$-5.95\times   10^{-1}$&$-$\\
1.5&$-7.30\times 10^{-1}$&$-$\\
1.54&$-8.82\times 10^{-1}$&$-$\\
1.56&$-9.88\times   10^{-1}$&$-$\\
1.58&$-1.16$&$-$\\
1.59&$-1.35$&$-$\\
1.592&$-1.44$&$-$\\
1.594&$-1.59$&$-$\\\hline
&&\\[62.5mm]
\end{tabular}
\hspace{4mm}
\begin{tabular}{ccc}
\multicolumn{3}{c}{$L=2$} \\\hline
$g$ & $E_{\rm BTBA}^{\rm (data)}$ & $E_{\rm BTBA}^{\rm (fit)}$ \\\hline
0.6&$-1.02\times 10^{-3}$&$-1.02\times 10^{-3}$\\
0.8&$-5.85\times 10^{-3}$&$-5.85\times   10^{-3}$\\
1.&$-1.97\times 10^{-2}$&$-1.97\times 10^{-2}$\\
1.2&$-4.79\times 10^{-2}$&$-4.79\times 10^{-2}$\\
1.4&$-9.45\times   10^{-2}$&$-9.46\times 10^{-2}$\\
1.6&$-1.62\times 10^{-1}$&$-1.62\times 10^{-1}$\\
1.8&$-2.61\times 10^{-1}$&$-2.63\times   10^{-1}$\\
2.&$-4.40\times 10^{-1}$&$-4.59\times 10^{-1}$\\
2.1&$-5.41\times 10^{-1}$&$-5.86\times 10^{-1}$\\
2.2&$-6.62\times 10^{-1}$&$-7.83\times   10^{-1}$\\
2.3&$-8.13\times 10^{-1}$&$-$\\
2.4&$-1.00$&$-$\\
2.5&$-1.29$&$-$\\
2.55&$-1.53$&$-$\\
2.56&$-1.61$&$-$\\
2.565&$-1.68$&$-$\\
2.57&$-1.77$&$-$\\
2.572&$-1.86$&$-$\\
2.5722&$-1.87$&$-$\\
2.5724&$-1.89$&$-$\\
2.5726&$-1.90$&$-$\\
2.5728&$-1.94$&$-$\\
2.573&$-2.01$&$-$\\
2.5732&$-2.58$&$-$\\\hline
\end{tabular}

\vskip 6mm
\begin{tabular}{ccc}
\multicolumn{3}{c}{$L=\frac52$} \\\hline
$g$ & $E_{\rm BTBA}^{\rm (data)}$ & $E_{\rm BTBA}^{\rm (fit)}$ \\\hline
3.2&$-9.19\times 10^{-1}$&$-9.61\times 10^{-1}$\\
3.4&$-1.24$&$-$\\
3.6&$-1.63$&$-$\\
3.65&$-1.78$&$-$\\
3.7&$-2.00$&$-$\\
3.71&$-2.06$&$-$\\
3.72&$-2.13$&$-$\\
3.73&$-2.23$&$-$\\
3.74&$-2.64$&$-$\\\hline
&&\\[14.5mm]
\end{tabular}
\hspace{4mm}
\begin{tabular}{ccc}
\multicolumn{3}{c}{$L=3$} \\\hline
$g$ & $E_{\rm BTBA}^{\rm (data)}$ & $E_{\rm BTBA}^{\rm (fit)}$ \\\hline
2.4&$-1.62\times 10^{-1}$&$-1.62\times 10^{-1}$\\
2.8&$-2.86\times 10^{-1}$&$-2.86\times 10^{-1}$\\
3.2&$-4.50\times   10^{-1}$&$-4.51\times 10^{-1}$\\
3.6&$-6.69\times 10^{-1}$&$-6.74\times 10^{-1}$\\
3.8&$-8.00\times 10^{-1}$&$-8.09\times   10^{-1}$\\
4.&$-9.57\times 10^{-1}$&$-9.78\times   10^{-1}$\\
4.2&$-1.14$&$-1.18$\\
4.4&$-1.31$&$-1.42$\\
4.6&$-1.58$&$-$\\
4.8&$-1.87$&$-$\\
4.9&$-2.07$&$-$\\
5.& $-2.44$&$-$\\
5.03&$-2.73$&$-$\\\hline 
\end{tabular}

\bigskip
\caption{Numerical data for the BTBA energy of the $Y\olY$ ground state with the R-charge $L$. The symbol `$-$' means that the extrapolated energy hits the lower bound and thus diverges.}
\label{tab:raw}
\end{center}
\end{table}
\normalsize

\clearpage

\end{fmffile}

\bibliographystyle{JHEP}
\bibliography{refs_all}

\end{document}